\documentstyle[12pt,epsf]{article}
\title
{\large {\bf
The diagonal spin basis and calculation of processes \\involving
polarized particles \thanks{Published in: Physics of
Particles and Nuclei, {\bf 29} (1998) 469;  Fiz. Elem. Chastits
At. Yadra {\bf 29} (1998) 1133 }} }
\author {M.V. Galynsky \thanks{\bf E-mail: galynski@dragon.bas-net.by}
~ and S.M. Sikach \\  {\it Stepanov Institute of Physics,
Belarusian Academy of Sciences, Minsk} }
\date{}
\begin{document}
\def\eqnum#1{\eqno (#1)}
\maketitle

\begin{abstract}
The review of recently developed by the authors new techniques for covariant
calculation of matrix elements in QED, the so-called formalism of "Diagonal
Spin Basis" (DSB), is  presented. In DSB spin 4-vectors of {\em in-}\ and
{\em out-} fermions are expressed just in terms of their 4-momenta. In this
approach the little Lorentz group, common for the initial and final states,
is realized. This brings the spin operators of {\em in-}\ and {\em
out-} particles to coincidence, allowing to separate in a covariant way
the interactions with and without change of the spin states of the particles
involved in the reaction and to follow in details the
whole dynamics of the spin interactions. In contrast to methods of
CALCUL group and others, the developed approach is valid for both massive
fermions and massless ones. It is not necessary to introduce auxiliary
vectors in DSB. Just 4-momenta of particles participating in reactions are
required in it to construct the mathematical apparatus for calculations
of matrix elements. We apply this formalism to the following processes:
1) M\"{o}ller and Bhabha bremsstrahlung ($e^{\pm}e^{-} \to e^{\pm}e^{-}
\gamma $) in the ultrarelativistic (massless) limit when initial particles
and photon are helicity polarized; 2) Compton back-scattering of photons
of intensive circularly polarized laser wave focused on a beam of
longitudinally polarized ultrarelativistic electrons ($ e + n \gamma_{0}
\to e + \gamma $); 3) $e^{+} e^{-}$-pair production by a hard photon
in simultaneous collision with several laser beam photons ($\gamma  + n
\gamma_0 \to e^{+} + e^{-}$); 4) Bethe-Heitler process in the case of a
linearly polarized photon emission by an electron with account for proton
recoil and form factors; 5) the reaction $ e p \to  e p \gamma $ with
proton polarizability being taken into account in the kinematics when
proton bremsstrahlung dominates; 6) orthopositronium 3-photon annihilation
($ e^{+} e^{-} \to 3 \gamma $). The results obtained with the help of the
developed DSB-formalism certify its efficiency for calculating of
multiparticle processes when polarization is to be taken into account.
\end{abstract}

\section*{\Large \bf{Introduction}}

By now, the physics of spin phenomena has become an essential component of
the research program at many large accelerators of new generation [1-5].
This is a consequence of, first, the successful development of the
polarization technique, in particular, methods of obtaining polarized beams
and advances in the construction of polarized targets and polarimeters [3].
Second, the electroweak interactions play an important role in the energy
range of the current accelerators. They violate both $P$ and $C$ parity
[6,7], and also combined $CP$ invariance [8]. This violation manifests itself
in polarization effects, which are often used as precision tests of the
Standard Model with accuracy unattainable in other experiments [1,5]. Third,
it is necessary to go beyond the Standard Model to seek new particles and
new types of interactions, and here spin can play a very important role.

The advances in accelerator and polarization techniques have revealed new
possibilities in the study of polarized particle interaction processes.
Therefore, it is becoming more and more important to calculate theoretically
the probabilities for various elementary-particle interaction processes,
taking into account the particle polarizations and internal structure, and
also to develop new computational tools. When the standard approach [9-12]
is used to calculate the probabilities for various processes (i.e., to
calculate the squared moduli of matrix elements), the inclusion of the
particle polarizations greatly complicates both the calculations themselves
and the structure of the expressions obtained. Their covariance is often
lost.

A natural way to simplify the calculations for reactions involving polarized
particles is to calculate not the squared moduli of matrix elements, but
instead the matrix elements themselves. This can be done in several different
ways [13]. One way is to use the explicit form of the fundamental matrices
and state functions written in a particular basis of the representation space
of the Lorentz group in which they are defined. A noncovariant approach of
this type has already been used for spin-1/2 particles by Powell [14] in
1949. The general theoretical development of this method is due to Sokolov
[15]. This method continues to be used successfully to this day [16,17],
owing to the appearance of powerful computer programs for analytic
calculations.

However, the most widely used method for calculating the matrix elements
of QED processes is the covariant method not involving the use of the explicit
matrices and wave functions. It was proposed in 1961 independently by
Bellomo [18] and by Bogush and Fedorov [19]. This approach is based on the
method of projection operators in elementary-particle theory developed by
Fedorov [20].

The Bellomo method uses a trick which amounts to multiplying the matrix
elements $ M_{31}=\overline{\Psi}_{3} Q \Psi_{1}$ of the transition from
the initial state ($\Psi_{1}$) to the final state ($\Psi_{3}$),
where $Q$ is the interaction operator, by the quantity $ \overline{\Psi}_{1}
Z\Psi_{3} / \overline{\Psi}_{1} Z \Psi_{3}$, so that the amplitude $M_{31}$
can be reduced to calculation of a trace:\footnote{The indices have been
chosen in accordance with future application of the results to the reaction
$1 +2 \to 3 +4$}
$$
M_{31} = Tr ( P_{31} Q )\; , \; P_{31} = \Psi_{1} \; \overline{\Psi}_{3} \; ,
\eqnum {1}
$$
$$
P_{31} = \tau_{1} Z \tau_{3} / (\mid \overline{\Psi}_{1} Z \Psi_{3} \mid e^{i
\phi})\; , \; \overline{\Psi}_{1} Z \Psi _{3} = \mid \overline{\Psi}_{1} Z
 \Psi_{3} \mid e ^{ i \phi } \; .
\eqnum {2}
$$
Here $\tau_{1}$ and $\tau_{3}$ are the projection matrixs-diadics of the
initial and final states [20]: $ \tau_{i} = \Psi_{i} \; \overline{\Psi}_{i} ,
\; ( i = 1, 3) $. The operator $Z$ in (2) is arbitrary. In Ref. 18 it was
chosen to be: $Z = 1$. In recent years, the greatest progress in the
development of the Bellomo method (in the ultrarelativistic, massless case)
has been made by the CALCUL group [21]. The achievements of this
group are widely recognized and extensively used by scientists all over the
world. The CALCUL method has been generalized to fermions with nonzero mass
in Refs. 22 and 23, but this generalization requires the introduction of
additional vectors unrelated to the kinematics of the process under study,
making it inconvenient to use.

In the method proposed in Ref. 19, the operator $P_{31} = \Psi_{1} \;
\overline{\Psi}_{3} $ is constructed on the basis of a complex vector
parametrization of the Lorentz group [24-26] and the operators $T_{31}$
of representations of this group in the space of particle wave functions
[27,28], which play the role of operators for transitions from the initial
to the final state: $\Psi_{3} = T_{31} \Psi_{1} \; , \; \overline{\Psi}_{3} =
\overline{\Psi}_{1} T_{31}^{-1}$. Here the operator $P_{31} = \Psi_{1} \;
\overline{\Psi}_{3}$ is written as [13,19]:
$$
P_{31} = \Psi_{1} \; \overline{\Psi}_{3} = \tau_{1} T_{31} ^{-1} = T_{31}
 ^{-1} \tau_{3} \; .
\eqnum {3}
$$
This version was originally developed for longitudinally polarized Dirac
particles [29]. It was developed further by Fedorov [30-32] and his students
(see Ref. 13 and references therein). In principle, the method developed by
Fedorov (Ref. 13, Sec. 36) allows analytic expressions to be obtained for
the matrix elements of various QED processes for arbitrarily polarized Dirac
particles, either massive or massless, which is the main, decisive advantage
of this method over that of the CALCUL group. However, the striving for
generality is not always consistent with efficiency of the approach.

For a number of QED problems, the development of the approach of Refs. 13
and 19 for calculating matrix elements of multiparticle processes is largely
the result of progress in developing covariant methods of describing the
spin properties of two-particle systems based on the use of a vector
parametrization of the Lorentz little groups [13,37].

At the present time, the helicity basis introduced by Jacob and Wick [33]
is very popular in high-energy physics. This is a consequence of the
simplicity of the physical interpretation of helicity (the spin projection
on the direction of the particle momentum), the fact that the center of mass
of the system is distinguished in the helicity basis, and also the fact that
the helicity amplitudes admit a simple partial-wave analysis using the
$SO(3)$ group [33]. In addition, studying the helicities of moving particles
is analogous to studying the spins of particles at rest [13]. However, there
are several important factors which prevent helicity from playing the dominant
role in describing the spin projection of particles. One is that the helicity
is not a particle characteristic which is invariant under Lorentz
transformation [9,13]. Nevertheless, in the literature one can find articles
with titles like "A Covariant Method for Calculating Helicity Amplitudes"
(Ref. 34). In interpreting the dynamics of the spin interaction, amplitudes
with and without change of the particle helicity are often referred to as
amplitudes with and without spin flip. However, since the particle momentum
is changed by the interaction, it is clear that such a classification is
very arbitrary. Both types of amplitude actually describe a process with
a change in the particle spin state.

Many of these difficulties can be avoided for a particular choice of spin
basis of a reaction, namely, the diagonal spin basis (DSB), in which the
spin 4-vectors $s_{1}$ and $s_{3}$ of particles with 4-momenta $p_{1}$ and
$p_{3}$ ($s_{1} p_{1} = s_{3} p_{3} = 0 , s_{1} ^{2} = s_{3} ^{2} = - 1 $)
belong to the hyperplane formed by the 4-vectors $p_{1}$ and $p_{3}$
(Refs. 35 and 36):
$$
s_{1} = - \; { (v_{1} v_{3}) v_{1} - v_{3} \over \sqrt{ ( v_{1}v_{3}
)^{2} - 1 }} \; \; , \; \; s_{3} =  { ( v_{1} v_{3}) v_{3} - v_{1}
\over \sqrt{ ( v_{1}v_{3} )^{2} - 1 }} \; \; ,
\eqnum {4}
$$
where $v_{1} = p_{1}/m_1$ and $ v_{3} = p_{3}/m_3 $. The spin vectors (4)
obviously do not change under transformations of the Lorentz little group
common to particles with 4-momenta $p_{1}$ and $p_{3}$ [37]:
$L_{p_1,p_3} p_1 =p_1, L_{p_1,p_3} p_3 =p_3$. We note that it will be a
one-parameter subgroup of the rotation group with axis whose direction is
determined by the vector [13,37]:
$$
\vec a = c \; ( \vec p_{1}/p_{10} - \vec p_{3}/p_{30} ) \; ,
\eqnum {5}
$$
where $c$ is an arbitrary real number. The direction of $\vec a$ (5)
possesses the property that the projections of the spins of both particles
on it will have definite values even when the particles have different masses.
Therefore, the DSB naturally makes it possible to describe the spin states
of systems of any two particles (including ones with different masses) by
means of the spin projections on the single common direction
given by the vector (5).\footnote{The geometrical image of the difference
of two vectors is the diagonal of a parallelogram, hence the name "diagonal
spin basis" given by Fedorov.}

The fundamental fact that the Lorentz little group common to particles with
momenta $p_{1}$ and $p_{3}$ is realized in the DSB leads to a number of
remarkable consequences [35-42]. First, in this basis particles with
4-momenta $p_{1}$ (before the interaction) and $p_{3}$ (after the
interaction) have the same spin operators [38-40], which allows the covariant
separation of the interactions with and without change of the spin states
of the particles involving in the reaction, making it possible to trace the
dynamics of the spin interaction.

Second, in the DSB (4) the mathematical structure of the amplitudes is
maximally simplified, owing to the coincidence of the particle  spin operators,
the separation of Wigner rotations from the amplitudes [35,36], and the
decrease in the number of various scalar products of 4-vectors which
characterize the reaction. Third, in the DSB the spin states of massless
particless ($p_{1}^{2} = p_{3}^{2} = 0$) coincide up to a sign with the
helicity states [40-42].

Use of the DSB does not lead to loss of generality, because the transformation
to an arbitrary spin basis is carried out by means of Wigner $D$ functions
[43]. In the new expressions for the amplitudes, the original amplitudes
give the best representation of the dynamics of spin phenomena, and the $D$
functions are purely kinematical in nature.

Therefore, the DSB reveals new possibilities for developing methods to
calculate matrix elements and increasing the efficiency of such methods when
the Bogush-Fedorov approach is used [13,19].

The first calculation of matrix elements in the DSB was performed in Refs.
35 and 36, using the spinor formalism. The amplitudes were calculated for
the complete set of Dirac matrices $\Gamma_{i} \; (i =1,2 \ldots 16 )$
in which the arbitrary operator $Q$ entering into (1) is expanded. We also
note that the methods proposed in Refs. 18 and 19 are quite closely related,
as was first shown in [44], where the various methods of calculating matrix
elements were classified.

\section*{\Large \bf Notation and abbreviations }

$\vec{x}=(x_{a}) $ is a three-dimensional vector, and $ x_{a} \; (a=1,2,3)$
are its components. \\
$p=(p^{k}) = (p_{0}, \vec{p})$ is a four-dimensional vector in
Minkowski space. \\
$\vec{x} \vec{y} = x_{1}y_{1}+x_{2}y_{2}+x_{3}y_{3}$ is the
scalar product of the vectors $\vec{x}$ and $\vec{y}$. \\
$p_1 p_2 = p_{1\mu} p_{2\mu} =p_{10} p_{20} - \vec{p_1} \vec{p_2}$ is
the scalar product of the 4-vectors $p_1$ and $p_{2}$. \\
$ [\vec{x} \vec{y} ]$ is the vector product of the three-dimensional
vectors $\vec{x}$ and $\vec{y}$. \\
$\varepsilon_{abc}$ is the three-dimensional Levi-Civita symbol. \\
$[\vec{x} \vec{y}]_{a} = \varepsilon_{abc} x_{b} y_{c}$. \\
$ ( \vec{c} ) ^ {\times} _{ab} = \varepsilon_{adb} \; c_{d} , \;
\vec{c} ^ {\times}  \vec{x} = [ \vec{c} \; \vec{x} ] \; ,
\; \vec{x} \vec{c} ^ {\times} = \; [ \vec{x} \; \vec{c} ] $ . \\
$ \vec{x} \cdot \vec{y} = ( x_{a} y_{b} )$ is the dyadic formed from the
vectors $\vec{x}$ and $\vec{y}$ . \\
$ x \cdot y = ( x _{\mu} y _{\nu} ) $ is the dyadic formed from the 4-vectors
$ x = ( x_{\mu} )$ and $y = ( y_{\nu} ) $. \\
$ ( \vec{x} \cdot \vec{y} ) = \vec{x} \cdot \vec{y} + \vec{y} \cdot \vec{x}
\; , \; ( x \cdot  y ) = x \cdot y + y \cdot x $ are symmetrized dyadics. \\
$ [ \vec{x} \cdot \vec{y} ] = \vec{x} \cdot \vec{y} -\vec{y} \cdot \vec{x}
\; , \;  [ x \cdot y ] = x \cdot y - y \cdot x $ are alternating dyadics. \\
$ (\alpha^ {\times})_{\mu \nu} = 1/2 \;  \varepsilon_{\mu \nu \rho \sigma}
\alpha^{\rho \sigma} \; , \; \alpha^{\mu \nu}=-\alpha^{\nu \mu} \; ,
\; (\tilde \alpha = - \alpha ) .$ \\
$ ( [ a \cdot b ] ^{\times})_{\mu \nu}= \varepsilon_{\mu \nu \rho \sigma}
a^{\rho} b^{\sigma} .$ \\
$ [a , b , c]_{\mu} = ( [ a \cdot b ]^{\times} c)_ {\mu} =
\varepsilon_{\mu \nu \rho \sigma} a^{\nu} b^{\rho} c^{\sigma} .$ \\
$g_{\mu \nu}$ is the metric tensor in Minkowski space with
signature $(+, -, -, -)$ \\
$\varepsilon _{\mu \nu \rho \sigma } $ is the four-dimensional Levi-Civita
symbol, $\varepsilon _{0123}= - 1 $.    \\
$\gamma^{\mu} $ are the Dirac matrices , $\; \hat a = a_{\mu} \gamma^{\mu}
= (\gamma  a), \\
\gamma^{5} = -i\;\gamma^{0} \gamma^{1} \gamma^{2} \gamma^{3} \; ,
 \gamma^{5+} = \gamma^{5} .$ \\
The algebra of the Dirac matrices is: $\hat a \hat b + \hat b \hat a = 2 ab.\\
\hat a \hat b \hat c = \hat d -i \gamma^{5} \hat f,\; d=(ab +[a \cdot b])c,
\; f=( [a \cdot b])^{\times}c$.

\noindent
For algebraic operations we use the notations: \\
$ \ast \; \;  $ for complex conjugation. \\
$ + \;   $ for Hermitian conjugation. \\
$ \sim  \; $ for the transpose. \\
$ \times  \; $ for the dual \\
$ \cdot  \; \; \; $ for the dyadic product. \\

\noindent
QED stands for quantum electrodynamics. \\
DSB stands for diagonal spin basis. \\
OVB stands for orthonormal vector basis. \\
RCS stands for real Compton scattering. \\
VCS stands for virtual Compton scattering. \\
CBS stands for Compton back-scattering. \\

\noindent
Everywhere we use the system of units in which the speed of light $c$ and
Planck's constant $\hbar$ are equal to unity: $ c = \hbar = 1$ .

\section{\bf Spin operators in the DSB }

We shall use the following approaches in describing the spin properties of
particles: (a) the approach proposed by Bargmann and Wigner, in which the
spin-projection operators are determined by using the generators of the
Lorentz little groups. These are known in the literature as the
Pauli-Bargmann-Lyubanskii operators [43,45]. (b) The covariant spin theory
developed by Fedorov on the basis of vector parametrization of the Lorentz
little groups and their representations [13]. These approaches are
essentially equivalent. However, vector parametrization
of the Lorentz group not only allows simplification of the theory of the spin
properties of elementary particles, but also disposes of (see Ref. 13)
commonly encountered, incorrect statements about some approaches [9,10],
such as "for a given momentum the spin projection on an arbitrary axis
cannot have a definite value".

We shall start from the fact that in momentum space the free state of a
particle with 4-momentum $p$ and spin projection $\delta$ on the $\vec c$
axis is described by the state vector $|p,\delta>$ (we drop the indices
denoting the spin $j$, the mass $m$, and other particle characteristics).
The particle spin $j$ is defined as the angular momentum in the rest frame,
where the orbital angular momentum is zero. It is therefore convenient to
define the state vector $|p,\delta>$ in terms of the state vector in the
rest frame $|p^0,\delta>$, where $p^0 = (m,0)$. Here we shall assume that
the vector on which the spin is projected (i.e., the axis of spin projections
$\vec c$) in the particle rest frame is the spatial part of the spin 4-vector
$s^0 = (0, \vec c)$, satisfying the conditions: $s^0 p^0 =0, s^{02} = -
\vec c~^{2}= -1$. Let $\Lambda_{p}$ be a boost, i.e., a Lorentz transformation
such that $p = \Lambda_{p} \; p^0$, $s=\Lambda_{p} s^0, sp=0, s^2= -1,
s = (s_0, \vec s)$, where
$$
\vec s = \left( 1 + { \vec p \cdot \vec p \over m ( p_{0} + m )} \right)
 \; \vec c  \; , \; s_{0} \; = \; {\vec p \vec c \over m} \; .
\eqnum {1.1}
$$
Then
$$
|p,\delta> = T_p \; |p^0,\delta> \; \; ,
\eqnum {1.2}
$$
where $T_p = T(\Lambda_p)$ is the operator for this transformation acting
in the space of state vectors. The state vector $|p,\delta>$ satisfies the
equations:
$$
P^{\mu} \; |p,\delta> = p^{\mu} \; |p,\delta>\; ,
\eqnum {1.3}
$$
$$
\sigma \; |p,\delta> = \delta \; |p,\delta> \; ,
\eqnum {1.4}
$$
$$
w^2 \; |p,\delta> = -j(j+1) \; |p,\delta> \;.
\eqnum {1.5}
$$
Here $P^{\mu}$ and $\sigma$ are the energy-momentum and spin-projection
operators:
$$
\sigma = s^{\mu} \; w_{\mu}\; ,
\eqnum {1.6}
$$
where $w^{\mu}$ is the Pauli-Lyubanskii 4-vector [43]:
$$
w_{\mu} = - \; {1 \over 2m} \; \varepsilon _{\mu \nu \lambda \rho} \;
M^{\nu \lambda} \; p^{\rho} \;,
\eqnum {1.7}
$$
and $M^{\nu \lambda}$ are the angular-momentum operators. Using (4),
(1.6), and (1.7), we find that the spin-projection operators for the initial
and final particles $\sigma_1 = ws_1$ and $\sigma_3 = ws_3$ in the DSB (4)
coincide and have the form [38,39]:
$$
\sigma_1 = \sigma_3 = {1 \over 2 \sqrt{(v_1 v_3)^2 - 1}} \;
\varepsilon_{\mu \nu \rho \sigma } M^{\mu \nu} v_1^{\rho} v_3^{\sigma} \; .
\eqnum {1.8}
$$
It should be noted that in any other basis different from the diagonal one,
the operators $\sigma_1$ and $\sigma_3$ do not coincide and, therefore,
do not commute with each other.

The requirement that the Lorentz little groups coincide for particles with
momenta $p_{1}$ and $p_{3}$ imposes rigorous constraints not only on the
 choice of particle spin vectors $s_{1}$ and $s_{3}$, but also on the
spin-projection axes $\vec c_{1}$ and $\vec c_{3}$ (see (1.1)). As was shown
in Refs. 35 and 36, $\vec c_{1}$ and $\vec c_{3}$ have the form:
$$
\vec c_{1} = {\vec v_{31}\over \mid \vec v_{31}\mid } \; , \; \vec
c_{3} = -{ \vec v_{13}\over \mid\vec v_{13}\mid } \; ,
\eqnum {1.9}
$$
where $\vec v_{13} \; ( \vec v_{31} )$ is the spatial part of the relativistic
difference of the 4-velocities of the first and third (third and first)
particles ($v_{ij} = v_{i} \ominus v_{j} = ( v_{ij0},  \vec v_{ij})$), defined
as the velocity of the i-th particle in the rest frame of the j-th particle
[36]:
$$
v_{ij} = v_{i} \ominus v_{j} = \Lambda_{p_j}^{-1} \; v_{i} \; .
\eqnum {1.10}
$$
Here $ \Lambda_{p_j}^{-1}$ is the boost, $ \Lambda_{p_j}^{-1}  v_{j} =
v_{j}^{0} = (1, 0 )$, and $ v_{ij}^{2} = v_{ji}^{2} = 1, \;
v_{ij0} = v_{ji0} =  v_{i} v_{j}$, and $\mid \vec v_{ij} \mid = \mid \vec
v_{ji} \mid = \sqrt{ ( v_{i} v_{j} )^{2} - 1 }$. The vectors $\vec v_{13}$
and $\vec v_{31} $ have the form [36]:
$$
\vec v_{13} = \vec v_{1} - \vec v_{3} \left ( v_{10} - {\vec v_{1} \vec v_{3}
\over 1 + v_{10}} \right ) \; , \; \vec v_{31} = \vec v_{3} - \vec v_{1} \left
( v_{30} - {\vec v_{1} \vec v_{3} \over 1 + v_{30}} \right ) \; .
\eqnum {1.11}
$$

To illustrate the properties of the DSB, let us consider an interaction
process in the rest frames of the initial and final particles. In the first
case [$p_1 = (m_1, 0)$] the spin-projection axes $\vec c_{1}$ and $\vec
c_{3}$ are parallel to the momentum of the final particle [this follows from
 (1.9) and (1.11)]:
$$
\vec c_{1} = \vec c_{3} = \vec v_{3}/\mid\vec v_{3}\mid \; .
\eqnum {1.12}
$$
In the rest frame of the final particle [$p_3 = (m_3, 0)$] the spin-projection
axes are antiparallel to the momentum of the initial particle:
$$
\vec c_{1} = \vec c_{3} = - \vec v_{1}/\mid\vec v_{1}\mid \; .
\eqnum {1.13}
$$
Obviously, in these cases the Lorentz little group $L_{p_1p_3}$ is a subgroup
of the group of rotations about the direction of the momentum of the moving
particle, which is the spin-projection axis for both particles. This is a
special case of Eq. (5).

Let us give another, equivalent representation for the spin-projection
operator (1.6), expressed in terms of the antisymmetric matrix $\alpha(p) =
[v \cdot s]^{\times}, \; \alpha(p) \; p = 0$, and $M^{\mu \nu}$:
$$
\sigma = {1 \over 2}\; ([v \cdot s]^{\times})_{\mu \nu} \; M^{\mu \nu} \; .
\eqnum {1.14}
$$
In the DSB the alternating dyadics $[v_1 \cdot s_1]$ and $[v_3 \cdot s_3]$
coincide:
$$
[v_1 \cdot s_1] = [v_3 \cdot s_3] = {[v_1 \cdot v_3] \over \sqrt{ (v_1v_3)
^2 - 1}} \; ,
$$
which ensures that the spin operators $\sigma_1$ and $\sigma_3$ coincide.
We write the matrix $\alpha(p) = [v \cdot s]^{\times}$ in expanded form:
$$
\alpha (p) \; = { 1\over m} \; \left( \begin{array}{cc}
0  & [\vec c \vec p ] \\ - [ \vec c \vec p ] &
p_{0} ( 1 -  \vec p \cdot \vec p/ (( p_{0} + m) p_{0})\vec c ) ^{\times}
\end{array} \right) \; .
\eqnum {1.15}
$$
It is easily verifed that it has the same form in the rest frame ($\vec p =
0$) and for $\vec c \parallel \vec p$:
$$
\alpha (p) \; = \; \left( \begin{array}{cc} 0  & 0 \\ 0 & \vec c~^{\times}
\end{array} \right) \; ,
\eqnum {1.16}
$$
where $\vec c$ is an arbitrary unit vector in the first case and $\vec c
= \vec p / |\vec p |$ in the second. Therefore, study of the helicity states
of moving particles is analogous to study of the spins of particles at rest,
which is one reason for the popularity of the helicity basis.

Let us now turn to spin-1/2 particles, the states of which are described
by bispinors $u^{\delta}(p,s)$ satisfying the Dirac equation:
$$
(\hat p - m ) u^{\delta}(p,s) = 0, \; \overline{u}^{\delta}(p,s) (\hat p
- m ) = 0 \; ,
\eqnum {1.17}
$$
where $\overline{u} = u^{+} \gamma^0$ with $\overline{u}^{\delta} (p,s)
u^{\delta}(p,s) = m$. The Dirac matrices satisfy commutation and recursion
relations:
$$
\gamma^{\mu} \gamma^{\nu} + \gamma^{\nu} \gamma^{\mu} = 2 g^{\mu \nu} \; ,
\eqnum {1.18}
$$
$$
\gamma^{\mu} \gamma^{\nu} \gamma^{\rho} = g^{\mu \nu} \gamma^{\rho} +
g^{\nu \rho} \gamma^{\mu} - g^{\mu \rho} \gamma^{\nu} + i \gamma^5 \;
\varepsilon^{\mu \nu \rho \sigma} \gamma_{\sigma} \; ,
\eqnum {1.19}
$$
$$
\gamma^5 \gamma^{\mu} \gamma^{\nu} = g^{\mu \nu} \gamma^5 -
 i/2 \; \varepsilon^{\mu \nu \rho \sigma} \gamma_{\rho} \gamma_{\sigma} \; .
\eqnum {1.20}
$$
We write out these relations in the form without indices [13]:
$$
\hat a \hat b + \hat b \hat a = 2 ab, \;
\hat a \hat b \hat c = \hat d - i \gamma^{5} \hat f \; ,
\eqnum {1.21}
$$
$$
\gamma^{5} \hat a \hat b = a b \gamma^{5}  -i  \overline{\;  [ a \cdot b ]
^{\times} } \; .
\eqnum {1.22}
$$
$$
d = ( a b + [ a \cdot b ] ) c \; , \; f = [ a \cdot b ]^{\times} c =
[a,b,c], \; f_{\mu} = \varepsilon_{\mu \nu \rho \sigma} a^{\nu}
b^{\rho} c^{\sigma},
\eqnum {1.23}
$$
$$
\overline{\;  [a \cdot b]^{\times}} = 1/2 \;
([a \cdot b]^{\times})_{\mu \nu} \; \gamma^{\mu} \gamma^{\nu}, \;
\eqnum {1.24}
$$
Let us also give some expressions which will be useful later on [13]:
$$
\overline{\alpha} \hat a - \hat a \overline{\alpha} = 2 \; \widehat {\alpha a}
= 2 \; (\gamma \alpha a), \; \overline{\; \alpha^{\times}} \hat a -
\hat a \overline{\; \alpha^{\times}} = 2 \; \widehat { \alpha^{\times}a} =
2 \; (\gamma \alpha^{\times} a) \;,
\eqnum {1.25}
$$
where $ \overline{\alpha} =  1/2 \, \alpha_{\mu \nu } \gamma^{\mu}
\gamma^{\nu}$, and $ \alpha$ is an arbitrary antisymmetric matrix.
The first of these expressions can be obtained by multiplying (1.19) by
$1/2 \alpha_{\mu \nu}a_{\rho}$, then by $1/2 a_{\mu} \alpha_{\nu \rho}$,
and subtracting the results. (The second one is found similarly.)

In bispinor space the generators of the Loretz group $M^{\mu \nu}$ have the
form [43]:
$$
M^{\mu \nu} = i/4 \; (\gamma^{\mu} \gamma^{\nu} - \gamma^{\nu} \gamma^{\mu}).
\eqnum {1.26}
$$
Then the spin-projection operator (1.14) for a spin-1/2 particle can be
written as follows [32], using (1.26) and (1.22):
$$
\sigma = {i \over 2} \; \overline{\;[v \cdot s]^{\times}} \; = {1 \over 2}\;
\gamma^{5} \hat s \hat v, \; \sigma \hat p = \hat p \sigma .
\eqnum {1.27}
$$
Therefore, the covariant electron spin-projection operator (1.14), which
is directly related to the Loretz little group [13], differs by only the
factor $\hat v $ from the widely used operator $\sigma^{'}$ (Refs. 9-12):
$$
\sigma^{'} \; = \; {1 \over 2} \gamma^{5} \hat s \; , \; [\sigma^{'}
\hat p]_{-} \; = \; 0 \; , \; s^{2} = - 1 \; , \; s p = 0 \; .
\eqnum {1.28}
$$
Here the commutation condition for the operator $\sigma$ and $\hat p$ is
satisfied automatically, as is easily verifed by using (1.25) and the
equation $[v \cdot s]^{\times} p = 0$. Therefore, for both moving particles
and particles at rest the spin projection on an arbitrary axis can have a
definite value [13]. The actions of the operators $\sigma $ (1.27) and
$\sigma^{'}$ (1.28) on the particle state vector coincide because the Dirac
equation is valid.

Let us consider a binary reaction $p_{1} + p_{2} \to p_{3} + p_{4}$ in
which particles 1 and 3 are of the same type, as are particles 2 and 4
(for example, electron-nucleon scattering $ep \to ep$, and so on). Given
the spin structure of the matrix elements of this process, it is most
convenient to use the DSB in which particles 1 and 3 and particles 2 and 4
have the same spin-projection operators. In order to construct the raising
and lowering spin operators of the particles, we introduce the orthonormal
vector basis (OVB) $n_{A}$, with $n_{A} n_{B} = g_{AB} \; (A,B = 0,1,2,3)$
(Ref. 36):
$$
n_{1} = [ n_{0}\cdot n_{3} ]^{\times} n_{2} \; ,\; n_{2} = [ p_{1}\cdot p_{3}
]^{\times} r /\rho \; ,
$$
$$
n_{3} = {( p_{3} - p_{1} )\over \sqrt{-(p_{3} - p_{1})^{2}}} \; , \;
 n_{0} = {( p_{3} + p_{1} )\over \sqrt{ (p_{3} + p_{1})^{2}}} \; ,
\eqnum {1.29}
$$
where $r$ is the 4-momentum of a particle participating in the reaction
different from $p_{1}$ and $p_{3}$, and $\rho$ is determined from the
normalization conditions $n_{1}^{2}=n_{2}^{2}=n_{3}^{2}=-n_{0}^{2}=-1$.
Therefore, the axes $n_{0}$ and $n_{3}$ belong to the hyperplane formed
by the 4-momenta $p_{1}$ and $p_{3}$, and $n_{1}$ and $n_{2}$ are orthogonal
to them. The four vectors $n_{A}$  satisfy the relations [36]:
$$
[n_{A}\cdot n_{B}]^{\times} \; = \; 1/2 \; \varepsilon_{AB}~^{CD} [n_{C}\cdot
n_{D}]\; , \; [n_{A} , n_{B} , n_{C}] = -\varepsilon_{ABC}~^{D} n_{D} \; .
\eqnum {1.30}
$$
They also satisfy the completeness relation:
$$
n_{0}\cdot n_{0} - n_{1}\cdot n_{1} - n_{2}\cdot n_{2} - n_{3}\cdot n_{3} = g\; ,
\eqnum {1.31}
$$
by means of which an arbitrary 4-vector $p$ can be written as:
$$
p = pn_{0}\cdot n_{0} - pn_{1}\cdot n_{1} - pn_{2}\cdot n_{2} -
pn_{3}\cdot n_{3}   \; .
$$
In the DSB, not only the spin-projection operators $\sigma_{1}$ and
$\sigma_{3}$ (1.27), but also the raising and lowering operators
$\sigma_{1}^{\pm\delta}$ and $\sigma_{3}^{\pm\delta}$ for particles 1 and
3 coincide. In the OVB (1.29) they have the form [39,40]:
$$
\sigma = \sigma_{1} = \sigma_{3} =1/2 \gamma^{5} \hat s_{1} \hat v_{1} =
1/2 \gamma^{5} \hat s_{3} \hat v_{3} = 1/2\gamma^{5} \hat n_{0} \hat n_{3}
= i/2 \hat n_{1} \hat n_{2} \; ,
\eqnum {1.32}
$$
$$
\sigma^{\pm\delta} = \sigma_{1}^{\pm\delta} = \sigma_{3}^{\pm\delta} = -1/2
\gamma^{5} \hat n_{\pm\delta} \; ,  n_{\pm\delta} = n_{1} \pm i \delta n_{2}
\; , \; \delta = \pm 1 \; ,
\eqnum {1.33}
$$
$$
\sigma u^{\delta}(p_{i}) = \delta /2 u^{\delta}(p_{i}) \; , \;
\sigma^{\pm\delta} u^{\mp\delta}(p_{i}) = u^{\pm\delta}(p_{i}) \; ,
\sigma^{\pm\delta} u^{\pm\delta}(p_{i}) = 0 \; ,
\eqnum {1.34}
$$
$$
[\sigma \sigma^{\pm\delta}]_{-} = \pm \delta \sigma^{\pm\delta} , \;
[\hat p_1 \sigma^{\pm\delta}]_{-} = [\hat p_3 \sigma^{\pm\delta}]_{-} = 0 \; ,
\eqnum {1.35}
$$
where $u^{\delta}(p_{i})=u^{\delta}(p_{i},s_{i})$ are the bispinors of the
first and third particles.

Let us consider the projection operators for spin-1/2 particles,
$\tau^{\delta} = u^{\delta}(p_{i}) \; \overline {u}^{\delta}(p_{i})$
[13,20]:
$$
\tau^{\delta} = 1/4 (\hat p + m) ( 1 - \delta \gamma^{5} \hat s ) \; .
\eqnum {1.36}
$$
In the DSB the operators $\tau^{\delta}_{i}$ (1.36) have the form [39,40]:
$$
\tau^{\delta}_{1} = 1/4 \; ( m + ( \xi_{+} \hat n_{0} - \xi_{-} \hat n_{3} )
+ \delta \gamma^{5} \; ( \xi_{-} \hat n_{0} - \xi_{+} \hat n_{3} - m \;
\hat n_{3} \hat n_{0} ) ) \; ,
\eqnum {1.37}
$$
$$
\tau^{\delta}_{3} = 1/4 \; ( m + ( \xi_{+} \hat n_{0} + \xi_{-} \hat n_{3} )
- \delta \gamma^{5} \; ( \xi_{-} \hat n_{0} + \xi_{+} \hat n_{3} +  m \;
\hat n_{3} \hat n_{0} ) ) \; ,
\eqnum {1.38}
$$
where $\xi_{\pm} = \sqrt{ ( p_{1} p_{3} \pm m^{2} )/2 }$. Owing to (1.32),
the spin parts of the projection operators for particles 1 and 3 can be made
identical in the DSB, and so we have [46,47]:
$$
\tau^{\delta}_{i} = - 1/8 \; (\hat p_{i} + m ) \; \hat n_{\delta} \;
 \hat n_{\delta}^{\ast} \; ,
\eqnum {1.39}
$$
where $n_{\delta}^{\ast} = n_{1} - i \delta n_{2} \; = n_{-\delta}$ and
$n_{\delta} n^{\ast}_{\delta} = -2$.

The bispinors of the initial and final states of the particles,
$u^{\delta}(p_{1})$ and $u^{\delta}( p_{3})$, can be related to each other
by using the transition operators $T_{31}$ and $T_{13} = T_{31}^{-1}$ [13,19]:
$$
u^{\delta}( p_{3} ) = T_{31} \; u^{\delta}(p_{1}) \; , \overline {u}^{\delta}
 (p_{3}) = \overline {u}^{\delta}(p_{1}) \; T_{13} \; ,
\eqnum {1.40}
$$
which in the DSB have the form [39,40]:
$$
T_{31} = { 1 + \hat v_{3} \hat v_{1}\over \sqrt{ 2 (v_{1} v_{3} + 1)}} \; ,
\; T_{13} = { 1 + \hat v_{1} \hat v_{3}\over \sqrt{ 2 ( v_{1} v_{3} + 1)}} \; .
\eqnum {1.41}
$$
Rewriting (1.41) in the OVB (1.29) and isolating the spin-projection operator
$\sigma$ (1.32), we obtain [40]:
$$
T_{31} = \xi_{+}^{'} -2 \xi_{-}^{'} \gamma^{5} \sigma \; , \; T_{13} =
\xi_{+}^{'} + 2 \xi_{-}^{'} \gamma^{5} \sigma \; ,
\eqnum {1.42}
$$
from which we find the relation between the bispinors $u^{\delta}(p_{3})$ and
$u^{\delta}( p_{1})$ (Ref. 41):
$$
u^{\delta}( p_{3} ) = ( \xi_{+}^{'} - \delta \gamma^{5} \xi_{-}^{'} ) \; u^{\delta}( p_{1}) \; ,
\; u^{\delta}( p_{1} ) = ( \xi_{+}^{'} + \delta \gamma^{5} \xi_{-}^{'} ) \; u^{\delta}( p_{3}) \; ,
\eqnum {1.43}
$$
where $\xi_{\pm}^{'} = \xi_{\pm} /m$. We also note that the Dirac equation
can be used to reduce the transition operators $T_{31}$ and $T_{13}$ (1.41)
to the same form [40]:
$$
T_{31} = T_{13} = \hat n_{0} \; .
\eqnum {1.44}
$$

In the massless case the projection operators $ \tau^{\delta}_{1}$ and
$ \tau^{\delta}_{3}$ (1.38) and (1.39) take the form [40-42]:
$$
\tau^{\delta}_{1} = \hat p_{1} \; ( 1 - \delta \gamma^{5})/4 \; ,
\tau^{\delta}_{3} = \hat p_{3} \; ( 1 + \delta \gamma^{5})/4 \; .
\eqnum {1.45}
$$
It is easy to show that the operators $\tau^{\delta}_{1}$ and
$\tau^{\delta}_{3}$ (1.45) satisfy the relations:
$$
\gamma^{5} \tau^{\delta}_{1} = \delta \; \tau^{\delta}_{1} \; , \gamma^{5}
\tau^{\delta}_{3} = - \delta \; \tau^{\delta}_{3} \; ,
\eqnum {1.46}
$$
$$
\tau^{\delta}_{1} \; \gamma^{5} = - \delta \; \tau^{\delta}_{1}  \; ,\;
\tau^{\delta}_{3} \; \gamma^{5} = \delta \; \tau^{\delta}_{3} \; ,
\eqnum {1.47}
$$
which imply that in the massless case the initial state is a helicity state,
and the final state has negative helicity.

Therefore, the DSB possesses a number of remarkable features which allow
great simplification of the covariant calculation of the matrix elements
for QED processes, to which we now turn.

\section{\bf Calculation of matrix elements using the DSB }

The study of multiparticle reactions and the polarization phenomena arising
in them requires effective computational tools. One is based on the use of
the DSB (4). In the DSB the particle spin operators coincide. This allows
the covariant separation of interactions with and without change of the spin
states of the particles involved in the reaction. In the DSB, Wigner rotations
[36,38], which are purely kinematical in nature, are separated from the
amplitudes. This leads to maximal simplification of the mathematical structure
of the diagonal amplitudes, and the resulting expressions give the truest
reflection of the physical essential of spin phenomena.

Let us turn to the calculation of the matrix elements of QED processes.
They have the form
$$
M^{\pm\delta,\delta} = \overline {u}^{\pm \delta}(p_{3}) Q u^{\delta}
(p_{1}) \; ,
\eqnum {2.1}
$$
where $Q$ is the interaction operator, and $u^{\delta}(p_{1})$ and
$u^{\pm\delta}(p_{3})$ are the bispinors of the initial and final states,
with $\overline {u}^{\delta}(p_{i}) \; u^{\delta}(p_{i})=m\;,
\; p_{i}^{2}=m^{2} \; ,\; (i=1, 3)$.

In the Bogush-Fedorov covariant approach [13,19] the calculation of matrix
elements of the form (2.1) reduces to finding the trace:
$$
M^{\pm\delta,\delta} = Tr ( P_{31}^{\pm\delta,\delta} Q ) \; , \; P_{31}^
{\pm\delta,\delta} = u^{\delta}(p_{1}) \; \overline {u}^{~\pm \delta}(p_{3}) \; ,
\eqnum {2.2}
$$
$$
P_{31}^{\delta,\delta} = u^{\delta}(p_{1}) \; \overline {u}^{\delta}(p_{3})
= u^{\delta}(p_{1}) \; \overline {u}^{\delta}(p_{1}) T_{13} = \tau_{1}
 ^{\delta} T_{13} \; ,
\eqnum {2.3}
$$
$$
P_{31}^{-\delta,\delta} = u^{\delta}(p_{1}) \; \overline {u}^{~-\delta}(p_{3})
= \sigma^{+\delta} u^{-\delta}(p_{1}) \; \overline {u}^{~-\delta}(p_{3}) =
\sigma^{+\delta} P_{31}^{-\delta,-\delta} \; .
\eqnum {2.4}
$$
The operators $P_{31}^{\pm\delta,\delta}$ determine the structure of the
spin dependence of the matrix elements (2.1) in the case of transitions
without spin flip $M^{\delta,\delta}$ and with spin flip $M^{-\delta,\delta}$.
Theyr explicit form in the DSB can easily be obtained by using (1.33),
(1.37)-(1.41), and (1.44) (Refs. 39 and 40):
$$
4 P_{31}^{\delta,\delta} = ( \; \xi_{+} + m \; \hat n_{0} - \xi_{-} \; \hat n_{3}
 \hat n_{0} + \delta \gamma^{5} \; ( \; \xi_{-} - m \; \hat n_{3} - \xi_{+}
\; \hat n_{3} \hat n_{0} ) ) \; ,
\eqnum {2.5}
$$
$$
4 P_{31}^{-\delta,\delta} = - \delta \; ( \; \xi_{-} + m \; \hat n_{3}
 + \xi_{+} \; \delta \; \gamma^{5} \; )\; \hat n_{\delta} \; .
\eqnum {2.6}
$$
Equations (2.5) and (2.6) can be used to calculate the matrix elements, both
with and without spin flip, for arbitrary $Q$. In particular, if the
interaction operator reduces to the form
$$
Q = \hat a + \gamma^{5} \; \hat b \; ,
\eqnum {2.7}
$$
where $a$ and $b$ are 4-vectors, then for the matrix elements (2.1) we will
have [39,40]:
$$
M^{\delta,\delta} = m \; ( a n_{0} + \delta \; b n_{3} \; ) \; ,
\eqnum {2.8}
$$
$$
M^{-\delta,\delta} = -\delta \xi_{-} \; (a n_{\delta}) + \xi_{+} \;( b n_{\delta}) \; .
\eqnum {2.9}
$$
Equations (2.5) and (2.6) can be written more compactly by using the operators
(1.39) and (1.44), and also the expressions [39,40]:
$$
\hat n_{3} \hat n_{0} \hat n_{\delta} = - \delta \gamma^{5} \hat n_{\delta} \; ,
\; \gamma^{5} \hat n_{\delta} \hat n_{0} = \delta \hat n_{3} \hat n_{\delta} \; ,
\gamma^{5} \hat n_{\delta} \hat n_{3} = \delta \hat n_{0} \hat n_{\delta} \; .
\eqnum {2.10}
$$
As a result, for the operators $P_{31}^{\pm\delta,\delta}$ we will have
[46,47]:
$$
4 P_{31}^{\delta,\delta} = ( \hat p_{1} + m ) \; \hat n_{\delta} \;
 \hat n_{0} \; \hat n_{\delta}^{\ast} /2 \; ,
\eqnum {2.11}
$$
$$
4 P_{31}^{-\delta,\delta} = \delta \; ( \hat p_{1} + m) \;
 \hat n_{\delta} \; \hat n_{3} \;  .
\eqnum {2.12}
$$
Let us give yet another representation for the operators $P_{31}^{\pm
\delta,\delta}$ in (2.3), (2.4) in the DSB [42]:
$$
4 P_{31}^{ \delta,\delta} = (\hat p_1 + m ) \left ( {1 \over \sqrt{2(p_1p_3
+ m^2)}} - {\delta \gamma^5 \over \sqrt{2(p_1 p_3 - m^2)}} \right )
( \hat p_3 + m ) \; ,
\eqnum {2.5a}
$$
$$
4 P_{31}^{- \delta,\delta}= - \; {\delta ( \hat p_1 + m ) \over rn_1 }
\; \Biggl \{ \; {1 \over \sqrt{2(p_1 p_3 - m^2)}} \; \left ( \hat r
- m \; {\; (p_1 + p_3) r \; \over p_1p_3 + m^2} \right )
$$
$$
+ \; {\delta \gamma^5 \over \sqrt{2(p_1p_3 + m^2)}} \;
\left ( \hat r + m \; {\; (p_3 -p_1) r \; \over p_1p_3 - m^2} \right )
\; \Biggr \} \; ( \hat p_3 + m ) \; ,
\eqnum {2.6a}
$$
where $rn_1$ is calculated by using the completeness relation (1.31):
$$
(rn_1)^2 = (rn_0)^2 - (rn_3)^2 - r^2 \; , \; r n_2 = 0\; .
$$
Thus, the representation (2.5a) and (2.6a) is attractive in that it contains
the Dirac operators only on terms of the particle 4-momenta $p_1, p_3$, and
$r$, in contrast to (2.5), (2.6), (2.11) and (2.12), which involve
$\hat n_{\delta}$ and $\hat n_{\delta}^{\ast}$. Moreover, the structure of
the operators $P_{31}^{\pm \delta,\delta}$ in (2.5a), (2.6a) is such that
they automatically satisfy the Dirac equations: $(\hat p_1 - m)P_{31}^
{\pm \delta,\delta} = P_{31}^{\pm \delta,\delta}(\hat p_3-m)=0$. This was
used to derive (2.5a) and (2.6a) from (2.5) and (2.6).

Let us explain the choice of 4-vector $r$ in terms of which the axes $n_1$
and $n_2$ (1.29) appearing in (2.5) and (2.6) are defined. First, it is chosen
from the 4-momenta of the particles in the reaction under study, in contrast
to the CALCUL approach, in which it is defined from considerations of
convenience. Let us illustrate this for the example of the reaction
$e^{-}(p_1) + \mu^{-}(p_2) \to e^{-}(p_3) + \mu^{-}(p_4) + \gamma (k)$, which
corresponds to Feynman graphs containing two fermion lines. For each of these
lines it is necessary to construct the corresponding operators
$P_{31}^{\pm \delta,\delta}$ and $P_{42}^{\pm \delta',\delta'}$ in (2.5a),
and (2.6a), expressed in terms of $p_1, p_3, r_1$ and $p_2, p_4, r_2$. For
this process it is very convenient to make the choice: $r_1=r_2=k$, so that
$k n_2 = k n_2^{'} = 0$ and $(k n_1)^2=(kn_0)^2-(kn_3)^2$. The vectors
$r_1$ and $r_2$ can also be chosen to be the 4-momenta belonging to different
fermion lines: $r_1 = p_2, \;r_2 = p_1$ (here we are considering transitions
$p_1 \to p_3$ and $p_2 \to p_4$). We note that the arbitrariness in the choice
of the 4-vector $r$ in (1.29), i.e., replacement of $r$ by $r'$, leads to
the expression [42]:
$$
n_1' +i \delta n_2' = e^{i\delta \phi} ( n_1 +i \delta n_2) \; , \;
e^{i\delta \phi}={r(n_1 \cdot n_1 + i \delta [n_0 \cdot n_3]^{\times}) r'
\over (r n_1) (r'n_1')}
$$
and affects only the phase factor of the matrix elements $M^{\pm\delta,
\delta}$.

Processes involving identical particles (for example, $ee \to ee, \;
e e \to e e \gamma$, and so on) correspond to direct and exchange graphs
[10]. They are associated with matrix elements $M_1$ and $M_2$ of the form:
$$
M_{1} = \overline {u}^{\pm \delta}(p_{3}) Q_{1} u^{\delta}(p_{1}) \;
\cdot \overline {u}^{\pm \delta^{'}}(p_{4}) Q_{2} u^{\delta^{'}}(p_{2}) \; ,
$$
$$
M_{2} = \overline {u}^{\pm \delta}(p_{3}) Q_{3} u^{\delta'}(p_{2}) \;
\cdot \overline {u}^{\pm \delta'}(p_{4}) Q_{4} u^{\delta}(p_{1}) \; ,
$$
which are calculated as:
$$
M_1 = Tr( P_{31}^{\pm\delta,\delta} Q_1 ) \; Tr( P_{42}^{\pm\delta',\delta'} Q_2 ) \; ,
M_2 = Tr ( P_{31}^{\pm\delta,\delta} Q_3 \; P_{42}^{\pm\delta',\delta'} Q_4 ) \; .
$$
Therefore, the calculation of the direct graphs reduces to a product of traces,
while that of the exchange graphs reduces to a trace extended by the product
of the corresponding operators [13].

Let us give some useful expressions which are valid in the DSB [40]:
$$
\hat a \; u^{\delta}(p_{1}) = ( a n_{0} + a n_{3} \delta \gamma^{5} ) \;
u^{\delta}(p_{3}) + a n_{\delta} \gamma^{5} u^{-\delta}(p_{1}) \; ,
$$
$$
\hat a \; u^{\delta}(p_{3}) = ( a n_{0} + a n_{3} \delta \gamma^{5} ) \;
u^{\delta}(p_{1}) + a n_{\delta} \gamma^{5} u^{-\delta}(p_{3}) \; ,
$$
$$
\overline {u}^{\delta}(p_{1}) \; \hat a = \overline {u}^{\delta}(p_{3})
( a n_{0} - a n_{3} \delta \gamma^{5} ) - a n_{\delta}^{\ast} \;
\overline {u}^{-\delta}(p_{1})\; \gamma^{5} \; ,
\eqnum {2.13}
$$
$$
\overline {u}^{\delta}(p_{3}) \; \hat a = \overline {u}^{\delta}(p_{1})
( a n_{0} - a n_{3} \delta \gamma^{5} ) - a n_{\delta}^{\ast} \;
\overline {u}^{-\delta}(p_{3})\; \gamma^{5} \; ,
$$
where $a$ is an arbitrary 4-vector ($n_{\delta}^{\ast} = n_{-\delta}$).

In the DSB (4) the particle spin vectors are expressed in terms of the
4-momenta, so that the number of independent scalar products entering into
the final expressions for the matrix elements after calculation of the traces
(2.2) is decreased. For the same reason the circular-polarization vector
$e_{\lambda}$ of a photon with 4-momentum $k$ emitted by a particle in the
transition $p_{1} \to p_{3}$ is conveniently defined by using the 4-vectors
$ p_{1}, p_{3}$ and $k$ (Refs. 21 and 40):
$$
e_{\lambda} = { [ n_{0} \cdot n_{3} ] k + i \lambda \; [ n_{0} \cdot n_{3} ]
^{\times} k \over  \sqrt{2} \rho } \; , \; [ n_{0} \cdot n_{3} ] = { [ p_{1}
 \cdot p_{3} ] \over 2 \xi_{+} \xi_{-} } \; ,
\eqnum {2.14}
$$
where $\rho = \sqrt{ -([ p_{1} \cdot p_{3} ] k )^{2}}/2 \xi_{+} \xi_{-}$. Then
for the dyadic $e_{\lambda} \cdot e_{\lambda}^{\ast}$ we easily find:
$$
e_{\lambda} \cdot e_{\lambda}^{\ast} = {1\over 2} \; \left ( - g + { k \cdot n_{1}
+ n_{1} \cdot k \over k n_{1}} + {k \cdot k \over k n_{1}^{2}} + i \lambda
\; { [ k \cdot n_{1} ]^{\times} \over k n_{1} } \right ) \; .
\eqnum {2.15}
$$
Using (1.25) and (1.22) the operators $\hat e_{\pm \lambda}$ ($\hat e_
{\lambda}^{\ast} = \hat e_{- \lambda}$) can be written as follows [40]:
$$
\hat e_{\pm \lambda} = N_{13} ( \hat k \hat p_{3} \hat p_1 ( 1 \mp \lambda
\gamma^{5} ) - \hat p_3 \hat p_{1} \hat k ( 1 \pm \lambda \gamma^{5} )
 \mp 2 p_{1} p_{3} \lambda \gamma^{5} \hat k ) \; ,
\eqnum {2.16}
$$
$$
N_{13}^{-1} = 2^{1/2} ( 8 p_{1} p_{3} \cdot p_{1} k \cdot p_{3} k - m^{2}
( ( 2 p_{1} k )^{2} + (2 p_{3} k )^{2} ) )^{1/2} \; .
$$
In the massless case ($p_{1}^{2}=p_{3}^{2}=0$) the operators $P_{31}^{\pm
 \delta,\delta}$ in (2.5) and (2.6) take the form [40]:
$$
4 P_{31}^{\delta,\delta} =  \xi ( 1 + \delta \gamma^{5} ) ( 1 + \hat n_{0}
\hat n_{3} ) \; , \; 4 P_{31}^{-\delta,\delta} = - \delta \xi ( 1 + \delta
 \gamma^{5} ) \hat n_{\delta} \; ,
\eqnum {2.17}
$$
where $ \xi = \xi_{+} = \xi_{-} = \sqrt{ p_{1} p_{3} / 2 }$. Similarly,
from (2.5a) and (2.6a) we have:
$$
4 P_{31}^{\delta,\delta} =  { ( 1 + \delta \gamma^{5} )\;  \hat p_1 \; \hat p_3
\over \sqrt{ 2 p_1 p_3}}\; , \;
4 P_{31}^{-\delta,\delta} = - \delta \; { ( 1 - \delta \gamma^{5} ) \; \hat p_1 \; \hat r \; \hat p_3
\over rn_1 \sqrt{ 2 p_1 p_3}}\; .
\eqnum {2.17a}
$$
Using (1.25), it is easy to show that the representations (2.17) and (2.17a)
are equivalent. As noted above, in calculating processes with the emissions
or absorption of a real photon with 4-momentum $k$, it is convenient to make
the choice $r=k$ for the 4-vector $r$ entering into (2.17a). Then the
denominator of the operator $P_{31}^{-\delta,\delta}$ in (2.17a) takes the
form $ rn_1\sqrt{ 2 p_1 p_3} = \sqrt{2p_1 k \cdot 2 p_3 k}$, and we obtain
a result similar to that of Ref. 23, except that our expressions involve the
4-momentum of a real photon, and not an auxiliary lightlike 4-vector $k$.
These points are very important for our approach, in which we use only the
4-momenta of the particles participating in the reaction.

Using (1.46) and (1.47), Eq.(2.13) can be written as [40]:
$$
\hat a \; u^{\delta}(p_{1}) = ( a n_{0} - a n_{3} ) \; u^{\delta}(p_{3}) -
 \delta a n_{\delta} u^{-\delta}(p_{1}) \; ,
$$
$$
\hat a \; u^{\delta}(p_{3}) = ( a n_{0} + a n_{3} ) \; u^{\delta}(p_{1}) +
\delta a n_{\delta} u^{-\delta}(p_{3}) \; ,
$$
$$
\overline {u}^{\delta}(p_{1}) \; \hat a = ( a n_{0} - a n_{3} ) \; \overline
{u}^{\delta}(p_{3}) - \delta a n_{\delta}^{\ast} \;
\overline {u}^{-\delta}(p_{1}) \; ,
\eqnum {2.18}
$$
$$
\overline {u}^{\delta}(p_{3}) \; \hat a = ( a n_{0} + a n_{3} ) \; \overline
{u}^{\delta}(p_{1}) + \delta a n_{\delta}^{\ast} \;
\overline {u}^{-\delta}(p_{3}) \; .
$$
In the massless case the relation between the bispinors of the initial and
final states takes a particularly simple form [see (1.44)]:
$$
u^{\delta}(p_{3}) = { \hat p_{3} \over \sqrt{ 2 p_{1} p_{3}}} \; \;
 u^{\delta}(p_{1}) \; , \; u^{\delta}(p_{1}) = { \hat p_{1}
 \over \sqrt{ 2 p_{1} p_{3}}} \; \; u^{\delta}(p_{3}) \; .
\eqnum {2.19}
$$
In this massless limit, the terms containing $\gamma^{5} \hat k$ in (2.16)
can be dropped, owing to gauge invariance. As a result, for the operators
$\hat e_{\pm \lambda}$ we obtain the expressions used by the CALCUL group
[21,48]:
$$
\hat e_{\pm \lambda} = N_{13} ( \hat k \hat p_{3} \hat p_1 ( 1 \mp \lambda
\gamma^{5} ) - \hat p_3 \hat p_{1} \hat k ( 1 \pm \lambda \gamma^{5} ) ) \; ,
\eqnum {2.20}
$$
$$
N_{13}^{-1} = 4 \; ( p_{1} p_{3} \cdot p_{1} k \cdot p_{3} k ) ^{1/2}\; .
$$
Using (2.18)-(2.20), we can easily verify the correctness of the expressions
[21,40,48]:
$$
\hat e_{\lambda} \; u^{\delta}(p_{1}) = - ( 1 + \delta \lambda ) \; 2 p_{1} k \; N_{13}
\hat p_{3} \; u^{\delta}(p_{1}) \; ,
$$
$$
\overline {u}^{\delta}(p_{3}) \; \hat e_{\lambda}^{\ast} = ( 1 - \delta
 \lambda ) \; 2 p_{3} k \; N_{13} \overline {u}^{\delta}(p_{3}) \; \hat p_{1} \; .
\eqnum {2.21}
$$
If photon emission occur in transition $p_{A} \to p_{B}$, then, making
the replacement $(p_{1}, p_{3}) \to (p_{A},p_{B})$ in (2.20), we obtain
the operators $\hat e_{\lambda AB}$, whose action on bispinors is the same
as that of $\hat e_{\lambda} = \hat e_{\lambda 13}$ except for a phase
[21,40]:
$$
\hat e_{\lambda 13} = \hat e_{\lambda AB} \; exp\; (i \phi_{AB}) \; , \;
exp\; (i \phi_{AB}) = i \lambda 2^{1/2} e_{\lambda 13} n_{2(AB)} \; ,
\eqnum {2.22}
$$
where $n_{2(AB)}$ are unit vectors:
$$
n_{2(AB)} = [ p_{A} \cdot p_{B} ]^{\times} k /\rho_{(AB)} \; , \;
\rho_{(AB)} = ( 2 p_{A} p_{B} \cdot p_{A} k \cdot p_{B} k )^{1/2} \; .
$$
Up to now our discussion has pertained to the case with only electrons in
the initial and final states. If one state is an electron and the other a
positron, the amplitude of the process will have the form [9]:
$$
M^{\pm\delta,\delta}_{31} = \Bigl\{ \begin{array}{l}
\overline {u}^{\pm \delta}(-p_3) Q u^{\delta}(p_{1})   \\
 \overline {u}^{\pm \delta}(p_3) Q u^{\delta}(-p_{1}) \; \; ,
\end{array} \Bigr.
\eqnum {2.23}
$$
where $u^{\delta}(-p_{1})$ and $ \overline {u}^{\pm\delta}(-p_{3}) $ are
the positron bispinors in the final and initial states, with $\overline
{u}^{\delta}(-p_{i})u^{\delta}(-p_{i})=-m \; \; (i= 1, 3)$. The upper
amplitude in (2.23) corresponds to pair annihilation, and the lower one to
pair production. To construct the operators
$$
P_{31}^{\pm\delta,\delta} = u^{\delta}(p_{1}) \; \overline {u}^{\pm\delta}
(-p_{3}) \; \; , \; P_{31}^{\pm\delta,\delta} = u^{\delta}(-p_{1}) \;
\overline {u}^{\pm\delta}(p_{3}) \; ,
\eqnum {2.24}
$$
used to reduce the determination of the matrix elements (2.23) to calculation
of the traces $M^{\pm\delta,\delta}=Tr(P_{31}^{\pm\delta,\delta} Q)$, we need
to use the relation between the positron and electron in the DSB [36,38]:
$$
u^{\delta} (-p) = - \delta \; \gamma^{5} \; u^{-\delta} (p) \; , \;
\overline {u}^{\delta}(-p) = \overline {u}^{-\delta}(p) \; \delta \; \gamma^{5} \; .
\eqnum {2.25}
$$
As a result, for the operators $P_{31}^{\pm\delta,\delta}$ used to calculate
the amplitudes for pair annihilation, we obtain:
$$
4 P_{31}^{\delta,\delta} = \delta \; ( \hat p_{1} + m ) \; \hat n_{0} \;
\hat n_{\delta} \; , \; 4 P_{31}^{-\delta,\delta} = - \; ( \hat p_{1} + m )
\; \hat n_{\delta} \; \hat n_{3} \; \hat n_{\delta}^{\ast} /2 \; .
\eqnum {2.26}
$$
Similar expressions can be obtained for the operators $P_{31}^{\pm\delta,
\delta}$ in the case of pair production.

We have used this formalism for calculating matrix elements in the DSB to
obtain the cross sections for several real QED processes, to which we now
turn.

\section{\bf The cross sections for the processes $e^{\pm}e^{-}
\to e^{\pm} e^{-} \gamma $ \newline in the  ultrarelativistic massless case}

M\"{o}ller and Bhabha bremsstrahlung $e^{\pm}e^{-} \to e^{\pm} e^{-} \gamma$
are background processes in studying hadron states. Moreover, the study of
these processes allows verification of QED in higher orders of perturbation
theory. The cross sections for these processes are quite awkward, even in
the ultrarelativistic limit. Only relatively recently has it been possible
to write them down in a compact form for unpolarized [49] and transversely
polarized initial particles [50]. Using the methods described above [Eqs.
(2.17)-(2.22)], the present authors have obtained [40] compact expressions,
in the ultrarelativistic, massless limit, for the differential cross sections
 of the processes $e^{\pm}e^{-} \to e^{\pm} e^{-}\gamma$ for the case where
not only the initial particles but also the photon are helically polarized.
As was shown in Ref. 40, the cross sections for these processes are written
as the product of two factors, one universal and coinciding with that obtained
earlier [49] for unpolarized particles. Let us consider M\"{o}ller and Bhabha
bremsstrahlung,
$$
e^{-}(p_{1}) + e^{\pm }(p_{2}) \rightarrow e^{-}(p_{3}) + e^{\pm}(p_{4}) + \gamma(k) \; ,
\eqnum {3.1}
$$
assuming that the initial and final $e^{\pm}$ particles are massless
($p_{i}^{2}=0, i=1,2,3,4 $). The details of the calculations of the matrix
elements for (3.1), which correspond to eight Feynman diagrams [9,10], are
given in Ref. 40, and so we shall not dwell on them here. We introduce the
invariant variables [49,50]:
$$
s = ( p_{1} + p_{2} )^{2} \; , \; t = ( p_{1} - p_{3} )^{2} \; , \; u = ( p_{1} - p_{4} )^{2} \; ,
$$
$$
s^{'} = ( p_{3} + p_{4} )^{2} \; , \; t^{'} = ( p_{2} - p_{4} )^{2} \; , \; u^{'} = ( p_{2} - p_{3} )^{2} \; ,
\eqnum {3.2}
$$
and also the notation $ \delta, \delta^{'}$ and $\lambda$ for the helicities
of the initial particles and the photon, respectively. Then the differential
cross sections for the processes $e^{-}e^{\pm} \to e^{-} e^{\pm} \gamma$ in
the case helically polarized initial leptons and photon have the form [40]:
$$
d \sigma_{M} = { \alpha^{3} \over \pi^{2} s } \; A_{M} \; W_{M} \; d \Gamma \; , \;
d \sigma_{B} = { \alpha^{3} \over \pi^{2} s } \; A_{B} \; W_{B} \; d \Gamma \; ,
\eqnum {3.3}
$$
$$
A_{M} = A_{MB} / t \; t^{'} \; u \; u^{'} \; , \;
A_{B} = A_{MB} / t \; t^{'} \; s \; s^{'} \; \; ,
\eqnum {3.4}
$$
$$
A_{MB} = 1/2 \; \{ s s^{'} ( s^{2} + s^{'2} ) + t t^{'} ( t^{2} + t^{'2} ) + u u^{'} ( u^{2} + u^{'2} ) ~~~~~~~~~~~~
$$
$$
 + \; \delta \; \delta^{'} \; ( s s^{'} ( s^{2} + s^{'2} ) - t t^{'} ( t^{2} + t^{'2} ) - u u^{'} ( u^{2} + u^{'2} ) )~~~~
\eqnum {3.5}
$$
$$
 + \; \delta \; \lambda \; ( - s s^{'} ( s^{2} - s^{'2} ) - t t^{'} ( t^{2} - t^{'2} ) - u u^{'} ( u^{2} - u^{'2} ) )~~
$$
$$
 + \; \delta^{'} \lambda \; ( - s s^{'} ( s^{2} - s^{'2} ) + t t^{'} ( t^{2} - t^{'2} ) + u u^{'} ( u^{2} - u^{'2} ) ) \} \; ,
$$
$$
W_{M} = - \left ( {p_{1}\over p_{1} k} + {p_{2}\over p_{2} k} - {p_{3}\over p_{3} k} - {p_{4}\over p_{4} k} \right )^{2} \; ,
\eqnum {3.6}
$$
$$
W_{B} = - \left ( {p_{1}\over p_{1} k} + {p_{4}\over p_{4} k} - {p_{3}\over p_{3} k} - {p_{2}\over p_{2} k} \right )^{2} \; ,
\eqnum {3.7}
$$
$$
d \Gamma = \delta^{4} ( p_{1} + p_{2} -p_{3} - p_{4} - k ) {d^{3} \vec p_{3}
 \over 2 p_{30}} \; {d^{3} \vec p_{4} \over 2 p_{40}} \; {d^{3} \vec k \over 2
\; \omega} \; \; ,
$$
where $ x_{A} = p_{A} k \; , \; ( A = 1, 2, 3, 4)$, and $\alpha $ is the
fine-structure constant. We note that Eq. (3.5) for $A_{MB}$ is invariant
under crossing transformations:
$$
p_{2} \leftrightarrow - p_{4} \; , \; x_{2} \leftrightarrow - x_{4} \; , \;
s \leftrightarrow u \; , \; s^{'} \leftrightarrow u^{'} \; , \; \delta^{'}
\leftrightarrow - \delta^{'} \; .
\eqnum {3.8}
$$
The expressions for $A_{MB}, W_{M}$, and $W_{B}$ can be written in a different
form [40]:
$$
A_{MB} = 1/2 \; \{ ( 1 + \delta \delta^{'} ) \; ( ( 1 + \delta \lambda )
 s s^{'} s^{'2} + ( 1 - \delta \lambda ) s s^{'} s^{2} ) +
$$
$$
 + ( 1 - \delta \delta^{'} ) \; ( ( 1 + \delta^{'} \lambda ) \;  ( t t^{'}
t^{2} + u u^{'} u^{2} ) + ( 1 - \delta^{'} \lambda ) \;  ( t t^{'} t^{'2} +
u u^{'} u^{'2} ) ) \} \; ,
\eqnum {3.9}
$$
$$
 - W_{M} =  {s \over x_{1} x_{2}} + {s^{'} \over x_{3} x_{4}} +
{t \over x_{1} x_{3}} + {t^{'} \over x_{2} x_{4}} +
{u \over x_{1} x_{4}} + {u^{'} \over x_{2} x_{3}} \; ,
\eqnum {3.10}
$$
$$
W_{B} = {s \over x_{1} x_{2}} + {s^{'} \over x_{3} x_{4}} -
{t \over x_{1} x_{3}} - {t^{'} \over x_{2} x_{4}} +
{u \over x_{1} x_{4}} + {u^{'} \over x_{2} x_{3}} \; .
\eqnum {3.11}
$$
When soft photons are emitted ($ s=s^{'}, \;t=t^{'}, \; u= u^{'} $), $A_{M}$
and $A_{B}$ take the form [40]: ($\delta=\pm1, \delta'=\pm1$)
$$
A_{M} = {s^{2} + u^{2}\over t^{2}} + {s^{2} + t^{2}\over u^{2}} + {2 s^{2}
 \over t u } + \delta \delta^{'} \left ( {s^{2} - u^{2}\over t^{2}} + {s^{2} -
 t^{2}\over u^{2}} + {2 s^{2}\over t u } \right ) \; \; ,
\eqnum {3.12}
$$
$$
A_{B} = {u^{2} + s^{2} \over t^{2}} + {u^{2} + t^{2}\over s^{2}} + {2 u^{2}
 \over s t } - \delta \delta^{'} \left ( {u^{2} - s^{2}\over t^{2}} + {u^{2} +
 t^{2}\over s^{2}} + {2 u^{2}\over s t  } \right ) \; \; .
\eqnum {3.13}
$$
They differ only by overall factors from the cross sections for elastic
processes $e^{\pm}e^{-} \to e^{\pm} e^{-} $ when the initial particles are
longitudinally polarized (see Ref. 10).

For unpolarized photons, from (3.9) we have:
$$
A_{MB} = ( 1 + \delta \delta^{'} ) \;  s s^{'} \; ( s^{2} + s^{'2} ) +
 ( 1 - \delta \delta^{'} ) \; ( t t^{'} ( t^{2} + t^{'2} ) + u u^{'} ( u^{2} +
 u^{'2} ) ) \; .
\eqnum {3.14}
$$
Therefore, the ratio of the cross sections for particles with parallel and
antiparallel spins have the same form [40] for the two reactions
$e^{\pm}e^{-} \to e^{\pm} e^{-} \gamma $ (as in the case of the elastic
processes $e^{\pm}e^{-} \to e^{\pm} e^{-} $; see Ref. 10):
$$
{d \sigma_{\uparrow \uparrow} \over d \sigma_{\uparrow \downarrow}} = {
 t t^{'} ( t^{2} + t^{'2} ) + u u^{'} ( u^{2} + u^{'2} ) \over  s s^{'} \;
 ( s^{2} + s^{'2} ) } \; .
\eqnum {3.15}
$$

\section {\bf Polarization phenomena in the three-photon
annihilation of orthopositronium}

In recent years the three-photon annihilation of orthopositronium $^{3}S_{1}
\to 3 \gamma$ has attracted a great deal of attention, because experiments
to measure the decay width of orthopositronium revealed a discrepancy with
the theoretical predictions [51]. Several attempts have been made to resolve
this contradiction. In Ref. 52, relativistic corrections were included in
the cross section for the annihilation of a slow $e^{+}e^{-}$ pair into two
or three photons, but this did not solve the problem. The contribution of
the five-photon decay mode of orthopositronium, calculated in Ref. 53,
indicates that this mechanism also cannot eliminate the discrepancy in the
width. All these problems, including the results of Refs. 51 and 53,
require further analysis and confirmation. The work of Ref. 54 does
not represent an attempt to resolve the orthopositronium problem. Almost
all the known results pertaining to polarization phenomena in the reaction
$e^{+}e^{-} \to 3 \gamma $ were obtained there, but by calculating the
matrix elements in the DSB. The purpose of that study was to demonstrate
the effectiveness of that method for a process to which the CALCUL
method is inapplicable. A key feature of the technique is the very
specific choice of the photon polarization vectors (2.20), which
is valid only for the massless case.

The main process determining the positronium lifetime is three-photon
annihilation. Here the decay probability can be related to the cross section
for annihilation of a free pair [10]:
$$
e^{-}(p_{1}) + e^{+}(p_{3}) \rightarrow \gamma(k_{1}) + \gamma(k_{2}) + \gamma(k_{3})  \; .
\eqnum {4.1}
$$
Since the momenta of the electron and positron in positronium are small
[10] ($\mid \vec p_{1} \mid, \mid \vec p_{3} \mid \; \sim m \alpha$, where
$\alpha$ is the fine-structure constant), in calculating the annihilation
cross section they can be considered to be at rest at the origin [i.e., we
assume that $p_{1} = p_{3} = p = (m,0,0,0)]$. In this case the matrix element
of the reaction (4.1) takes the form:
$$
M^{\pm\delta,\delta}_{31}=\overline {u}^{\pm \delta}(-p) Q u^{\delta}(p) \; ,
\eqnum {4.2}
$$
where $u^{\delta}(\pm p)$ are the electron and positron bispinors,
$\overline{u}^ {\delta}(\pm p) u^{\delta}(\pm p) = \pm m$, and $Q$ is the
interaction operator, which corresponds to six Feynman diagrams [10]. Let
us consider the kinematics of the process $e^{+}e^{-} \to 3 \gamma $ in the
$e^{+}e^{-}$ c.m. frame, in which the momenta $p_{1}$ and $p_{3}$ have the
form $p_{1} = (p_{0}, 0, 0, -m\alpha)$ and $p_{3}=(p_{0},0,0, m\alpha),
\, p_{0}=m \sqrt{1 + \alpha^{2}}$. We introduce the OVB $a_{A}$:
$$
a_{0} = (1, 0 , 0 , 0) \; ,  \; a_{1} = (0, 1 , 0 , 0), \; a_{2} = (0, 0 , 1 , 0 ), \;
a_{3} = (0, 0 , 0 , 1) \; ,
$$
using which we find  :
$$
p_{1} = - \xi_{-} a_{3} + \xi_{+} a_{0} \; , \; p_{3} = \xi_{-} a_{3} +
 \xi_{+} a_{0} \; ,
\eqnum {4.3}
$$
$$
s_{1} = \xi^{'}_{+} a_{3} - \xi^{'}_{-} a_{0} \; , \; s_{3} = \xi^{'}_{+} a_{3} +
 \xi^{'}_{-} a_{0} \; ,
\eqnum {4.4}
$$
where $\xi^{'}_{\pm} = \xi_{\pm} / m, \; \xi^{'}_{+} = \sqrt{1 + \alpha^{2}}$
, and $ \xi^{'}_{-} = \alpha $, with $s_{1}p_{1} = s_{3}p_{3} =0$,
and $s_{1}^{2} = s_{2}^{2} = - 1$. Therefore, in the limit $\alpha \to 0$ the
electron and positron spin vectors $s_{1}$ and $s_{3}$ (4.4) in the DSB (4)
coincide:
$$
s_{1} = s_{3} = a_{3} \; ,
\eqnum {4.5}
$$
i.e., the direction of the positron motion is singled out as the common
axis of spin projection. The momentum conservation law
$$
\vec k_{1} + \vec k_{2} + \vec k_{3} = 0
\eqnum {4.6}
$$
determines the annihilation plane in which the photon momenta lie. We shall
also assume that the vectors $\vec a_{1}$ and $\vec a_{3}$ lie in this plane,
while the vector $\vec a_{2}$ is normal to it, i.e., $\vec a_{2} \vec n_{i}
=0 \;, \; \vec n_{i}=\vec k_{i} /\omega_{i}$, and $\vec n_{i}~^{2}=1 ,\;
(i=1,2,3)$.

\noindent
Let us construct the photon circular-polarization vectors $ e_{\lambda i}
=(0, \vec e_{\lambda i})$:
$$
\vec e_{\lambda i} = ( [\vec a_{2} \vec n_{i} ] + i \lambda_{i} \; \vec a_{2} )/
\sqrt{2} \; , \; \vec e_{\lambda i} \; \vec n_{i} = 0 \; , \; \vec e_{\lambda i}
 \vec e_{\lambda i}^{~\ast} = 1 \; ,
\eqnum {4.7}
$$
where $\lambda_{i}$ are the photon helicities, $\lambda_{i} = \pm 1$.

In the limiting case that we are considering, the operators (2.26) used to
calculate the matrix elements (4.2) have the form:
$$
4 P_{31}^{\delta,\delta} = \delta \; ( m + \hat p ) \; \hat a_{\delta} \; , \;
4 P_{31}^{-\delta,\delta} = - \; ( m + \hat p ) \; \hat a_{\delta} \;
\hat a_{3} \; \hat a_{\delta}^{\ast} /2 \; ,
\eqnum {4.8}
$$
where $a_{\pm \delta} = a_{1} \pm i \delta a_{2} \; , \; \delta = \pm
 \; 1 \; ( a_{\delta}^{\ast} = a_{-\delta} )$.

The explicit form of the matrix elements $M^{\pm\delta,\delta}$ for the
process (4.1) in the case of circularly polarized photons was obtained in
Ref. 54:
$$
2^{3/2} M^{\delta,\delta} = \delta \; \sum_{i=1}^3 \alpha_{i} \; (\; \delta
\lambda_{i} + c_{i}\;) \;( n_{jk} -1 ) / m \; ,
\eqnum {4.9}
$$
$$
2^{3/2} M^{-\delta,\delta} = \sum_{i=1}^3 \alpha_{i} \; s_{i} \;
( n_{jk} -1 )/m \; ,
\eqnum {4.10}
$$
where $\alpha_{i}$ (i=1,2,3) are the polarization factors:
$$
\alpha_{1}=(1+\lambda_{2} \lambda_{3})(1-\lambda_{1} \lambda_{2}), \;
\alpha_{2}=(1+\lambda_{1} \lambda_{3})(1-\lambda_{2} \lambda_{3}), \;
\alpha_{3}=(1+\lambda_{1} \lambda_{2})(1-\lambda_{1} \lambda_{3}), \;
\eqnum {4.11}
$$
$$
\alpha_{1} \; \alpha_{2} = \alpha_{1} \; \alpha_{3} = \alpha_{2} \; \alpha_{3} = 0 \; , \;
\alpha_{i}^{2} = 4 \;\alpha_{i} \; ,
\eqnum {4.12}
$$
and the quantities $c_{i}, \; s_{i}, \; n_{jk} = n_{kj} \;(i,j,k=1,2,3 \;)$
are: $ n_{jk}  =  \vec n_{j} \vec n_{k} = c_{j} c_{k} + s_{j} s_{k} ,
\; c_{i} = \vec a_{3} \vec n_{i} , \; s_{i} = \vec a_{1} \vec n_{i} ,
\; s_{i}^2 + c_{i}^2 = 1$, with the indices $i,\; j$, and $k$ in (4.9) and
(4.10) representing a cyclic permutation of the numbers 1,2,3.

The matrix elements (4.9) and (4.10) determine the annihilation of a free
$e^{+} e^{-}$ pair in the case of parallel ($M^{\delta,\delta}$) and
antiparallel ($M^{-\delta,\delta}$) spins of the electron and positron.
They are real and vanish if all the photons have the same helicity, i.e.,
when $ \lambda_{1} = \lambda_{2} = \lambda_{3}$.

The differential cross section for the process (4.1) is expressed in terms
of the matrix elements $M^{\pm\delta,\delta}$ in (4.9) and (4.10) as
$$
d \sigma_{3 \gamma} = { \alpha^{3} ( M^{\pm\delta,\delta} )^{2} \over ( 2 \pi )^{2}
\; 4 m^{2} v } \; \left ( \prod_{i=1}^{3} { d^{3} \vec k_{i} \over \omega_{i}}
\; \right ) \delta^{4} ( 2 p - k_{1} - k_{2} - k_{3} ) \; ,
\eqnum {4.13}
$$
where $v$ is the relative velocity of the $e^{+}$ and $e^{-}$ in the c.m.
frame ($v \sim \alpha$). We introduce the notation $ \sigma^{\lambda_{1},
\lambda_{2}, \lambda_{3}}_{~~\delta e^{+}} = 1/2 \; ( \; ( M^{\delta,\delta}
 )^{2} + ( M^{\delta,-\delta} )^{2} \; )$. Then for $ \sigma^{\lambda_{1} ,
 \lambda_{2} , \lambda_{3}}_{~~\delta e^{+}} $ we find [54]:
$$
\sigma^{\lambda_{1} , \lambda_{2} , \lambda_{3}}_{~~\delta e^{+}} =
\sum_{i=1}^3 \alpha_{i} \; (\;1 + \delta \lambda_{i} \;  c_{i}\;) \;
( 1 - n_{jk} )^{2}/2 m^2 \; ,
\eqnum {4.14}
$$
which determines the annihilation cross section when all the particles except
the electron are helically polarized. For the quantities
$$
 \sigma^{\lambda_{1} , \lambda_{2} , \lambda_{3}} = 1/4 \sum
_{\delta} (( M^{\delta,\delta} )^{2} + ( M^{\delta,-\delta} )^{2} ),
 \; \sigma^{\lambda_{1}}_{\delta e^{+}} = 1/2 \sum_{\lambda_{2},\lambda_{3}}
(( M^{\delta,\delta} )^{2} + ( M^{\delta,-\delta} )^{2}) \; ,
$$
the meaning of which is clear from the notation, we find
$$
\sigma^{\lambda_{1} , \lambda_{2} , \lambda_{3}} = ( \alpha_{1} \;
( 1 - n_{23} )^{2} + \alpha_{2} \; ( 1 - n_{13} )^{2} + \alpha_{3} \;
( 1 - n_{12} )^{2} )/2 m^2 \; ,
\eqnum {4.15}
$$
$$
\sigma^{\lambda_{1}}_{\delta e^{+}}=2 \; (( 1+ \delta \lambda_{1} c_{1})(1-n_{23})^{2} +
(1-\delta \lambda_{1} c_{2})(1-n_{13})^{2}+(1-\delta \lambda_{1}
 c_{3})(1-n_{12})^{2} ) /m^{2} \; .
\eqnum {4.16}
$$
In the case of unpolarized particles we obtain the well known result [10]
$$
\overline{\sigma} = \sum_{\lambda_{1},\lambda_{2},\lambda_{3} } \;
\sigma^{\lambda_{1} , \lambda_{2} , \lambda_{3}}  = 4 \; ( \; ( 1 - n_{12} )^{2} + ( 1 - n_{13} )^{2} +
 ( 1 - n_{23} )^{2} \; ) /m^{2} \; .
\eqnum{4.17}
$$

Let us calculate the probability for the process (4.1) when one of the photons
is linearly polarized in the annihilation plane ($\sigma_{x}$) or perpendicular
to it ($\sigma_{y}$) (and the other two are unpolarized), and also the degree
of linear polarization $p_{l}$:
$$
p_{l} = (\sigma_{y} - \sigma_{x})/(\sigma_{y} + \sigma_{x}) \; .
\eqnum {4.18}
$$
For this we go from the helicity states $\mid + 1 >$ and $\mid - 1 >$ of
a photon of momentum $\vec k_1$
$$
\mid + 1> = ( \mid x> + i \mid y> ) /\sqrt {2} \; , \; \mid - 1> =
( \mid x> - i \mid y> ) /\sqrt {2} \; ,
$$
to states with linear polarization $\mid x>$ and $\mid y>$:
$$
<x \mid = ( <+1\mid + <-1\mid ) / \sqrt{2} \; , \;
<y \mid = i \; ( <+1\mid - <- 1\mid ) / \sqrt{2} \; .
$$
Then for the amplitudes and probabilities we find [54]
$$
M^{\delta}_{x} = ( M^{~\delta , ~\delta}_{\lambda 2,\lambda 3} +
M^{~\delta,~\delta}_{-\lambda 2,\lambda 3} ) / \sqrt{2} \; ,
M^{-\delta}_{~x} = ( M^{-\delta,~\delta}_{\lambda 2,\lambda 3} +
M^{-\delta,~\delta}_{-\lambda 2,\lambda 3} ) / \sqrt{2} \; ,
$$
$$
M^{\delta}_{y} = i \; ( M^{~\delta , ~\delta}_{\lambda 2,\lambda 3} -
M^{~\delta,~\delta}_{-\lambda 2,\lambda 3} ) / \sqrt{2} \; ,
M^{-\delta}_{~y} = i \; ( M^{-\delta,~\delta}_{\lambda 2,\lambda 3} -
M^{-\delta,~\delta}_{-\lambda 2,\lambda 3} ) / \sqrt{2} \; ,
$$
$$
\sigma_{x} = 1/4 \sum_{\delta \lambda_{2} \lambda_{3}} ( \mid M^{\delta}_{x} \mid^{2} + \mid M^{-\delta}_{x} \mid^{2} ) , \;
\sigma_{y} = 1/4 \sum_{\delta \lambda_{2} \lambda_{3}} ( \mid M^{\delta}_{y} \mid^{2} + \mid M^{-\delta}_{y} \mid^{2} ) \; .
$$
Computing $\sigma_{x} , \; \sigma_{y}$, and $p_{l}$, we obtain [54]
$$
\sigma_{x} = 2/m^{2} \; ( A - B ) , \; \sigma_{y} = 2/m^{2} \; ( A + B )
 , \; p_{l} = B / A \; ,
\eqnum {4.19}
$$
$$
A = ( 1 - n_{12} )^{2} + ( 1 - n_{13} )^{2} + ( 1 - n_{23} )^{2}, \;
B = ( 1 - n_{12} ) ( 1 - n_{13} ) ( 1 - n_{23} ) \; .
$$
Equation (4.19) coincides with the result of Refs. 55 and 56.

Let us use (4.9) and (4.10) to construct the amplitudes for orthopositronium
annihilation [10]: $X_{1,1} = M^{++} , X_{1,0} = ( M^{+-} + M^{-+} )
/\sqrt{2}$, and $X_{1,-1} = M^{--} (X_{1,\pm 1} = X_{1,\delta})$, corresponding
to the projections of the total spin of the system on the direction of
$\vec a_{3}$ equal to +1, 0, and -1, and the same for parapositronium
(with total spin and projection equal to zero): $X_{0,0}=(M^{+-}-M^{-+})
/ \sqrt{2}$. We find [54]:
$$
X_{1,\delta} = M^{\delta,\delta} \; , \; X_{1,0} = \sqrt{2} \; M^{\delta,
-\delta} \; , \; X_{0,0} = 0 \; .
\eqnum {4.20}
$$
Summation of $X_{1,\delta}^{2}$ and $X_{1,0}^{2}$ over the photon
polarizations $\lambda_{2}$ and $\lambda_{3}$ gives
$$
\sum_{\lambda_2,\lambda_3} X_{1,\delta}^{2} = {2 \over m^2} \; ((1+\delta
\lambda_{1} c_{1})^2 (1-n_{23})^{2}+(1-\delta \lambda_{1} c_{2})^2
(1-n_{13} )^{2}  +(1-\delta \lambda_{1} c_{3})^2 (1-n_{12})^{2}) ,~~~
$$
$$
\sum_{\lambda_2,\lambda_ 3} X_{1,0}^{2} = 4 \; ( (1 - n_{12})^{2}(1-c_{3}^{2})
+ (1 - n_{13})^{2}(1-c_{2}^{2}) + (1 - n_{23})^{2}(1-c_{1}^{2}) ) /m^{2} \;  .
$$
Averaging the squares, we again obtain the well known result [10]
$$
\sum_{\lambda_1,\lambda_2,\lambda_3} \; ( X_{1,+1}^{2} + X_{1,0}^{2} + X_{1,-1}^{2} ) =
4 \; \overline{\sigma} \; ,
$$
where $\overline{\sigma}$ is given by (4.17).

It was shown in Ref. 55 that the amplitudes for the three-photon annihilation
of orthopositronium $H_{fi} \; ( X_{1,+1}, \; X_{1,0}, \; X_{1,-1})$ can
be written as
$$
H_{fi} = \vec t \vec u \; , \; \vec u = \vec u_{1} + \vec u_{2} + \vec u_{3} \; ,
\eqnum {4.21}
$$
where the vector $\vec u_{1}$ is a function of the photon polarization
vectors:
$$
\vec u_{1} = \vec e_{1} \; ( \vec e_{2} \vec e_{3} - \vec e_{2}^{~'}
\vec e_{3}^{~'} ) + \vec e_{1}^{~'} \; ( \vec e_{2} \vec e_{3}^{~'} +
\vec e_{3} \vec e_{2}^{~'} ) \; ,
\eqnum {4.22}
$$
$$
\vec e_{i} = \vec e_{\lambda i} \; ,\; \vec e_{i}^{~'} =
[ \vec e_{i} \vec n_{i} ] \; ( i = 1,2,3 ) \; ,
$$
and the vectors $\vec u_{2}$ and $\vec u_{3}$ are obtained from $\vec u_{1}$
by cyclic permutation of the indices. The complex vector $\vec t$
characterizes the triplet state of orthopositronium.

Let us construct the tensor $\Phi = \vec u \cdot \vec u^{~\ast}$ in terms
of which the three-photon annihilation probability is expressed. According
to our calculations, the tensor $\Phi$ can be written as three terms [54]:
$$
\Phi = a_{1} \Phi_{1} + a_{2} \Phi_{2} + a_{3} \Phi_{3} \; ,\; Tr(\Phi) =
a_{1} + a_{2} + a_{3} = m^{2} \; \sigma^{\lambda_{1},\lambda_{2},
\lambda_{3}} /2 \; ,
\eqnum {4.23}
$$
$$
a_{1} = \alpha_{1} (1-n_{23})^{2} \; , \; a_{2} = \alpha_{2} (1-n_{13})^{2} \; , \;
 a_{3} = \alpha_{3} (1-n_{12})^{2} \; ,
\eqnum {4.24}
$$
where each of the tensors $\Phi_{i}$ (i=1,2,3) is just the "beam tensor"
(the three-dimensionally covariant polarization density matrix) of the
corresponding circularly polarized photon [57]:
$$
\Phi_{i} = \vec e_{\lambda i} \cdot \vec e_{\lambda i}^{~\ast} = (1 -
\vec n_{i}\cdot\vec n_{i})/2 + i/2 \lambda_{i} \; \vec n_{i}^{~\times} \; ,
\; \Phi_{i} \vec n_{i} = 0 \; , \; Tr(\Phi_{i}) = 1 \; .
\eqnum {4.25}
$$
Since the tensor $\Phi$ corresponds to the sum of three waves and its
trace $Tr(\Phi) = a_{1} + a_{2} + a_{3}$ coincides up to an overall
coefficient with the probability $\sigma^{\lambda_{1}, \lambda_{2},
\lambda_{3}}$, each of the $a_{i} \; (i=1,2,3)$ in (4.24) determines the
probability for the appearance of a single photon having polarization vector
$\vec e_{\lambda i}$ and direction of motion $\vec n_{i}$.

\section{\bf The reaction $ep \to ep \gamma$ and the proton polarizability}

There has recently been much interest in studying Compton scattering on
nucleons at low and intermediate energies. This is because the fundamental
structure constant of the nucleon-the electric and magnetic polarizabilities-
can be determined in this process. The nucleon polarizabilities contain
important information about the nucleon structure at large and intermediate
distances, in particular, the radius of the quark core, the meson cloud,
and so on (see the detailed discussion of these questions in Refs. 58 and
59). Knowledge of the amplitudes for Compton scattering on nucleons is also
required to interpret the data on photon scattering on nuclei. For example,
studies of this type can answer the question of how greatly the
electromagnetic properties of free and bound nucleons differ.

All the experimental results on the proton polarizabilities have been obtained
from data on elastic $\gamma p$ scattering below the pion photoproduction
threshold. However, it has recently been shown that measurement of the proton
polarizabilities at the Novosibirsk storage ring with electron beam energy
of 200 MeV using an internal jet target appears very promising. As proposed
in [60], this can be done using the reaction
$$
e^{-}(p_{1})+p^{+}(q_{1}) \to e^{-}(p_{2}) + p^{+}(q_{2}) + \gamma(k)
\eqnum {5.1}
$$
in the kinematics corresponding to electron scattering at small angles and
photon scattering at rather large angles, which corresponds to small
4-momentum transfer from the initial electron to the final $\gamma$ and
$p$. In the lowest order of perturbation theory, the process (5.1) is
described by the three graphs shown in Fig. 1.

\vspace{1.7cm}
\begin{figure}[hbt]
\centerline{\epsfbox[10 10 550 700]{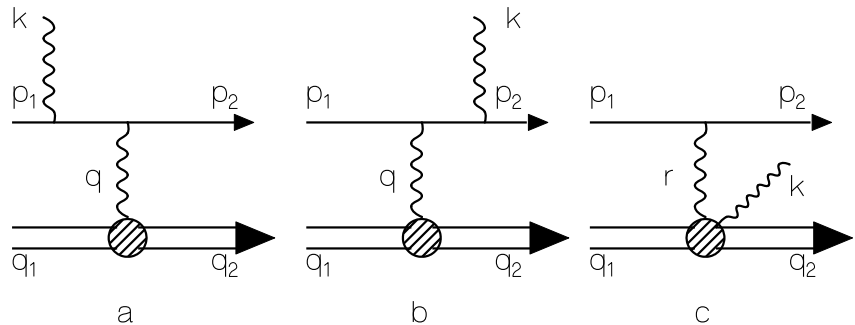}}
\vspace{-22cm}
\caption{ Graphs corresponding to the reaction $ep \to ep\gamma$. }
\end{figure}

The first two (a) and (b) correspond to electron bremsstrahlung
(Bethe-Heitler graphs), and the third (c) corresponds to proton
bremsstrahlung [graph with virtual Compton scattering (VCS) on a proton].
The kinematics described above was chosen for the following reasons.
First, the subprocess of real Compton scattering (RCS) on the proton
is realized in it, because at small electron scattering angles
the virtual photon with 4-momentum $r = p_{1} - p_{2}$ (see Fig. 1) becomes
almost real. Here the quantity $|r| = \sqrt{-(p_{1} - p_{2})^{2}}$ turns
out to be small, $|r| \sim m$, where $m$ is the electron mass. Second, for
electron scattering at small angles and photon scattering at fairly large
angles, the contribution of the graph corresponding to proton bremsstrahlung
dominates, i.e., it is several orders of magnitude larger than the
contribution of the Bethe-Heitler graphs to the cross section for the process
(5.1) (Ref. 61). This is the main requirement needed to isolate the subprocess
of Compton scattering on a proton [60] from the reaction $e p\to ep \gamma$.

The estimates made in Ref. 60 using the method of equivalent photons and the
scalar model showed that the reaction (5.1) offers a good possibility of
obtaining high-statistics data on the Compton scattering cross section and
the proton polarizability. Measurement of the electric and magnetic
polarizabilities of the proton ($\alpha_{p}$ and $\beta_{p}$) with higher
accuracy than in earlier studies is one of the most important problems to
be solved by experiments in the near future [62,63].

However, to obtain high-statistics data on the cross section for $\gamma
p$ scattering and the proton polarizability it is essential to use a
theoretical model more accurate than that in Ref. 60. It must include both
the spin properties of the particles and the main structural parameters
characterizing  the electromagnetic structure of the hadron. The model can
be based on the result of Ref. 64, where a general calculation of the reaction
$e p\to e p \gamma$ was performed. The cross section was expressed in terms
of 12 form factors corresponding to the VCS subprocess on the proton
(i.e., the contribution of the graph in Fig. 1c) and two form factors
corresponding to the Bethe-Heitler graphs.

The differential cross section for the reaction $e p\to e p \gamma $ in the
above kinematics was calculated in Ref. 65. It was expressed in terms of the
six invariant amplitudes for RCS [58,66], and also the electric and magnetic
form factors of the proton [10].

Let us consider the amplitudes corresponding to the graphs of Fig. 1. The
sum of the two Bethe-Heitler graphs (a) and (b) corresponds to the matrix
element
$$
M_{1} = \overline{u}(p_{2}) Q^{\mu}_{e} u(p_{1}) \cdot \overline{u}(q_{2})
\Gamma_{\mu}(q^{2}) u(q_{1}) \; {1 \over q^{2}} \; ,
\eqnum {5.2}
$$
$$
Q^{\mu}_{e} = \gamma^{\mu} \; {\hat p_{1} - \hat k + m \over - 2 p_{1} k }
\; \hat e + \hat e \; {\hat p_{2} + \hat k + m \over  2 p_{2} k } \;
\gamma^{\mu} \; ,
\eqnum {5.3}
$$
$$
\Gamma_{\mu}(q^{2}) = f_{1} \; \gamma_{\mu} + {\mu_{p} \over 4M} \; f_{2} \;
( \; \hat q \gamma_{\mu} - \gamma_{\mu} \hat q \; ) \; ,
\eqnum {5.4}
$$
where $u(p_{i})$ and $u(q_{i})$ are the bispinors of electrons and protons
with 4-momenta $p_{i}$ and $q_{i}$, $p_{i}^{2} = m^{2}, \; q_{i}^{2}
= M^{2}, \; \overline{u}(p_{i})\; u(p_{i})=2m, \; \overline{u}(q_{i}) \;
u(q_{i})= 2M, \; (i = 1,2), \; \mu_{p}, \;  f_{1}$, and $f_{2}$ are
respectively the anomalous magnetic moment and the Dirac and Pauli form factors
of the proton [10], $q = q_{2}-q_{1}$ is the momentum transfer, $e$ is the
polarization 4-vector of a photon with momentum $k, \; e k=k^{2}=0$, and
$M$ is the proton mass.

In the limit of interest $ |r| \sim m$, the matrix element corresponding
to the graph of Fig. 1c will be expressed in terms of the six invariant RCS
amplitudes $T_{i} \; (i = 1,2 \ldots 6) $ obtained from the theory of
dispersion relations and the data on $\pi$-meson photoproduction on
nucleons [66]. It has the form [64]
$$
M_{2} = \overline{u}(p_{2}) \gamma^{\mu} u(p_{1}) \cdot \overline{u}(q_{2})
M_{\mu \nu} e^{\nu} u(q_{1}) \; {1 \over r^{2}} \; ,
\eqnum {5.5}
$$
$$
M_{\mu \nu} = { C_{\mu} C_{\nu} \over C^{2}} \; ( T_{1} + T_{2} \hat K ) +
{ D{\mu} D{\nu} \over D^{2}} \; ( T_{3} + T_{4} \hat K ) +
$$
$$
+ { (C_{\mu} D_{\nu} - C_{\nu} D_{\mu}) \over D^{2}} \; \gamma^{5} \;
T_{5} \;  + \; { (C_{\mu} D_{\nu} + C_{\nu} D_{\mu}) \over D^{2}} \;
T_{6} \hat D \; .
\eqnum {5.6}
$$
The tensor $M_{\mu \nu}$ is constructed using a set of four mutually
orthogonal 4-vectors $C, D , B$, and $K$:
$$
K = 1/2 \; ( r + k ) \; , \; Q = 1/2 ( r - k ) \; , \; R = 1/2 ( q_{1} + q_{2} ) \; ,
$$
$$
C = R - { ( R K )\over K^{2}} \; K - { ( R B )\over B^{2}} \; B \; , \;
B = Q - { ( Q K )\over K^{2}} \; K  \; ,
\eqnum {5.7}
$$
$$
D_{\mu} = \varepsilon_{\mu \nu \rho \sigma} K^{\nu} B^{\rho} C^{\sigma} \; ,
$$
and it satisfies the requirements of parity conservation and gauge
invariance:
$$
M_{\mu \nu} k^{\nu} = r^{\mu} M_{\mu \nu} = 0 \; .
\eqnum {5.8}
$$
To calculate the matrix elements (5.2) and (5.5) in the DSB, we introduce
two OVBs $a_{A}$ and $b_{A} \; (A = 0,1,2,3)$ using the 4-momenta $p_{1},
\; p_{2}, \; k $ and $ q_{1}, \; q_{2}, \; k $:
$$
a_{0} = p_{+}/\sqrt{p_{+}^{2}} \; , \; a_{3} = p_{-}/\sqrt{-p_{-}^{2}} \; ,
\; a_{2} = [a_{0} \cdot a_{3}]^{\times} k /\rho \; , \;  a_{1} = [a_{0}
\cdot a_{3}]^{\times} a_{2} \; ,
\eqnum {5.9}
$$
$$
p_{\pm} = p_{2} \pm p_{1} \; , \; a_{\pm \delta} = a_{1} \pm i \delta a_{2} \; ,
 \; \delta = \pm 1 \; , \; a_{2} k = 0 \; , \; a_{1}^{2} =
a_{2}^{2} = a_{3}^{2} = - a_{0}^{2} = - 1 \; .
$$
$$
b_{0} = q_{+}/\sqrt{q_{+}^{2}} \; , \; b_{3} = q_{-}/\sqrt{-q_{-}^{2}} \; ,
\; b_{2} = [b_{0} \cdot b_{3}]^{\times} k /\rho^{'} \; , \;  b_{1} = [b_{0}
\cdot b_{3}]^{\times} b_{2} \; ,
\eqnum {5.10}
$$
$$
q_{\pm} = q_{2} \pm q_{1} \; , \; b_{\pm \delta^{'}} = b_{1} \pm i \delta^{'} b_{2} \; ,
 \; \delta^{'} = \pm 1 \; , \; b_{2} k = 0 \; , \; b_{1}^{2} =
b_{2}^{2} = b_{3}^{2} = - b_{0}^{2} = - 1 \; .
$$
where $\rho$ and $\rho^{'}$ are determined from the normalization conditions.
Then the electron and proton operators $P_{21}^{\pm \delta,\delta}$ and
$Q_{21}^{\pm \delta^{'}, \delta^{'}}$ [see (2.11) and (2.12)] will have the
form
$$
P_{21}^{\delta,\delta} = 1/4 \; ( m + \hat p_{1} \; ) \; \hat a_{\delta} \;
\hat a_{0} \; \hat a_{\delta}^{\ast} \; , \; P_{21}^{-\delta,\delta} =
\delta/2 \; ( m +  \hat p_{1} \; ) \; \hat a_{\delta} \; \hat a_{3} \; \; ,
\eqnum {5.11}
$$
$$
Q_{21}^{\delta ',\delta '} = 1/4 \; ( M + \hat q_{1} \; ) \;
\hat b_{\delta '} \; \hat b_{0} \; \hat b_{\delta '}^{\ast} \; , \;
Q_{21}^{-\delta ',\delta '} = \delta^{'}/2 \; ( M +  \hat q_{1} \; ) \;
\hat b_{\delta '} \; \hat b_{3} \; \; ,
\eqnum {5.12}
$$
while the matrix elements (5.2) and (5.5) in the case of various combinations
of electron and proton spin states reduce to a product of traces:
$$
M_{1}={1\over q^{2}} \;Tr ( \; P_{21}^{\pm\delta,\delta} Q^{\mu}_{e} \;)\;
Tr(\; Q_{21}^{\pm\delta ',\delta '} \Gamma_{\mu}(q^{2}) \; ) \; ,
\eqnum {5.13}
$$
$$
M_{2} = {1\over r^{2}} \; Tr ( \; P_{21}^{\pm\delta,\delta} \gamma^{\mu}
\; )\; \; Tr( \; Q_{21}^{\pm\delta ',\delta '} M_{\mu \nu} e^{\nu} \; ) \; .
\eqnum {5.14}
$$
In the unpolarized case it is most efficient to use the calculation of the
matrix elements in the DSB in conjunction with the standard approach [10].
The calculations performed by the first [i.e., using (5.13) and (5.14)] and
second methods give identical result. Nevertheless, the second method, which
will also be discussed below, is preferable, because it gives results
considerably more quickly. To find the probability for the process (5.1)
it is sufficient to calculate only the matrix elements of the electron and
proton currents:
$$
( J^{\pm\delta,\delta}_{e} )_{\mu} = \overline{u}^{\pm\delta}(p_{2})
\gamma_{\mu} u(p_{1})^{\delta}  = Tr ( \; P_{21}^{\pm\delta,\delta}
\gamma_{\mu} \; ) \; ,
\eqnum {5.15}
$$
$$
( J^{\pm\delta ',\delta '}_{p} )_{\mu} = \overline{u}^{\pm \delta '}(q_{2})
\Gamma_{\mu}(q^{2}) u^{\delta '}(q_{1}) = Tr( \; Q_{21}^{\pm\delta ',
\delta '} \Gamma_{\mu}(q^{2}) \; ) \; ,
\eqnum {5.16}
$$
and also the quantity
$$
X^{\pm\delta ',\delta '}_{\mu} = \overline{u}^{\pm \delta '}(q_{2})
M_{\mu \nu} e^{\nu} u^{\delta '}(q_{1}) =
Tr( \; Q_{21}^{\pm\delta ',\delta '} M_{\mu \nu} e^{\nu} \; ) \; \; .
\eqnum {5.17}
$$
The calculations give [11,36,47]
$$
( J^{\delta,\delta}_{e} )_{\mu} = 2 m ( a_{0} )_{\mu} \; , \;
( J^{-\delta,\delta}_{e} )_{\mu} = - 2 \delta y_{-} ( a_{\delta} )_{\mu} \; ,
\eqnum {5.18}
$$
$$
( J^{\delta ',\delta '}_{p} )_{\mu} = 2 g_{e} M ( b_{0} )_{\mu} \; , \;
( J^{-\delta ',\delta '}_{p} )_{\mu} = - 2 \delta^{'} y_{-}^{'} g_{m} (
b_{\delta '} )_{\mu} \; ,
\eqnum {5.19}
$$
where $y_{-}=\sqrt{-p_{-}^{2}} /2 \; ,\; y_{-}^{'}=\sqrt{-q_{-}^{2}}/2$,
and $g_{e}$ and $g_{m}$ are just the proton electric and magnetic form
factors [10]:
$$
g_{e} = f_{1} + \mu_{p} {q^{2}\over 4 M^{2}} \; f_{2} \; , \; g_{m} =
f_{1} + \mu_{p} \; f_{2} \; .
\eqnum {5.20}
$$
Therefore, in the DSB the matrix elements of the proton current corresponding
to transitions without spin flip are expressed in terms of the electric form
factor $g_{e}$, and the interaction with spin flip is expressed in terms
of the magnetic form factor $g_{m}$.

After the matrix elements of the proton current (5.16) are determined, the
calculation of the contribution of the two Bethe-Heitler graphs reduces to
the calculation of VCS on the electron [47,65]:
$$
|M_{1}^{\pm\delta ',\delta '}|^{2} = {1\over q^{4}} |\overline{u}(p_{2})
\left (\hat J_{p}^{\pm\delta ',\delta '} {\hat p_{1} - \hat k + m \over
- 2 p_{1} k } \hat e + \hat e \; {\hat p_{2} + \hat k + m \over  2 p_{2} k }
\hat J_{p}^{\pm\delta ',\delta '} \right ) u(p_{1}) |^{2} \; .
\eqnum {5.21}
$$
Denoting the result of averaging and summing the expression
$|M_{1}^{\pm\delta ',\delta '}|^{2}$ over the polarizations of the initial
and final particles by $Y_{ee}$, we obtain [47,65]:
$$
Y_{ee} = 1/4 \; \sum_{\delta ' e} \; Tr\{ \; ( \hat p_{2} + m ) \;
\widehat Q_{e}^{ \pm\delta ',\delta '} \; ( \hat p_{1} + m ) \;
\widehat{\overline{Q}}_{e}^{\pm\delta ',\delta '} \; \} / q^{4} \; ,
\eqnum {5.22}
$$
where $ \widehat Q_{e}^{ \pm\delta ',\delta '} = ( Q_{e}^{\mu} ) \, (
 J_{p}^{\pm\delta ',\delta '} )_{\mu}$ is the operator in parentheses between
the electron bispinors $ \overline{u}(p_{2})$ and $u(p_{1})$ in Eq. (5.21),
and $\widehat{\overline{Q}}_{e}^{\pm\delta ',\delta '} = \gamma_{0}\, (
\widehat Q_{e}^{ \pm\delta ',\delta '} )^{+} \gamma_{0}$.
Owing to the factorization of the electric and magnetic form factors
$g_{e}$ and $g_{m}$ in (5.19), the Bethe-Heitler term in the cross section
for the reaction $ e p \to e p \gamma \; Y_{ee}$ (5.22) will contain only
the squares of the Sachs form factors (see Refs. 36,47,65,67, and 68).

Similarly, the calculation of the contribution of the graph in Fig. 1c reduces
to the calculation of quasireal Compton scattering on the proton. Using the
expressions for the electron current (5.18), we have
$$
\mid M_{2}^{\pm\delta,\delta}\mid^{2} = {1\over r^{4}} \mid \overline{u}
(q_{2}) \; \widehat Q_{p}^{ \pm\delta , \delta } \; u(q_{1}) \mid^{2} \; ,
\eqnum {5.23}
$$
where $\widehat Q_{p}^{ \pm\delta, \delta }=(J^{\pm\delta,\delta}_{e})^{\mu}
 \;M_{\mu \nu} e^{\nu}$. Denoting the result of averaging and summing Eq.
(5.23) over the polarizations of the initial and final particles by $Y_{pp}$,
we obtain [65]
$$
Y_{pp} = 1/4 \; \sum_{\delta e} \; Tr \{ \; ( \hat q_{2} + M ) \;
\widehat Q_{p}^{ \pm\delta , \delta } \; ( \hat q_{1} + M ) \;
\widehat{\overline{Q}}_{p}^{\pm\delta , \delta } \; \} / r^{4} \; ,
\eqnum {5.24}
$$
where $ \; \widehat{\overline{Q}}_{p}^{~\pm\delta, \delta }=\gamma^{0}\;
(\widehat Q_{p}^ { \pm\delta , \delta } )^{+} \gamma^{0}$. Finally, to
calculate the interference term in the case of unpolarized particles
$$
Y_{ep} = 1/4 \; \sum_{\delta,\delta ',e} \; 2 Re \; M_{1} \; M_{2}^{\ast}
\eqnum {5.25}
$$
we shall use the matrix elements of the proton current (5.19) and also the
4-vectors $X^{\pm\delta ',\delta '}_{\mu}$ (5.17), which have the form [65]
$$
X^{-\delta ',\delta '}_{\mu} = -2 \delta^{'} y_{-}^{'} b_{1} k \left (
{ C_{\mu} C_{\nu} \over C^{2}} T_{2} + { D{\mu} D{\nu} \over D^{2}} T_{4}
 + i \delta^{'} y_{+}^{'} y_{-}^{'} { (C_{\mu} D_{\nu} + C_{\nu} D_{\mu})
 \over D^{2}} T_{6} \right ) e^{\nu} \; ,
$$
$$
X^{\delta ',\delta '}_{\mu} = 2 \left ( y_{+}^{'} \left (
 { C_{\mu} C_{\nu} \over C^{2}} \left ( T_{1} + {\nu_{1} M \over 1 - \tau} T_{2} \right )
 + { D{\mu} D{\nu} \over D^{2}} \left ( T_{3} + {\nu_{1} M \over 1 - \tau} T_{4} \right )
 \right ) \right. +
\eqnum {5.26}
$$
$$
+ \left. \delta^{'} y_{-}^{'} { (C_{\mu} D_{\nu} - C_{\nu} D_{\mu})
\over D^{2}} \; T_{5}  \right )  e^{\nu} \; ,
$$
where $y_{+}^{'}=\sqrt{q_{+}^{2}}/2=M \sqrt{1-\tau}\;,\;\tau =q^{2}/4 M^{2}$
and $\nu_{1}=k q_{+}/2M^{2}$. As a result, for the matrix element $M_{2}$
(5.5) we find
$$
M_{2} = \overline{u}(p_{2}) \hat{X}^{\pm \delta ',\delta '} u(p_{1}) / r^{2} \; ,
\eqnum {5.27}
$$
and Eq. (5.25) reduces to calculation of the trace [65]:
$$
Y_{ep} = 1/4 \sum_{\delta,\delta ',e} 2 Re \; \{Tr\; ( ( \hat p_{2} + m )
 \widehat Q_{e}^{ \pm\delta ',\delta '}  ( \hat p_{1} + m )
 \hat{\overline{X}}^{\pm\delta ',\delta '} )\} / q^{2} /r^{2} \; ,
\eqnum {5.28}
$$
where $\hat{X}^{\pm \delta ',\delta '} = \gamma^{\mu} \; X^{\pm\delta ',
\delta '}_{\mu}$ and $\hat{\overline{X}}^{\pm\delta ',\delta '} =
( X^{\pm\delta ',\delta '}_ {\mu} )^{\ast} \gamma^{\mu} $. The interference
term $Y_{ep}$ (5.28) is a linear combination of the proton electric and
magnetic form factors, because the operators $\widehat Q_{e}^{\pm\delta ',
\delta '}$ are expressed linearly in terms of the matrix elements of the
proton current: $\widehat Q_{e}^{ \pm\delta ',\delta '} = (Q_{e})^{\mu}
( J^{\pm\delta ',\delta '}_{p} )_{\mu} \; $, [see Eqs. (5.3) and (5.19)].

Therefore, the problem of finding the probability for the reaction $ e p
\to e p \gamma$ in this approach has been reduced to calculation of the
traces (5.22), (5.24), and (5.28), which was done by means of the program
REDUCE. For the differential cross section we then obtained [65]:
$$
d \sigma = { \alpha^{3} \mid T \mid ^{2} \delta^{4} ( p_{1} + q_{1} - p_{2}
- q_{2} - k ) \over 2 \pi^{2} \sqrt{ (p_{1} q_{1})^{2} - m^{2} M^{2}}} \;
\; { d^{3} \vec p_{2} \over 2 p_{20}} \; {d^{3} \vec q_{2} \over 2 q_{20}}
{d^{3} \vec k \over 2 \omega} \; \; ,
\eqnum {5.29}
$$
$$
\mid T \mid^{2} = 1/4 \; \sum_{pol} \mid M_{fi} \mid^{2} = Y_{ee} + Y_{ep}
+ Y_{pp} \; ,
\eqnum {5.30}
$$
$$
Y_{ee} = {8 M^{2}\over q^{4}} \; ( \; g_{e}^{2} \; Y_{I} + \tau \; g_{m}^{2} \; Y_{II} \; ) \; ,
\eqnum {5.31}
$$
$$
Y_I= -\; \frac{\lambda_1}{\lambda_2}-\frac{\lambda_2}{\lambda_1}-
\frac{m^2q^2}2\left(\frac 1{\lambda_1}-\frac 1{\lambda_2}\right)^2-
\frac {r^2q^2}{2\lambda_1\lambda_2}
$$
$$
-\; {m^{2} \over 2 M^2(1-\tau)} \left ( {p_{1} q_{+} \over \lambda_{2}} -
{p_{2} q_{+} \over \lambda_{1}} \right )^{2} - {\tau \over (1 - \tau)}
{( (p_{1} q_{+})^{2} + (p_{2} q_{+})^{2} ) \over \lambda_{1} \lambda_{2}} \; ,
\eqnum {5.32}
$$
$$
Y_{II}= -\; \frac{\lambda_1}{\lambda_2}-\frac{\lambda_2}{\lambda_1}-
\frac{m^2q^2}2\left(\frac 1{\lambda_1}+\frac 1{\lambda_2}\right)^2-
\frac {r^2q^2}{2\lambda_1\lambda_2}
$$
$$
+\; {m^{2} \over 2 M^2(1-\tau)} \left ( {p_{1} q_{+} \over \lambda_{2}} -
{p_{2} q_{+} \over \lambda_{1}} \right )^{2} + {\tau \over (1 - \tau)}
{( (p_{1} q_{+})^{2} + (p_{2} q_{+})^{2} ) \over \lambda_{1} \lambda_{2}}
\eqnum {5.33}
$$
$$
- \; 2 \left ( {m^2 \over \lambda_{1}} - {m^2 \over \lambda_{2}} \right )^{2} + 4
\; m^2 \left ( {1 \over \lambda_{1}} - {1 \over \lambda_{2}} \right ) \; ,
\nonumber
$$
$$
Y_{ep}=-\; {32 M^3\over r^2 q^2 (4 \nu_{4}^2 - \nu_{2}^2)} \Biggl \{
g_{e} Re \left [ y_{1} \left ( T_{1} + {\nu_{1} M\over 1 - \tau }
 T_{2} \right ) + y_{2} \left ( T_{3} + {\nu_{1} M\over 1 - \tau } T_{4}
 \right ) \right ] \biggr.
$$
$$
\biggl. + \tau g_{m} \left [-{\nu_{1} M\over 1 - \tau }
Re ( y_{1} T_{2} + y_{2} T_{4} ) + 4 M Re
( z_{1} T_{2} + z_{2} T_{4} + z_{3} T_{6} ) \right ] \Biggr \},
\eqnum {5.34}
$$
$$
Y_{pp}=-\; \Biggl \{ (\alpha_{1}^2 \alpha_{3} +
\nu_{3}) [(1-\tau) |T_{1}|^2+2 \nu_{1} M Re(T_{1}
T_{2}^{\ast}) + M^2 (\nu_{1}^2 - \nu_{2}^2) |T_{2}|^2]
$$
$$
+\; (\alpha_{2}+\nu_{3}) [(1-\tau) |T_{3}|^2+2 \nu_{1} M
Re(T_{3} T_{4}^{\ast})+M^2(\nu_{1}^2-\nu_{2}^2) |T_{4}|^2]
\eqnum {5.35}
$$
$$
~ + \; ( \alpha_{1}^2 \alpha_{3} + \alpha_{2} + 2 \nu_{3} ) \tau \left (
 - {|T_{5}|^2 \over M^4 \nu_{2}^2} + { M^2 \over \alpha_{3} }
|T_{6}|^2 \right ) \Biggr \} \; {16 M^4 \over r^{4}} \; .
$$

For the invariant variables in Eqs. (5.30)-(5.35) used in determining the
Bethe-Heitler term ($Y_{ee}$), the interference term ($Y_{ep}$), and the
term corresponding to proton bremsstrahlung ($Y_{pp}$), we used the notation
adopted in Ref. 64:
\begin{eqnarray}
&&y_{1} = 2 \alpha_{1} [ \alpha_{1} \alpha_{3} ( \nu_{2} \nu_{5} - \nu_{1}
 \nu_{4} ) + 2 \nu_{4}^2 + \nu_{2} \nu_{3} ] , \; \nu_{1} = k q_{+}/2M^{2}\; ,
 \; \nu_{2} = - kq_{-}/ 2 M^2 \; ,\nonumber \\
&&y_{2} = 2 \alpha_{2} \; ( \nu_{2} \nu_{5} - \nu_{1} \nu_{4} ) - \alpha_{1}
 \nu_{2}^2  , \; \nu_{3} = r^2 / 4 M^2  , \; \nu_{4} = kq_{+}/ 4 M^2 \; ,
\; \nu_{5} = p_{+} q_{+} / 4 M^2  ,\nonumber \\
&&y_{3}=-(4 \nu_{3} / \nu_{2}^2 ) \; [\alpha_{1} \alpha_{3} (\nu_{1} \nu_{2}
 ( \nu_{2} + \nu_{3} ) - 2 \nu_{4} ( \nu_{1} \nu_{4} - \nu_{2} \nu_{5} ) ) +
\nu_{4} ( 4 \nu_{4}^2 - \nu_{2}^2 ) ] \; , \nonumber \\
&&\alpha_{1}=\nu_{5}+\nu_{1}\nu_{4}(2 \nu_{3}+\nu_{2}) / \nu_{2}^2 \; ,
\; \alpha_{3}=\nu_{2}^2 / (\nu_{2}^2+(\nu_{2}+\nu_{3})(\nu_{1}^2 -
\nu_{2}^2 ) ) \; , \nonumber \\
&&\alpha_{2} = m^2/M^2-\nu_{3}+M^6/D^2[-(\nu_{1} \nu_{4}+\nu_{2}
\nu_{5})^2+4\nu_{3}(\nu_{4}^2-\nu_{1} \nu_{4} \nu_{5})-4\nu_{3}
\nu_{4}^2(\nu_{2}+\nu_{3})] \; ,\nonumber \\
&&D^2 = M^6 \; ( \nu_{2}^2 + ( \nu_{2} + \nu_{3} ) ( \nu_{1}^2 - \nu_{2}^2 ) ) =
M^6 \nu_{2}^2 / \alpha_{3} \; , \; \lambda_{1} = p_{1} k \; , \; \lambda_{2}
= p_{2} k \; ,\nonumber \\
&&z_{1} = \nu_{1} \nu_{4} \alpha_{1}^2 \alpha_{3} \; ,
 \; z_{2} = \nu_{2} \nu_{4} \alpha_{2} \; , \;
z_{3} = 1/4 \alpha_{1} \; ( 2 \nu_{2} ( 2 \alpha_{2} + \nu_{2} + \nu_{3} ) +
 4 \nu_{4}^2 - \nu_{2}^2 ) \; .\nonumber
\end{eqnarray}

We note that the expression obtained for the differential cross section
(5.29) coincides, apart from the definition of the initial quantities
(the tensor $M_{\mu \nu}$), with the result obtained in Ref. 64, if in the
latter $f_{1}$ and $f_{2}$ are expressed in terms of $g_{e}$ and $g_{m}$.
Nevertheless, the Bethe-Heitler term $Y_{ee}$ and the interference term
$Y_{ep}$ have a more compact form, owing to the factorization of the electric
and magnetic form factors.

Let us consider the effects due to contribution of all three graphs to the
cross section for the reaction (5.1) in the selected kinematics when the
initial proton is at test [$q_{1} = (M,0)$] and the electron beam energy
is $E_{e} = 200$ MeV. Performing the required integration over the phase
space in the rest frame of the initial proton, we obtain [65]:
$$
d \sigma = { \alpha^{3} \omega ^2 \mid \vec q_{2} \mid T \mid ^{2}
 \over 16 \pi^{2} M \mid \vec p_{1} \mid (p_{2} k) } \;
\;  d E_{pk} \; d \Omega_{q_2} \; d \Omega_{\gamma} \; ,
\eqnum {5.36}
$$
where $d \Omega_{\gamma}$ and $d \Omega_{q_2}$ are the elements of the photon
and proton solid angles, and $E_{pk}$ is the kinetic energy of the recoil
proton.

Let calculate the differential cross section (5.36) numerically in the region
$5 \leq E_{pk} \leq 35$ MeV with the sum and the difference of the electric
($\alpha_{p}$) and magnetic ($\beta_{p}$) polarizabilities equal to
$\alpha_{p}+ \beta_{p} = 14$ and $\alpha_{p} - \beta_{p} = 10$ (in units
of $10^{-4}fm^3$) [58-60]. We assume that the reaction kinematics is
planar, and that the photon emission and proton scattering angles are
$\vartheta_{\gamma}=135^{0}$ and $\vartheta_{p}=- 20.5^{0}$. (All angles
are measured from the direction of motion of the primary electron beam).
The calculation [65] show that in the entire range of proton kinetic energy
considered, $5 \leq E_{pk} \leq 35$ MeV, for the selected angles
$\vartheta_{\gamma} =135^{0}$ and $\vartheta_{p} = - 20.5^{0}$ the electron
scattering angle $\vartheta_{e}$ and the 4-momentum transfer $ |r| =
\sqrt{-(p_{2}-p_{1})^2}$ are bounded by the values $| \vartheta_{e}|
\leq 6.4^{0}$ and $|r| \leq 7.3$ MeV, with the minimum value of $|r|$
corresponding to forward electron scattering.

The results of numerical calculations of the differential cross section
(5.36), $d \sigma / d E_{pk} / d \Omega_{q_2} / d \Omega_{\gamma}$ in the
kinematics described above are shown graphically in Fig. 2. We see that in
the angular range studied the cross section for the reaction $ e p \to
e p \gamma$ has a sharp peak consisting of two maxima. This peak originates
from the factor $1/r^4$ in Eq. (5.35) for $Y_{pp}$. The two maxima have a
kinematical origin and arise from the interference of two pole graphs
corresponding to quasireal Compton scattering. The cross section (5.36)
has a strong angular dependence, which, in particular, causes the two maxima
to disappear when the proton (or photon) emission angle is changed by only
one a degree (i.e., for $\vartheta_{p} = - 19.5^{0}$), so that we have an
ordinary peak at $E_{pk} = 25$ MeV.

\begin{figure}[hbt]
\vspace{2.3cm}
\centerline{\epsfxsize=0.60\textwidth\epsfbox[10 10 550 600]{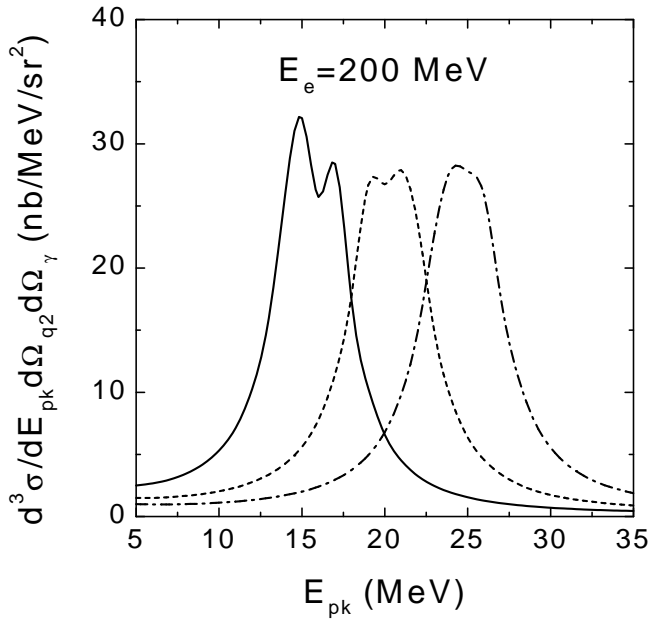}}
\vspace{-7.2cm}
\caption
{Differential cross section (5.36) for reaction $ep \to ep\gamma$
in the kinematics where proton bremsstrahlung dominates, see comments in the
text. Proton scattering and photon emission angles are $\vartheta_p=-20.5^0$
(solid line), $\vartheta_p= -20.0^0$ (dashed line), $\vartheta_p=-19.5^0$
(dot-dashed line), and $\vartheta_{\gamma}= 135^0$.}
\end{figure}

The differential cross section (5.36), shown by the graphs in Fig. 2, is
the sum of the Bethe-Heitler ($\sigma_{ee}$), the interference
($\sigma_{ep}$), and the proton ($\sigma_{pp}$) terms [see (5.30)], where
the symbol ($\sigma$) denotes a cross section of the form (5.36) with
$\mid T \mid^2$ replaced by $Y_{ee}$, $Y_{ep}$, and $Y_{pp}$, respectively.
Numerical calculations shown that in the entire range of proton kinetic
energy studied, $5 \leq E_{pk} \leq 35$ MeV, the ratios of the Bethe-Heitler
term $\sigma_{ee}$ and the interference term $\sigma_{ep}$ to the term
corresponding to proton emission $\sigma_{pp}$ are bounded by the values
$\sigma_{ee} /\sigma_{pp}< 0.02$ and $| \sigma_{ep} | / \sigma_{pp} < 0.05$.
The calculations carried out for another set of angles ($\vartheta_{\gamma}
=135^{0}$ and $\vartheta_{p} = - 20^{0}$) give results which are only
insignificantly different: $\sigma_{ee} / \sigma_{pp}< 0.05$ and $|
\sigma_{ep} | / \sigma_{pp} < 0.075$. Since these ratios are much smaller
than unity, the main requirement (see Ref. 60) for separation of the
background, which is mainly electron bremsstrahlung, is satisfied.

To explain the sensitivity of the reaction $ e p \to e p \gamma$ to the
proton polarizability we performed numerical calculations of the cross section
(5.36) for the same set of angles ($\vartheta_{\gamma} = 135^{0}$ and
$\vartheta_{p} = - 20^{0}$) for fixed sum of the electric and magnetic
polarizabilities $\alpha_{p} + \beta_{p} = 14$ but different values of the
difference: (a) $\alpha_{p} - \beta_{p} = 10$ and (b) $\alpha_{p} - \beta_{p}
= 6$. It turned out that the cross section (5.36) is about 8\% larger for
the smaller difference of polarizabilities. Therefore, in this kinematics
the cross section for the reaction $ e p \to e p \gamma$ is quite sensitive
to the proton polarizability [65].

\section{\bf Emission of a linearly polarized photon by an electron
in the reaction $ep \to ep \gamma$}

Let us consider the emission of a linearly polarized photon by an electron
in the reaction $ ep \to ep \gamma$, taking into account the proton recoil
and form factors. Our study will be limited to the contribution of the two
Bethe-Heitler graphs (a) and (b) in Fig. 1, which corresponds to the matrix
element (5.2). The contribution of the graph with VCS on a proton can be
neglegted when the initial electrons have ultrarelativistic energies, and
the photon and final electron are scattered at small forward angles
($\vartheta_{\gamma} \sim m/E_{e}, \; \vartheta_{e} \sim m/E_{e}, \;
m/E_{e} \ll 1 $).

We are interested in these effects for the following reasons. First, even
though the Bethe-Heitler process has been studied earlier in the case of
the emission of linearly polarized photons [69,70] and is widely used to
obtain them at accelerators [71], up to now the proton recoil and form factors
have not been accurately taken into account (in contrast to the unpolarized
case). Second, as was shown in Ref. 72, the inclusion of these factors in
the case of unpolarized photons leads to a strong change of the differential
cross section for the Bethe-Heitler process. Since the polarization
characteristic of the scattered radiation are expressed in terms of the
differential cross section for the emission of an unpolarized photon (see
below), it is clear that inclusion of the recoil and form factors is
essential.

The covariant expression for the differential cross section for the
Bethe-Heitler process (in the Born approximation) taking into account the
proton recoil and form factors in the case of emission of a linearly polarized
photon has been obtained by us in Ref. 73. It has the form
$$
d \sigma_{BH} = { \alpha^{3} \mid T_{e} \mid ^{2} \delta^{4} ( p_{1} + q_{1} - p_{2}
- q_{2} - k ) \over 2 \pi^{2} \sqrt{ (p_{1} q_{1})^{2} - m^{2} M^{2}}} \;
\; { d^{3} \vec p_{2} \over 2 p_{20}} \; {d^{3} \vec q_{2} \over 2 q_{20}}
{d^{3} \vec k \over 2 \omega} \; \; ,
\eqnum {6.1}
$$
$$
\mid T_{e} \mid^2 = {4 M^{2}\over q^{4}} \; ( \; g_{e}^{2} \; Y_{I}^{~e}
 + \tau \; g_{m}^{2} \; Y_{II}^{~e} \; ) \; ,
\eqnum {6.2}
$$
$$
Y_{I}^{~e} = 2 - {\lambda_{1} \over \lambda_{2}} - {\lambda_{2} \over \lambda_{1}}
- {\tau \over 1 - \tau} \; {(k q_{+})^2 \over \lambda_{1} \lambda_{2}} +
 q^2 \; (e a)^2 + 4 \; (e A)^2 \; ,
\eqnum {6.3}
$$
$$
Y_{II}^{~e} = - 2 - {\lambda_{1} \over \lambda_{2}} - {\lambda_{2} \over \lambda_{1}}
+ {\tau \over 1 - \tau} \; {(k q_{+})^2 \over \lambda_{1} \lambda_{2}} +
 ( q^2 + 4 m^2 ) \; (e a)^2 - 4 \; (e A)^2 \; ,
\eqnum {6.4}
$$
$$
a = {p_{1} \over \lambda_{1}} - {p_{2} \over \lambda_{2}} \; \; ,
\; \; A = b_{0} + {(b_{0} p_{2}) p_{1} \over \lambda_{1}} - {(b_{0} p_{1}) p_{2}
 \over \lambda_{2}} \; .
\eqnum {6.5}
$$
All the quantities entering into (6.1)-(6.5) are defined in the preceding
section. Thus, the differential cross section for the Bethe-Heitler process
in the case of emission of a linearly polarized photon $d \sigma_{BH}$
(6.1) naturally splits into the sum of two terms containing only the squares
of the Sachs form factors and corresponding to the contribution of transition
without ($\sim g_{e} ^2 \; Y_{I}^{~e}$) and with ($\sim \tau \; g_{m}^2
Y_{II}^{~e}$) proton spin flip.

Let us discuss the properties of the 4-vector $a$, which is well known from
the theory of emission of long-wavelength photons [10], and the 4-vector
$A$. They both satisfy a condition which follows naturally from the
requirement of gauge invariance:
$$
a\; k = A \; k = 0 \; ,
$$
and, in addition, they are spacelike vectors: $a^2<0 $ and $A^2<0$. This
is easily verifed by using the 4-momentum conservation law and the explicit
form of $a^2$ and $ A^2$:
$$
a^2 = m^2 \; \left ( {1 \over \lambda_{1}} - {1 \over \lambda_{2}} \right )^2
+ {r^2 \over \lambda_{1} \lambda_{2}} \; ,
\eqnum {6.6}
$$
$$
A^2 = 1 + {m^2 \over 4 M^2 (1-\tau)}\; \left ({q_{+}p_{1} \over \lambda_{2}}
- {q_{+}p_{2} \over \lambda_{1}} \right ) + {\tau \over 1-\tau} \; {q_{+}p_{1}
\cdot q_{+}p_{2} \over \lambda_{1} \lambda_{2}} \; .
\eqnum {6.7}
$$
We note that the 4-vector $A$ was first introduced in Ref. 73.

Using the electron 4-momenta $p_{1}$ and $p_{2}$ and the photon 4-momenta $k$,
we construct the 4-vectors of the photon linear polarization $e_{\parallel}$
and $e_{\perp} \;(e_{\parallel} k=e_{\perp}k=e_{\parallel}e_{\perp}=0$):
$$
e_{\parallel} = {(p_{2} k) p_{1} - (p_{1} k ) p_{2} \over \rho^{'} } \; , \; e_{\perp}
 = {[p_{1} \cdot p_{2}]^{\times} k \over \rho^{'}} \; ,
\eqnum {6.8}
$$
where $\rho^{'}$ is determined from the normalization conditions:
$e_{\parallel}^2 = e_{\perp}^2 = - 1$. Then the degree of photon linear
polarization will be given by the following expressions [73]:
$$
P_{\gamma} = { \mid T_{\perp} \mid^2 - \mid T_{\parallel} \mid^2 \over
\mid T_{\perp} \mid^2 + \mid T_{\parallel} \mid^2 } = {A_{1} \over A_{2}} \; ,
\eqnum {6.9}
$$
where
$$
A_{1} = {16 \;M^2 \over q^4} \; ( g_{e}^2 \; A_{11} + \tau \; g_{m}^2 \;
 A_{12} ) \; ,
\eqnum {6.10}
$$
$$
A_{2} = {8 \;M^2 \over q^4} \; ( g_{e}^2 \; Y_{1} + \tau \; g_{m}^2 \;
 Y_{2} ) \; ,
\eqnum {6.11}
$$
$$
A_{11} = A^2 + \tau \; M^2 \; a^2 + 2 (e_{\perp} b_{0})^2 \; ,
\eqnum {6.12}
$$
$$
A_{12} = - A^2 + \tau \; M^2 \; a^2 - 2 (e_{\perp} b_{0})^2 + m^2 \; a^2 \; ,
\eqnum {6.13}
$$
$$
(e_{\perp} b_{0})^2 = - \; {4 (SD)^2 \over M^2 (1-\tau) a^2 \lambda_{1}^2
 \lambda_{2}^2 } \; ,
\eqnum {6.14}
$$
$$
SD = 1/2 \; \epsilon_{\mu \nu \rho \sigma} (p_{1})^{\mu} (p_{2})^{\nu} (q_{1})
^{\rho} (q_{2})^{\sigma} \; ,
\eqnum {6.15}
$$
$$
Y_{1} = 2 - {\lambda_{1} \over \lambda_{2}} - {\lambda_{2} \over \lambda_{1}}
- {\tau \over 1 - \tau} \; {(k q_{+})^2 \over \lambda_{1} \lambda_{2}} -
 2 \; \tau M^2 \; a^2 - 2 \; A^2 \; ,
\eqnum {6.16}
$$
$$
Y_{2} = - 2 - {\lambda_{1} \over \lambda_{2}} - {\lambda_{2} \over \lambda_{1}}
+ {\tau \over 1 - \tau} \; {(k q_{+})^2 \over \lambda_{1} \lambda_{2}} -
 2 \; \tau \; M^2 \; a^2 + 2 \; A^2 - 2 m^2 \; a^2 \; .
\eqnum {6.17}
$$
It is easy to check that $A_{2}$ (6.11) coincides with the expression for
$Y_{ee}$ (5.31) determining the Bethe-Heitler cross section in the case of
unpolarized particles: $ A_{2} = Y_{ee}$, and also that $Y_{1} = Y_{I}$ and
$Y_{2} = Y_{II}$ [see (5.32) and (5.33)].

Therefore, owing to the factorization of the squares of the form factors
$g_{e}$ and $g_{m}$ and also the use of the 4-vectors $a$ and $A$ (6.5),
the differential cross section for the Bethe-Heitler process in both the
cases of linearly polarized photon (6.2) and unpolarized photon (6.11),
(5.31), can be written in a rather compact form.

Let us integrate Eq. (6.1) over $d^3 \vec q_2$ and $d p_{20}$ in the rest
frame of the initial proton, $q_1 =(M,0)$. As a result, we find:
$$
{ d \sigma_{BH} \over d \omega \; d \Omega_{\gamma} \; d \Omega_{e} } =
{\alpha^3 \; \omega \over (2\pi)^2 } \; {|\vec p_2| \over |\vec p_1| } \;
{| T |^2 \over q^4 \; } \; ,
\eqnum {6.18}
$$
$$
| T |^2 =  g_{e}^{2} \; Y_{I}^{~e} + \tau \; g_{m}^{2} \; Y_{II}^{~e} \; .
\eqnum {6.19}
$$
Let us consider the limit of the cross section (6.18) when the proton is
a point (structureless) particle with infinite mass, i.e., we assume that
$g_{e} = g_{m} = 1$ and $q_2 = (M, \vec q) \simeq (M,0)$, where $\vec q
= \vec p_1 - \vec p_2 - \vec k$ is the momentum transferred to the proton.
In this limit ($M \to \infty$), $E_{kp} = \vec q~^2/2M \to 0, \; \vec q /2M
\to 0$, and $b_0 = (1, \vec q/2M) \simeq (1,0)$. We choose the Coulomb gauge
for the photon polarization vectors: $e = (0, \vec e)$, as a result of which
we find
$$
eb_0 = 0 , \; ea = {p_1 e \over \lambda_1 } - {p_2 e \over \lambda_2 }, \;
eA = p_{20}\; {p_1 e \over \lambda_1 } - p_{10} \; {p_2 e \over
\lambda_2 }, \; \tau (q_{+} k)^2 = \omega^2 q^2\;.
$$
Using these expressions to take an limit in (6.19), we obtain:
$$
|T|^2 = 2 - {\lambda_{1} \over \lambda_{2}} - {\lambda_{2} \over \lambda_{1}}
- {\omega^2 q^2 \over \lambda_{1} \lambda_{2}} +
 q^2 \; (e a)^2 + 4 \; (e A)^2 \; ,
\eqnum {6.20}
$$
or, in expanded form,
$$
|T|^2 = 2 - {\lambda_{1} \over \lambda_{2}} - {\lambda_{2} \over \lambda_{1}}
- {\omega^2 q^2 \over \lambda_{1} \lambda_{2}} +
\; (\;4 p_{20}^2 + q^2\; ) \left ({p_1 e \over \lambda_1 }\right )^2
$$
$$
+ \; (\; 4 p_{10}^2 + q^2\;) \left ({p_2 e \over \lambda_2 }\right )^2 - 2 \; (\; 4 p_{10} p_{20} + q^2\; ) \;
{p_1e \cdot p_2e \over \lambda_{1} \lambda_{2} }\; .
\eqnum {6.21}
$$
The expressions (6.18) and (6.21) for the differential cross section for the
Bethe-Heitler process $d \sigma_{BH} / d \omega / d \Omega_{\gamma}
/ d \Omega_{e}$ in the limit where the proton is an infinitely heavy,
structureless particle coincide with the analogous expressions of Ref. [69].

\section{\bf Virtual-photon polarization in the reaction\\
$ep \to ep \gamma \; (ep \to eX) $ }

The reaction $ep \to ep \gamma$ and VCS on a proton have recently become
interesting not only at low and intermediate energies [60], but also at high
electron energies and 4-momenta transferred to the proton [63,74-77]. The
VCS process offers greater possibilities for studying hadronic structure
than the RCS process, because in it the energy and three-momentum transferred
to the target can be varied independently. These attractive properties of
VCS have led to the suggestion that it be used for experimental study of
the nucleon structure [74,75] and have made it necessary to perform a thorough
theoretical study of the reaction $ep \to ep \gamma$ (including the use of
the noncovariant method of calculating helicity amplitudes; (see Refs. 63,
76 and 77 and references therein)). To calculate VCS on a proton, it is
necessary to know the hadron ($W_{\mu \nu}$) and lepton ($L_{\mu \nu}$)
tensors [63,78]:
$$
L_{\mu \nu}=J_\mu J_\nu^*,~~J_\mu=\overline{u}(p_2)\gamma_\mu u(p_1) \; ,
\eqnum {7.1}
$$
where $u(p_{i})$ are electron bispinors, $\overline{u}(p_{i}) u(p_{i}) =
2 m$, and $m$ is the electron mass ($i =1,2$). The interpretation of the
results is considerably simplified if the tensor $L_{\mu \nu}$ is expressed
in terms of the longitudinal and transverse polarization vectors of the
virtual photon. The corresponding expressions can be found in Refs. 63 and
78. However, they have two defects: (1) the electron mass is neglegted, which
is of course justified at ultrarelativistic electron energies and large
squared 4-momentum of the virtual photon; (2) they have a noncovariant
form. A lepton tensor free of these defects was constructed in Ref. 79.

Let us consider the question of the polarization state of a virtual
$\gamma$ with 4-momentum $r=p_{1}-p_{2}$ which is exchanged between the
electron and proton in the reaction $ep \to ep \gamma$ (see Fig. 1c). Using
the vectors of the orthonormal basis $a_{A}$ (5.9) ($A=(0,1,2,3)$):
$$
a_{0} = p_{+}/\sqrt{p_{+}^{2}} \; , \; a_{3} = p_{-}/\sqrt{-p_{-}^{2}} \; ,
\; a_{2} = [a_{0} \cdot a_{3}]^{\times} q_{1} /\rho \; , \;  a_{1} = [a_{0}
\cdot a_{3}]^{\times} a_{2} \; ,
\eqnum {7.2}
$$
$$
p_{\pm} = p_{2} \pm p_{1} \; ,
 \; a_{2} q_{1} = 0 \; , \; a_{1}^{2} =
a_{2}^{2} = a_{3}^{2} = - a_{0}^{2} = - 1 \; ,
\eqnum {7.3}
$$
which satisfy the completeness relation
$$
a_{0} \cdot a_{0} - a_{1} \cdot a_{1} - a_{2} \cdot a_{2} - a_{3} \cdot a_{3}
= g \; ,
\eqnum {7.4}
$$
we construct the 4-vectors of the longitudinal ($e_{3}$) and transverse
($e_{1}, \; e_{2}$) polarization of a virtual photon with 4-momentum $r$
(Ref. 79):
$$
e_{1} = {[a_{0} \cdot a_{1}] q_{1} \over \sqrt{(a_{3}q_{1})^2 + q_{1}^2} } \; ,
\; e_{2} = a_{2} = { [a_{0} \cdot a_{3}]^{\times}q_{1} \over \rho} \; , \;
e_{3} = { (1 + a_{3} \cdot a_{3})q_{1} \over \sqrt{(a_{3}q_{1})^2 + q_{1}^2} } \; ,
\eqnum {7.5}
$$
where
$$
\rho^2 = (a_{1}q_{1})^2= { 2p_{1}p_{2} \cdot p_{1}q_{1} \cdot p_{2}q_{1}
- M^2 ((p_{1}p_{2})^2 - m^4) - m^2 ((p_{1}q_{1})^2 + (p_{2}q_{1})^2) \over
 (p_{1}p_{2})^2 - m^4} .
$$
It is easily verifed that the 4-vectors $e_{i} \; (i=1,2,3)$ are orthogonal
to each other ($e_{i} e_{j} = 0, \; i \neq j$), and also that $e_{i}r
= e_{i}a_{3}= 0$ and $e_{1}^2 = e_{2}^2 = - e_{3}^2 = -1$. The 4-vectors
$e_{i}$ (7.5) are not changed when the auxiliary 4-vector $q_{1}$ is replaced
by $q_{1} + p_{1} - p_{2} = q_{2} + k$ [because $p_{1} - p_{2} = r
=-2y a_{3}$, where $y= \sqrt{-r^2}/2$, and because the vectors $a_{A}$ (7.2)
are orthogonal]. For this reason, study of the virtual-photon polarization
vectors $e_{i}$ (7.5) in the rest frame of the incident proton or in the
c.m. frame of the final proton and photon is equivalent and leads to the usual
expressions. Here we shall restrict ourselves to the rest frame of the
incident proton [$q_{1}= (M,0,0,0)]$, where the 4-vectors $e_{i}$
have the form:
$$
e_{1} = (0,1,0,0), \; e_{2} = (0,0,1,0), \; e_{3} = {1 \over \sqrt{-r^2}}
(\mid \vec r \mid, r_{0} \vec n_{3}) \; .
\eqnum {7.6}
$$
Here $\vec n_{3}$ is a unit vector directed along $\vec r \;(\vec n_{3}^{~2}
=1)$, and $r_{0}$ is the time component of the 4-vector $r=(r_{0}, \vec r)$.

The four mutually orthogonal vectors $e_{1}, e_{2}, e_{3}$, and $a_{3}$
also satisfy the completeness relation:
$$
e_{3} \cdot e_{3} - e_{1} \cdot e_{1} - e_{2} \cdot e_{2} - a_{3} \cdot a_{3}
= g \; ,
\eqnum {7.7}
$$
which allows $a_{0}$ and $a_{1}$ to be expressed in terms of $e_{1}$ and
$e_{3}$:
$$
a_{1}=\alpha e_{3} - \beta e_{1} \; , \; a_{0} = \beta e_{3} - \alpha e_{1}\; ,
\; \beta^2=1 + \alpha^2 \; \; ,
\eqnum {7.8}
$$
$$
\alpha = e_{3} a_{1} = a_{0} e_{1} = { a_{1}q_{1} \over \sqrt{(a_{3}q_{1})^2
 + q_{1}^2} }\; , \; \beta = e_{1} a_{1} = e_{3} a_{0} = {a_{0} q_{1} \over
\sqrt{(a_{3}q_{1})^2 + q_{1}^2} } \; .
\eqnum {7.9}
$$
In the DSB (4) the matrix elements of the electron current have the form
of (5.18):
$$
( J^{\delta,\delta}_{e} )_{\mu} = 2 m ( a_{0} )_{\mu}, \;
( J^{-\delta,\delta}_{e} )_{\mu} = - 2 \delta y \; ( a_{\delta} )_{\mu},
\eqnum {7.10}
$$
where $ a_{\pm \delta} = a_{1} \pm i \delta a_{2} ,  \; \delta = \pm 1$.
Let us write them in terms of the 4-vectors $e_{i}$ (7.5) (Ref. 79):
$$
( J^{\delta,\delta}_{e} )_{\mu}= 2 m \;  (\beta e_{3} - \alpha e_{1})_{\mu} \; , \;
( J^{-\delta,\delta}_{e} )_{\mu}  = - 2 \delta y \; (\alpha e_{3} - \beta e_{1} + i \delta e_{2})_{\mu} \; .
\eqnum {7.11}
$$
Therefore, for transition without electron spin flip $(J^{\delta,\delta}_{e})$
the virtual-photon polarization vector is a superposition of the longitudinal
($\beta e_{3}$) and transverse linear ($-\alpha e_{1}$) polarizations, while
for transition with spin flip $(J^{-\delta,\delta}_{e})$ it is a superposition
of the longitudinal ($\alpha e_{3}$) and transverse elliptical [$e_{\delta}
=(0, \vec e_{\delta})= -\beta e_{1} + i \delta e_{2}$] polarizations. Here
the state of a photon with elliptical polarization vector $e_{\delta}
=(0, \vec e_{\delta})$ will have degree of linear polarization (equal to
the ratio of the difference and sum of the squared semiaxes [57]) [79]:
$$
\kappa_{\gamma} = { \beta^2 - 1 \over \beta^2 + 1} = { \alpha^2 \over \beta^2
 + 1} \; .
\eqnum {7.12}
$$
Inverting this relation, we obtain:
$$
\beta^2 = { 1 + \kappa_{\gamma} \over 1 - \kappa_{\gamma}} \; \; , \;
\alpha^2= {2 \kappa_{\gamma} \over 1 - \kappa_{\gamma}} \; \; .
\eqnum {7.13}
$$
Now we find the squared moduli of the vectors $\vec e_{\delta}$ and
$\vec a_{\delta}$:
$$
\mid \vec e_{\delta} \mid^2 = 1 + \beta^2 = {2 \over 1 - \kappa_{\gamma}} \; ,
\; \mid \vec a_{\delta} \mid^2 = ( 1 + \beta^2 ) \; (1 + \kappa_{L}) \; ,
\eqnum {7.14}
$$
$$
\kappa_{L} = \kappa_{\gamma} \vec e_{3}^{~2} = \kappa_{\gamma} { r_{0}^2 \over
(-r^2)} \; , \; \vec e_{3}^{~2} = { r_{0}^2 \over (-r^2)} \; .
\eqnum {7.15}
$$
We introduce the normalized vectors $\vec e_{\delta}~'$ and $\vec a_{
\delta}~'$:
$$
\vec e_{\delta}~' = {\vec e_{\delta} \over \sqrt{1 + \beta^2}} =
\sqrt {{1 - \kappa_{\gamma} \over 2}} \; \vec e_{\delta} \; , \;
|\vec e_{\delta}^{~'}|^2 = 1 \; .
\eqnum {7.16}
$$
$$
\vec a_{\delta}~' = {\vec a_{\delta} \over \sqrt{1 + \beta^2}} =
\sqrt {{1-\kappa_{\gamma} \over 2}} \; \vec a_{\delta} \; , \;
|\vec a_{\delta}^{~'} |^2 = 1 + \kappa_{\gamma} \vec e_{3}^{~2} = 1 +
\kappa_{L} \; ,
\eqnum {7.17}
$$
Therefore, the elliptical-polarization vector $\vec e_{\delta}$ of a virtual
photon can be normalized to unity ($|\vec e_{\delta}~'|^2 = 1$), but the
presence of a longitudinal polarization makes this normalization impossible
for the total vector $\vec a_{\delta}~'$ simultaneously. The quantity
$\kappa_{L}$ (7.15) corresponding to the inequality $|\vec a_{\delta}~'
|^2 = 1 + \kappa_{L} \neq 1$ has the meaning of the degree of longitudinal
polarization of a virtual photon emitted in a transition with electron spin
flip. In the ultrarelativistic limit, when the electron mass can be neglected,
the quantities $\kappa_{\gamma}$ and $\kappa_{L}$ will be interpreted as the
total degrees of linear and longitudinal polarization of the virtual photon.
In this (massless) case we have:
$$
(a_{3}q_{1})^2 + q_{1}^2 = - M^2\; { \vec r^{~2} \over r^2} \; , \;
(a_{1} q_{1}) ^2 = M^2 \; ctg^2 \vartheta /2 \; ,
\eqnum {7.18}
$$
$$
\kappa_{\gamma}^{-1} = 1 - 2 \; {\vec r^{~2} \over r^2} \; tg^2
\vartheta /2 \; ,
\eqnum {7.19}
$$
where $\vartheta$ is the angle between the vectors $\vec p_{1}$ and
$\vec p_{2}$. Equation (7.19) for $\kappa_{\gamma}$ coincides with the result
of Ref. 78.

The vector $\vec a_{\delta}~'$ (7.17) can also be written as
$$
\vec a_{\delta}~' = \sqrt{\kappa_{L}} \; \vec n_{3} - \sqrt {{1 +
\kappa_{\gamma} \over 2}} \; \vec e_{1} + i \delta \; \sqrt {{1 -
\kappa_{\gamma} \over 2}} \; \vec e_{2} \; \; ,
\eqnum {7.20}
$$
which makes it easy to construct the polarization density matrix for a virtual
photon in the massless limit (both in the polarized case, which for massless
particles is helical polarization, and in the unpolarized case; see Ref.
78).

To obtain the complete expression for $\kappa_{\gamma}$ and $\kappa_{L}$
arising from the contributions of the matrix elements both without and with
spin flip, we construct the lepton tensor averaged over electron spin states.
Using the matrix elements (7.10) and (7.11), this can be done fairly simply
[79]:
$$
\overline{L}_{\mu \nu}= 4 m^2 \; (a_{0})_{\mu} (a_{0})_{\nu} + 4 y^2 \;
( (a_{1})_{\mu} (a_{1})_{\nu} + (a_{2})_{\mu} (a_{2})_{\nu} ) \; .
\eqnum {7.21}
$$
Using the completeness condition (7.3) and gauge invariance, the tensor
$\overline{L}_{\mu \nu}$ can be written as
$$
\overline{L}_{\mu \nu} = 4 x^2 \; (a_{0})_{\mu} (a_{0})_{\nu} - 4 y^2 \;
g_{\mu\nu} \; ,
\eqnum {7.22}
$$
where $x^2= m^2 + y^2$. The tensor $\overline{L}_{\mu \nu}$ (7.22) can be
used to reduce the calculation of the contribution of graphs with VCS on a
proton to the cross section for the reaction $ep \to ep \gamma$ to calculation
of the trace of a product of tensors:
$$
Y_{pp} = \overline{L}_{\mu \nu} \; W_{\mu \nu} \; , \; W_{\mu \nu} = V_{\mu}
V_{\nu}^{\ast} \; ,\; V_{\mu} = \overline{u}(q_{2}) \; M_{\mu \nu}
 e^{\nu} \; u(q_{1}) \; {1\over r^2} \; .
\eqnum {7.23}
$$
Let us express the tensor $\overline{L}_{\mu \nu}$ (7.21) in the terms of
the virtual-photon polarization vectors $e_{i}$ (7.5). As a result, it
naturally breaks up into the sum of three terms corresponding to the
contributions of transverse ($L_{T}$) and longitudinal ($L_{L}$) states and
their interference ($L_{LT}$) [79]:
$$
\overline{L} = 4y^2 \; ( L_{T} \; + \; L_{L} \; + \; L_{LT} \; ) \; ,~~~~~~~~~~~~
\eqnum {7.24}
$$
$$
L_{T} = e_{1} \cdot e_{1} \; (\beta^2 +  \alpha^2  m^2 / y^2 ) +
                 e_{2} \cdot e_{2}  \; , ~~
\eqnum {7.25}
$$
$$
L_{L} = e_{3} \cdot e_{3} \; ( \alpha^2 + \beta^2  m^2 / y^2 ) \; , ~~~~~~~~~~~~
\eqnum {7.26}
$$
$$
~~~~~L_{LT}= - \; ( e_{1} \cdot e_{3} + e_{3} \cdot e_{1} ) \;\alpha \beta \;
 (1 + m^2/y^2) \; .
\eqnum {7.27}
$$
Then the total degree of linear polarization of the virtual photon will be
given by
$$
\kappa_{\gamma} ' = { \beta^2 + \alpha^2 m^2/y^2 - 1 \over \beta^2 +
\alpha^2 m^2/y^2 +1} = { \alpha^2 \over \beta^2 +1 - 2 m^2/x^2} \; .
\eqnum {7.28}
$$
Since $\alpha$ and $\beta$ are the same in Eqs. (7.12) and (7.28) [see (7.9)],
the inclusion of the electron mass in the ultrarelativistic limit will lead
only to a slight increase of $\kappa_{\gamma}$ [79]:
$$
\kappa_{\gamma} ' \simeq  \kappa_{\gamma} \left (1 + {2m^2 \over x^2 (1
+ \beta^2)}  \right ) \; .
\eqnum {7.29}
$$
Inverting the relation in (7.28), we find
$$
\beta^2 + \alpha^2 m^2/y^2 = { 1 + \kappa_{\gamma} ' \over 1 -
\kappa_{\gamma} '} \; , \; \alpha^2 + \beta^2 m^2/y^2 = { 2 \kappa_{\gamma} '
 \over 1 - \kappa_{\gamma} '} + {m^2 \over y^2} \; .
\eqnum {7.30}
$$
We can separate the completely polarized and unpolarized parts in the
transverse tensor $L_{T}$ (7.25):
$$
L_{T} = e_{1} \cdot e_{1} \, (\beta^2 +  \alpha^2  m^2 / y^2 - 1 ) +
e_{1} \cdot e_{1} +  e_{2} \cdot e_{2}\, =
 {2 \over 1 - \kappa_{\gamma}'}\, (\, \kappa_{\gamma}'
\, e_{1} \cdot e_{1} + (1 - \kappa_{\gamma}') \, ( e_{1} \cdot e_{1}
+  e_{2} \cdot e_{2} )/2 \,)\,.
\eqnum {7.31}
$$
Therefore, the virtual-photon polarization density matrix $\rho_{ij}$ is
obtained from the tensor $\overline{L}_{ij}$ (7.24) just as in the massless
case (see Ref. 78):
$$
\rho_{ij} =(1-\kappa_{\gamma}') \;  \overline{L}_{ij}/8 y^2 \;.
\eqnum {7.32}
$$
For the degree of longitudinal polarization of the virtual photon we then
obtain:
$$
\kappa_{L} ' = {r_{0}^2 \over (-r^2)} \kappa_{\gamma} ' \; \left ( 1 +
{m^2 \over y^2} \; { (1 - \kappa_{\gamma} ') \over 2 \kappa_{\gamma} '} \right ) \; .
\eqnum {7.33}
$$
The expressions (7.28) and (7.33) for $\kappa_{\gamma} '$ and $\kappa_{L} '$
with $m = 0$ obviously become $\kappa_{\gamma}$ and $\kappa_{L}$ of (7.12)
and (7.15).

We conclude by noting that the region of applicability of the tensor
$\overline{L}_{\mu \nu}$ (7.24) is not limited to only VCS on a proton.
Since in fixed-target experimets the charged-lepton scattering
at available energies is mainly determined by virtual photon exchange, the
tensor $\overline{L}_{\mu \nu}$ (7.24) can also be used to study
deep-inelastic electron scattering ($e^{\pm} p \to e^{\pm} X$), and muon
scattering ($\mu^{\pm}p \to \mu^{\pm} X$), where inclusion of the mass
is more important.

\section{\bf Compton back-scattering of the photons of a circularly
\newline polarized laser wave on a beam of ultrarelativistic, \newline
longitudinally polarized electrons}

It was shown in Refs. 80 and 81 that, using existing (SLC) and planned
(VLEPP) accelerators with colliding $e^{+}e^{-}$ beams, it is possible to
obtain colliding $\gamma e$ and $\gamma \gamma$ beams of roughly the same
energy and luminosity as the original $e^{+}e^{-}$ beams. It have been
suggested that the intense beams of hard $\gamma$ rays needed for this be
obtained from the Compton back-scattering (CBS) of a powerful laser flash
focused on the electron beam [82]. For a sufficiently powerful flash in the
conversion region [81], processes with simultaneous absorption of several
laser photons from the wave become important:
$$
e^{-} + n \; \gamma_{0} \rightarrow e^{-} + \gamma \; , \; n \geq 1 \; ,
\eqnum {8.1}
$$
$$
\gamma + s \; \gamma_{0} \rightarrow e^{+} + e^{-} \; , \; s \geq 1 \; .
\eqnum {8.2}
$$
The first of these nonlinear processes leads to broadering of the spectrum
of high-energy photons [83], and the second effectively lowers the
$e^{+}e^{-}$-pair production threshold [84].

The process (8.1) and (8.2) were studied systematically in Ref. 85. In Ref.
16 they were studied from the view-point of providing sources of polarized
$\gamma$ and $e^{+}e^{-}$ beams. The phenomena arising in collisions of
polarized electrons with the photons of a circularly polarized electromagnetic
wave were analyzed in Ref. 86. Nonlinear effects were studied not only for
$\xi^2 < 1$, but also for $\xi^2 \geq 1$. Here $x^2$ is the wave intensity
parameter:
$$
\xi^2 = n_{\gamma} \left ({4 \pi \alpha \over m^2 \omega} \right ) \; ,
\eqnum {8.3}
$$
where $n_{\gamma}$ is the photon density in the wave and $\omega$ is the
photon energy, $\alpha$ is the fine structure constant, $m$ is the electron
mass. The emission spectra at high intensities ($\xi^2 \geq 1$) were first
calculated numerically in Ref. 83, but the particle polarization was not taken
into account.

Recently at the SLAC accelerator a series of experiments [87] are being
performed for $\xi \sim 1$ to verify nonlinear QED. This has become possible
owing to the use of supershort, strongly focused laser pulses.
The region of nonlinear effects for $\xi^2 \geq 1$ is very important here,
and it is of great interest because emission processes due to simultaneous
absorption of a large number of photons from the wave become important, and
the probabilities for these processes are essentially nonlinear functions of
the field strength.

As a rule, in the literature the laser wave is described as the field of
a planar electromagnetic wave [85,86]. The applicability of this model in
strong fields has been studied in Ref. 88.

According to Ref. 10, the $S$-matrix element for the transition of an electron
from the state $\psi_{p} = \psi^{\delta}(p,s)$ to the state $\psi_{p'}
= \psi^{\pm\delta}(p',s') \; , (\delta = \pm 1 )$, with the emission of a
photon of 4-momentum $k' = (\omega ', \vec k')$ and circular-polarization
vector $e_{\lambda '} $ is given by
$$
S_{fi} = -i\; e \; \int \overline{\psi}_{p'} \; \hat e ^{\ast}_{\lambda '}
\; \psi_{p} \; exp (i k'x) \; (2 \omega ')^{-1/2} \; d^4 x \; ,
\eqnum {8.4}
$$
where $\psi_{p}$ and $\overline{\psi}_{p'}$ are the exact wave functions
of electrons in the field of a circularly polarized electromagnetic wave,
corresponding to the vector potential
$$
A = a_{1} \cos(kx) + \lambda a_{2} \sin(kx), \; \lambda = \pm 1 .
\eqnum {8.5}
$$
Here $k$ is the wave vector, $k^2 = 0, a_{1}k=a_{2}k=a_{1} a_{2} = 0,
a_{1}^2 = a_{2}^2 = a^2$, and $\lambda$, $\lambda ^{'} $ are helicities
of a laser and emission photons. The explicit form of the matrix elements
(8.4) in the DSB was obtained in Refs. 38 and 86:
$$
S_{fi} = - \; {ie \; (4 \pi)^{1/2} \over (2\omega ' \; 2 q_{0}
\; 2q_{0} ')^{1/2}} \; \;\sum_{n=1}^{\infty} \; M^{~(n)}_{\pm \delta ,
\delta} \; (2\pi)^4 \; \delta^4 ( nk + q - q' - k') \; ,
\eqnum {8.6}
$$
$$
M^{~(n)}_{-\delta,\delta}=-\; \; {1\over 2}\; \; \lambda '\;(-\lambda)^{n}
 \; \xi \; \Biggl \{ -\; {2(1-u/u_{n}) \over \sqrt{v v' - 1}} \;
(J_{n-1} + J_{n+1} ) \Biggr.
$$
$$
 + \Biggl. \; {1 \over 2(u+1)} \; \left ( {(u+2)^2 \over \sqrt{v v ' + 1}} -
\delta \lambda ' \; {u^2 \over \sqrt{vv'-1}} \right ) \; J_{n + \lambda
\lambda '} \Biggr \} \; ,
\eqnum {8.7}
$$
$$
~M^{~(n)}_{\delta , \delta} = - \; \; {1\over 2} \; \lambda '\; (-
\lambda)^{n} \; \xi \; \sqrt{ {u\over u_{n}} \; \left (1- {u\over u_{n} }
\right ) } \; \left ( {u+2 \over u} \sqrt{{vv'-1\over vv'+1}} - \delta
\lambda '\right )
$$
$$
 \times \left ( \sqrt{ {vv' - 1 \over 1+\xi^2} } ( J_{n-1} + J_{n+1} ) -
{u u_{n} \sqrt{1 + \xi^2} \over 2 (u+1) \sqrt{vv' - 1} } \; J_{n + \lambda
 \lambda '} \right )\; ,
\eqnum {8.8}
$$
where
$$
q = p + {\xi^2 m^2 \over 2 kp} \; k \; , \; q' = p' + {\xi^2 m^2 \over 2 kp'} \;
k \; , \; q^2 = q'^2 = m_{\ast}^2 = m^2 \; ( 1 + \xi^2 ) \; ,
$$
$$
u = {kk'\over kp'} \; , \; u_{n} = {2 n kp \over m_{\ast}^2} \; , 2 (vv' -1)
= {u u_{n} \over u + 1 }  \left ( 1 + \xi^2 \; \left (1 - {u \over u_{n}}
\right ) \right ) \; ,
\eqnum {8.9}
$$
$$
J_{n + \lambda \lambda '} = {( 1 + \lambda \lambda ') \over 2} \; J_{n + 1} +
{( 1 - \lambda \lambda ') \over 2} \; J_{n - 1} \; , \; nk + q = k' + q' \; ,
$$
$$
z_{n} = { 2 n \xi \over \sqrt{1 + \xi^2} } \; \sqrt{ {u\over u_{n}} \left ( 1 -
{u \over u_{n}} \right ) } \; \; .
$$
Here $M^{~(n)}_{\delta , \delta}$ and $M^{~(n)}_{-\delta , \delta}$ are the
emission amplitudes of the $n$-th harmonic corresponding to transitions
without and with electron spin flip, $q$ and $q'$ are the electron
quasimomentum 4-vectors, $q = (q_{0}, \vec q), \; q' = (q'_{0}, \,
\vec q~')$, and $J_{n}$ are the $n$-th-order Bessel function of argument
$z_{n}$. It is easily verifed that the amplitudes $M^{~(n)}_{\pm \delta ,
\delta}$ have the following kinematical features. For $u = u_{n}$ and $n >1$
they vanish [$M^{~(n)}_{\pm \delta, \delta} (u = u_{n}) = 0 ]$. The reason
for this behavior of the amplitudes will be explained below. Knowledge of
the diagonal amplitudes (8.7) and (8.8) allows transformation to the
helicity amplitudes (see Ref. 86). As a result, we obtain the following
expressions for the differential cross section of the hard photon emission
by an electron in the field of circularly polarized electromagnetic wave
[86]:
$$
{d \sigma_c \over d u} = { \pi \alpha^2 \over x m^2 \xi^2 (u+1)^2} \;
\sum_{n = 1} ^{\infty} \; ( \; F_{1n} + \lambda \lambda_{e} \; F_{2n} +
\lambda \lambda '\; F_{3n} + \lambda_{e} \lambda ' \; F_{4n} \; ) \; ,
\eqnum{8.10}
$$
\begin{eqnarray}
F_{1n} &=& - 4 \; J_{n}^2 + \xi^2 \; \left ( 2 + {u^2 \over u + 1 }
\right ) \; ( J_{n-1}^2 + J_{n+1}^2 - 2 J_{n}^2 ) \; , \nonumber \\
F_{2n} &=& \xi^2 \; { (2+u)u \over u+1} \; \left ( 1 - 2 {u\over u_{n}}
\right ) \; ( J_{n-1}^2 - J_{n+1}^2 ) \; , \nonumber
\end{eqnarray}
$$
F_{3n} = \xi^2 \; \left ( 2 + {u^2 \over u + 1 } \right ) \; \left ( 1 - 2
\; {u\over u_{n}} \right ) \; ( J_{n-1}^2 - J_{n+1}^2 ) \; ,~~~~~~~
\eqnum {8.11}
$$
\begin{eqnarray}
~~~F_{4n} = {u\over u+1} \; ( - 4 \; J_{n}^2 + \xi^2 \; ( 2 + u ) \;
( J_{n-1}^2 + J_{n+1}^2 - 2 J_{n}^2 ) \; ) \; , \nonumber
\end{eqnarray}
here $x = 2kp/m^2$, $\lambda_e$ is the helicity of the electron, $\lambda_e
=\pm 1$. The expression inside the summation in (8.10) determines the emission
probability of the $n$-th harmonic when the polarization states of the laser
and the emitted photons and also the initial state of the electron are
helicity states. For $\xi^2 = 0$ Eq. (8.10) coincides with the result of
Ref. 89.

Using (8.10), the degree of circular polarization of a photon in the final
state $\lambda_{f}$ is defined as
$$
\lambda_{f} =  \sum_{n=1}^{\infty} \; ( \lambda \; F_{3n} + \lambda_{e} \;
F_{4n} \; ) \; / \; \sum_{n=1}^{\infty} \; ( \;  F_{1n} +  \lambda \lambda_{e} \;
F_{2n} \; )  \; .
\eqnum{8.12}
$$

For $\xi^2 < 1$ only the first few harmonics dominate in the cross section
(8.10) for the process (8.1). We expand the expressions (8.11) in the
parameter $\Delta = \xi^2 /(1 + \xi^2)$, expanding only the Bessel functions
and using the exact expressions for the $u_{n}$. As a result, for the first
three harmonics we have [86]:
\begin{eqnarray}
{F_{11}\over \xi^2} &=&2 + {u^2 \over 1 + u } - 4 {u\over u_{1}}
\left ( 1 - {u \over u_{1}} \right ) + 4 \Delta {u\over u_{1}}  \left ( 1 - {u \over u_{1}} \right )
\left [ 1 + {u^2 \over 1 + u } \right. \nonumber~~~ \\
&-&\left. {u\over u_{1}} \left ( 1 - {u \over u_{1}} \right ) \right ]
+ \Delta^2 {u^2\over u_{1}^2} \left ( 1 - {u \over u_{1}} \right )^2 \left [
{7\over 2} + {15\over 4} {u^2 \over 1+u} -{5\over 3} {u\over u_{1}} \left ( 1 - {u \over u_{1}} \right )
\right ] , ~~~\nonumber
\end{eqnarray}
$$
{F_{21}\over \xi^2} ={u(2+u)\over 1+u} \left ( 1 - 2{u\over u_{1}} \right )
\left [ 1 - 2 \Delta {u\over u_{1}} \left ( 1 - {u \over u_{1}} \right ) +
{5\over 4} \Delta^2 {u^2\over u_{1}^2} \left ( 1 - {u \over u_{1}} \right )^2
\right ] , ~~~
\eqnum {8.13}
$$
\begin{eqnarray}
{F_{31}\over \xi^2} &=&\left ( 2 + {u^2 \over 1 + u } \right ) \left ( 1
 - 2{u\over u_{1}} \right ) \left [ 1 - 2 \Delta {u\over u_{1}}
 \left ( 1 - {u \over u_{1}} \right ) + {5\over 4} \Delta^2
 {u^2\over u_{1}^2} \left ( 1 - {u \over u_{1}} \right )^2 \right ] ,
\nonumber \\
{F_{41}\over \xi^2} &=&{u\over 1+u} \biggl \{ 2 + u - 4
{u\over u_{1}} \left ( 1 - {u \over u_{1}} \right ) - 4 \Delta
{u\over u_{1}} \left ( 1 - {u \over u_{1}} \right )
\left [ 1 + u - {u\over u_{1}} \left ( 1 - {u \over u_{1}} \right )
\right ] \biggr. \nonumber \\
&+& \Delta^2 {u^2\over u_{1}^2} \left ( 1 - {u \over u_{1}} \right )^2
\left [{7\over 2} + {15\over 4} u -{5\over 3} {u\over u_{1}} \left
 ( 1 - {u \over u_{1}} \right )  \right ]  \biggr \}  , \nonumber
\end{eqnarray}
for the first harmonic;
\begin{eqnarray}
F_{12}& =& 4 \xi^2 \Delta {u\over u_{2}} \left ( 1 - {u \over u_{2}} \right )
\biggl \{ 2 + {u^2 \over 1 + u } - 4 {u\over u_{2}} \left ( 1 - {u
 \over u_{2}} \right ) \biggr. \nonumber \quad~~~~~~\\
&-& 2 \biggl.~ \Delta {u\over u_{2}} \left ( 1 - {u \over u_{2}} \right )
\left ( 4 + {3 u^2\over 1 + u} -{16\over 3} {u\over u_{2}}  \left ( 1
 - {u \over u_{2}} \right ) \right ) \biggr \} , \nonumber
\end{eqnarray}
$$
\hspace{3.2cm}F_{22} = 4 \xi^2 \Delta {u\over u_{2}} \left ( 1 - {u
\over u_{2}} \right ) {u (2+u)\over 1+u} \left ( 1 - 2{u\over u_{2}}
\right ) \left [1 - 4 \Delta {u\over u_{2}} \left (1-{u \over u_{2}} \right)
 \right ]  ,
\eqnum{8.14}
$$
\begin{eqnarray}
\hspace{2cm}~~F_{32} &=& 4 \xi^2 \Delta  {u\over u_{2}}  \left ( 1 - {u \over u_{2}} \right )
 \left ( 2 + {u^2 \over 1 + u } \right )  \left ( 1 - 2{u\over u_{2}} \right )
\left [1 - 4 \Delta {u\over u_{2}}  \left ( 1 - {u \over u_{2}} \right ) \right ]  ,
\nonumber \\
~~~~~~~~F_{42} &=& 4 \xi^2 \Delta  {u\over u_{2}}  \left ( 1 - {u \over u_{2}} \right )
 {u \over 1 + u}  \biggl \{ 2 + u - 4  {u\over u_{2}}  \left ( 1 - {u \over u_{2}} \right )
  \biggr. \nonumber \\
~~~~&-& 2 \Delta {u\over u_{2}} \left ( 1 - {u \over u_{2}} \right )
\biggl.\left [ 4 + 3 u - {16\over 3} {u\over u_{2}}  \left ( 1 - {u \over u_{2}} \right )
\right ] \biggr \}  , \nonumber
\end{eqnarray}
for the second harmonic; and
\begin{eqnarray}
\hspace{1.5cm}~~~F_{13} &=& {81\over 4} \xi^2 \Delta^2 \; {u^2\over u_{3}^2} \left ( 1 - {u \over u_{3}} \right )^2
\; \left ( \; 2 + {u^2 \over 1 + u } - 4 \; {u\over u_{3}} \; \left ( 1 - {u \over u_{3}} \right ) \;
\right ) \; , \nonumber \\
~~~F_{23} &=& {81\over 4} \xi^2 \Delta^2 \; {u^2\over u_{3}^2} \left ( 1 - {u \over u_{3}} \right )^2
\; {u\; (2+u)\over 1+u} \left ( 1 - 2{u\over u_{3}} \right ) \; , \nonumber
\end{eqnarray}
$$
\hspace{0.5cm}F_{33} \;\;= \;\; {81\over 4} \xi^2 \Delta^2 \; {u^2\over u_{3}^2} \left ( 1 - {u \over u_{3}} \right )^2
\; \left ( 2 + {u^2 \over 1 + u } \right ) \; \left ( 1 - 2{u\over u_{3}} \right ) \; ,
\eqnum{8.15}
$$
$$
\hspace{2.19cm}F_{43}\; \;= \; \;{81\over 4} \xi^2 \Delta^2 \; {u^2\over
u_{3}^2} \left ( 1 - {u \over u_{3}} \right )^2 \; {u \over 1 + u} \;
\left (\; 2 + u - 4 \; {u\over u_{3}} \; \left (1 - {u \over u_{3}} \right )
\; \right ) \; ,
$$
for the third harmonic.

The inclusion of the third harmonic, whose probability is proportional to
$\Delta^2$, leads to the appearance of terms containing $\Delta^2$ in Eqs.
(8.13) and (8.14). This is the main difference between the result obtained
in Ref. 86 for the emission probability of the first two harmonics and the
analogous expressions from Refs. 16 and 85.

Let us consider the case of a head-on collision of ultrarelativistic electrons
with the photons of a laser wave. To obtain the energy distribution of the
produced photons $d \sigma_c /dy$, where $y = \omega '/E$, and $E$ is the
electron energy, in (8.10) we must make the replacement $u \to y/(1-y)$
[85]. Here variation of the variable $u$ in the range $0 \leq u \leq u_{n}$
correspond to variation of $y$ in the range $y:0 \leq y \leq y_{n}$,
where
$$
u_n={n x \over 1+\xi^2}, \;
y_{n} = {u_{n} \over 1 + u_{n}} = {n x \over n x + 1 + \xi^2} \; , \;
x = {2 kp \over m^2} = {4 \omega E \over m^2} \; .
$$
Comparing the maximum possible energy of photons produced in ordinary Compton
scattering ($n = 1 , \xi^2 = 0$) with the energy calculated with inclusion
of nonlinear effects ($\xi^2 \neq 0$), we see that photons of the first
harmonic ($n = 1$) have lower maximum possible energy. However, the energy
of $\gamma$ quanta emitted in the absorption of several photons ($n > 1
+ \xi^2$) is greater than that available in ordinary Compton scattering.
Making the replacement: $u \to y/(1-y)$ in (8.10) and (8.11), we obtain the
distribution in the energy of the hard $\gamma$ quanta $y =\omega '/E$ [86]:
$$
{d \sigma_c \over d y} = { \pi \alpha^2 \over x m^2 \xi^2 } \;
\sum_{n = 1} ^{\infty} \; ( \; F_{1n} + \lambda \lambda_{e} \; F_{2n} +
\lambda \lambda '\; F_{3n} + \lambda_{e} \lambda ' \; F_{4n} \; ) \; ,
\eqnum{8.16}
$$
$$
F_{1n} = - 4 \; J_{n}^2 + \xi^2 \; \left ( 1 - y + {1 \over 1 - y } \right ) \;
( J_{n-1}^2 + J_{n+1}^2 - 2 J_{n}^2 ) \; , ~~~~~~~~
$$
$$
F_{2n} = \xi^2 \left ( - 1 + y + { 1 \over 1 - y } \right ) \;
\left ( 1 - 2 {y\over y_{n}}{(1-y_{n}) \over (1-y)} \right )
\; ( J_{n-1}^2 - J_{n+1}^2 ) \; ,
$$
$$
F_{3n} = \xi^2 \; \left ( 1-y + {1 \over 1-y } \right ) \; \left ( 1 - 2
\; {y\over y_{n}} {(1-y_{n}) \over (1-y)} \right ) \;
( J_{n-1}^2 - J_{n+1}^2 ) \; ,
\eqnum {8.17}
$$
$$
F_{4n} = - 4 y \; J_{n}^2 + \xi^2 \left ( -1+y +{1\over 1-y} \right ) \;
( J_{n-1}^2 + J_{n+1}^2 - 2 J_{n}^2  \; ) \; ,~~~~
$$
$$
z_{n} = {2 n \xi \over \sqrt{1+\xi^2}} \; \sqrt{\alpha_{n}} , \;
\alpha_{n} = {y \over y_{n}} \left (1 - {y \over y_{n}}\right )
{ (1-y_{n}) \over (1-y)^2} \; .
$$

Let us now turn to the more detailed analysis of the influence of nonlinear
effects on this process. We shall start from the following initial. We take
a head-on collision to be one in which the electrons have energy $E = 50$
and $300$ GeV, and $\omega = 1.17$ eV (a neodymium laser). We shall use the
expansions (8.13)-(8.15) for numerical calculations of the energy spectra
$(1/W) \; dW/dy$ (where $W = \sum_{n=1}^{n_{max}} \; W_{n}$ is the total
emission probability) and the degree of circular polarization $\lambda_{f}$
of an emitted photon for $\xi^2 < 1$. For $\xi^2 \geq 1$ we shall use the
exact expressions (8.16) and (8.17). In this case $n_{max}$ is determined
from the conditions for the series (8.16) to converge.

The results of numerical calculations of the energy spectra for various
polarizations of the initial electrons ($\lambda_{e}$) and laser photon
($\lambda$) are shown by the graphs in Figs. 3a, 3b, and 3c for
$\xi^2= 0.3, 1$, and $3$, respectively. We see from these figures that the
inclusion of nonlinear effects leads to a significant difference between
the calculated spectra and the spectra of ordinary Compton scattering.
First, the simultaneous absorption of several photons from the wave leads
to broadering of the hard-$\gamma$ spectrum and the appearance of additional
peaks corresponding to the emission of higher-order harmonics. For a given
electron energy this broadering is larger, the larger the wave intensity.
For example, for $E = 50$ GeV and $\xi^2 = 0.3$ the spectrum is bounded above
by the value $y \simeq 0.67$, while for $\xi^2 = 1$ it practically vanishes
at $y \simeq 0.8$, even though an insignificant fraction of the photons can
carry off up to 97\% of the electron energy. Second, the effective increase
of the electron mass [85] $m^2 \to m^2_{\ast} =m^2 (1 + \xi^2)$ leads to
compression of the spectra at smaller values of $y$, because for each $n$
the spectrum is bounded above by the value $y_{n} =n x /(1 + n x + \xi^2)$
and not by $n x /(1 + n x)$. The increase of the electron energy decreases
the relative compression of the first harmonic (see Fig. 3a). For relatively
low intensity of the laser wave ($\xi^2 = 0.3$) the main contribution to
the emission comes from photons of the first harmonic, and the yield of
photons from higher harmonics is insignificant. At intermediate intensity
($\xi^2 = 1$) the broadering of the spectrum due to nonlinear effects is
accompanied by an increase of the probability, and the yield of harder
photons becomes important. Finally, at high intensities ($\xi^2 = 3 $), as
seen from Fig. 3c, emission owing to nonlinear multiphoton absorption
processes becomes comparable to one-photon emission and even begins to
dominate (at $E = 50$ GeV). Therefore, emission of the first harmonic
dominates in the CBS spectra in the field of a circularly polarized
electromagnetic wave at $\xi^2 = 0.3$, while at $\xi^2 = 3$ the emission
in mainly due to higher harmonics, i.e., the emission of a hard photon
by an electron essentially becomes nonlinear [86].

\begin{figure}[hbt]
\hspace{-1.0cm}
\centerline{
\leavevmode\epsfxsize=0.4\textwidth \epsfbox{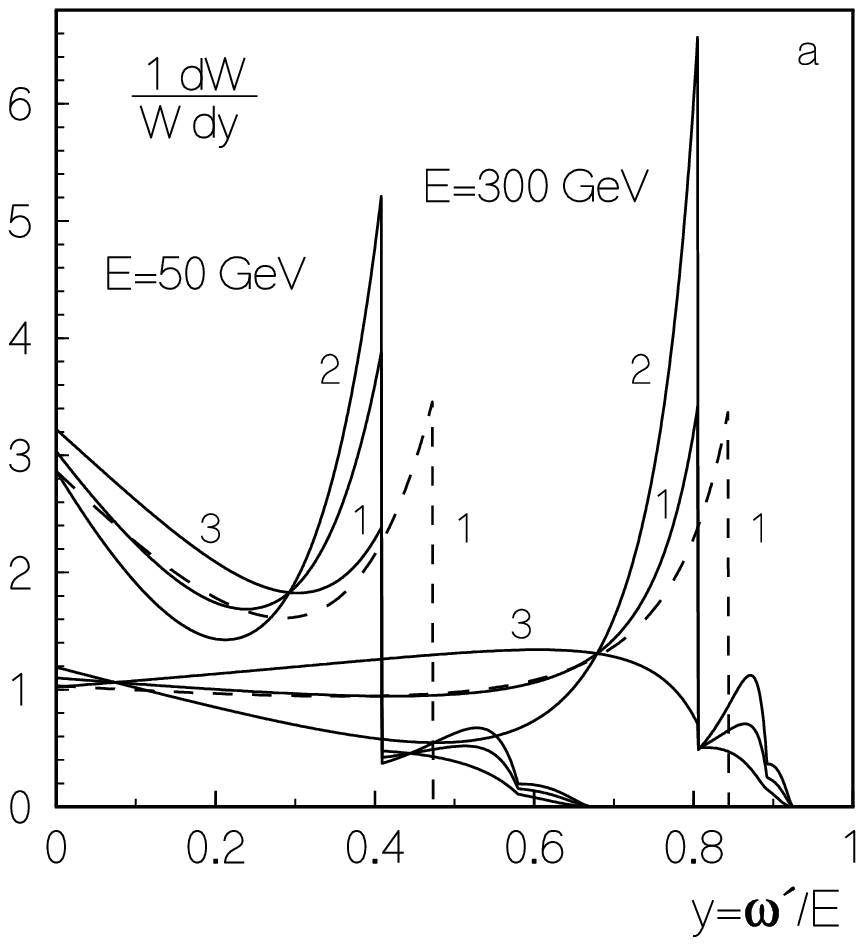}\hspace{-1.5cm}
\leavevmode\epsfxsize=0.4\textwidth \epsfbox{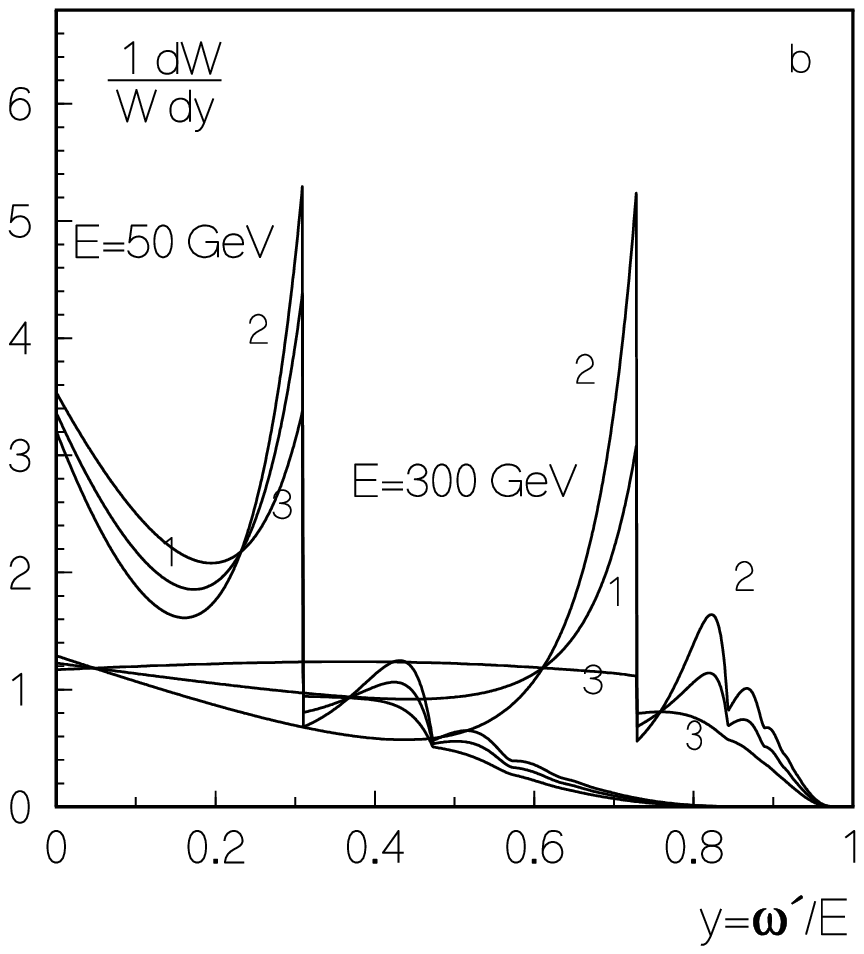}\hspace{-1.5cm}
\leavevmode\epsfxsize=0.4\textwidth \epsfbox{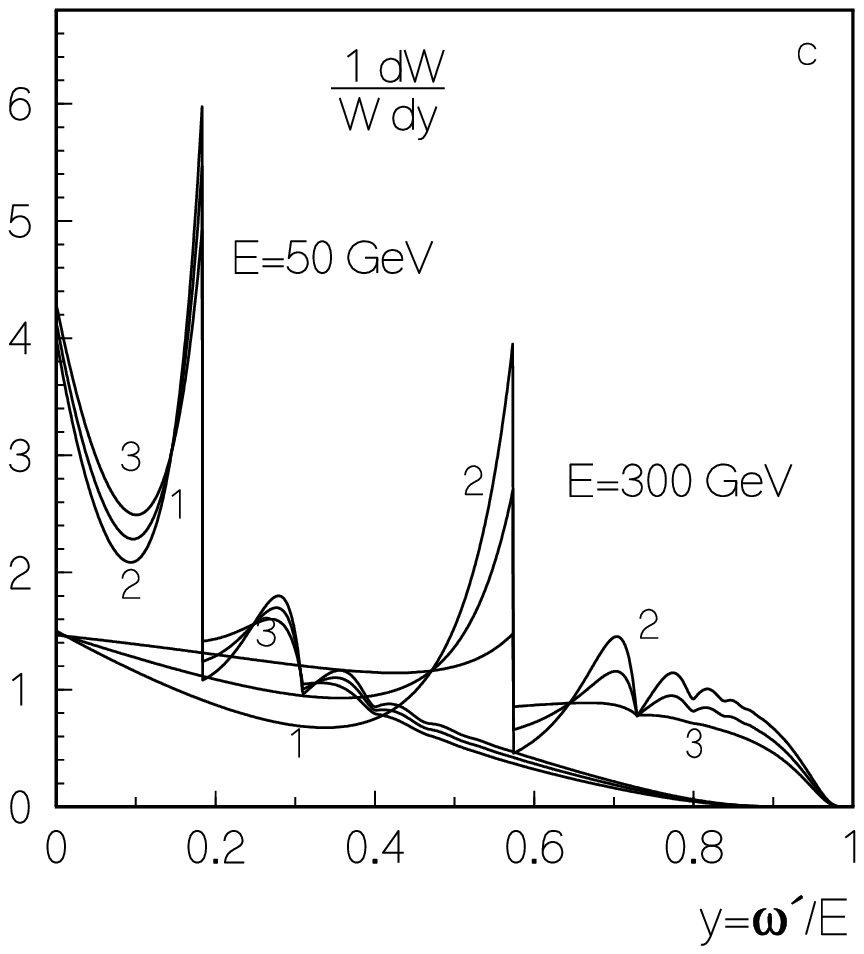} }
\vspace{-1cm}
\caption
{CBS spectra correspond for the following values of the intensity parameter
$\xi^2$: (a)$ \to 0.3; (b)\to 1; (c)\to 3$. The dashed lines correspond to
ordinary Compton scattering ($\xi^2=0$). The lines 1, 2, and 3 correspond
to the following choice of helicities of the electron and laser photon:
$1\to \lambda_{e} = 0 , \; \lambda = 1 ; \; 2\to \lambda_{e} = 1 ,
 \; \lambda = - 1; \; 3 \to \lambda_{e} = 1 , \; \lambda = 1$.}
\end{figure}

To study the polarization effects at each value of the energy $E$, we
calculated the energy spectra for the following polarization states of the
electron and laser photon:
$$
1 \to \lambda_{e} = 0 , \; \lambda = 1 ; \; 2 \to \lambda_{e} = 1 ,
 \; \lambda = - 1 ; \; 3 \to \lambda_{e} = 1 \; , \; \lambda = 1 .
$$
These correspond to lines 1, 2 and 3, respectively, in Fig. 3. Everything
said above about the behavior of the energy spectra pertained to these three
lines. Regarding their relative location, from Fig. 3 we see that the most
intense spectra correspond to the case where the electron and laser photon
spins are parallel ($\lambda \lambda_{e} = - 1$), while the least intense
ones correspond to antiparallel spins ($\lambda \lambda_{e}=1$), as in the
case of ordinary Compton back-scattering (see Ref. 89).

We also note that the difference between the spectra calculated for the
three polarization cases considered is very large at small values of the
intensity parameter ($\xi^2 = 0.3$), but insignificant at $\xi^2 = 3 \;
(E = 50$ GeV). It again arises only in connection with increasing electron
energy (see Fig. 3c for $E = 300$ GeV).

\begin{figure}[!hbt]
\hspace{-0.5cm}
\centerline{
\leavevmode\epsfxsize=0.4\textwidth\epsfbox{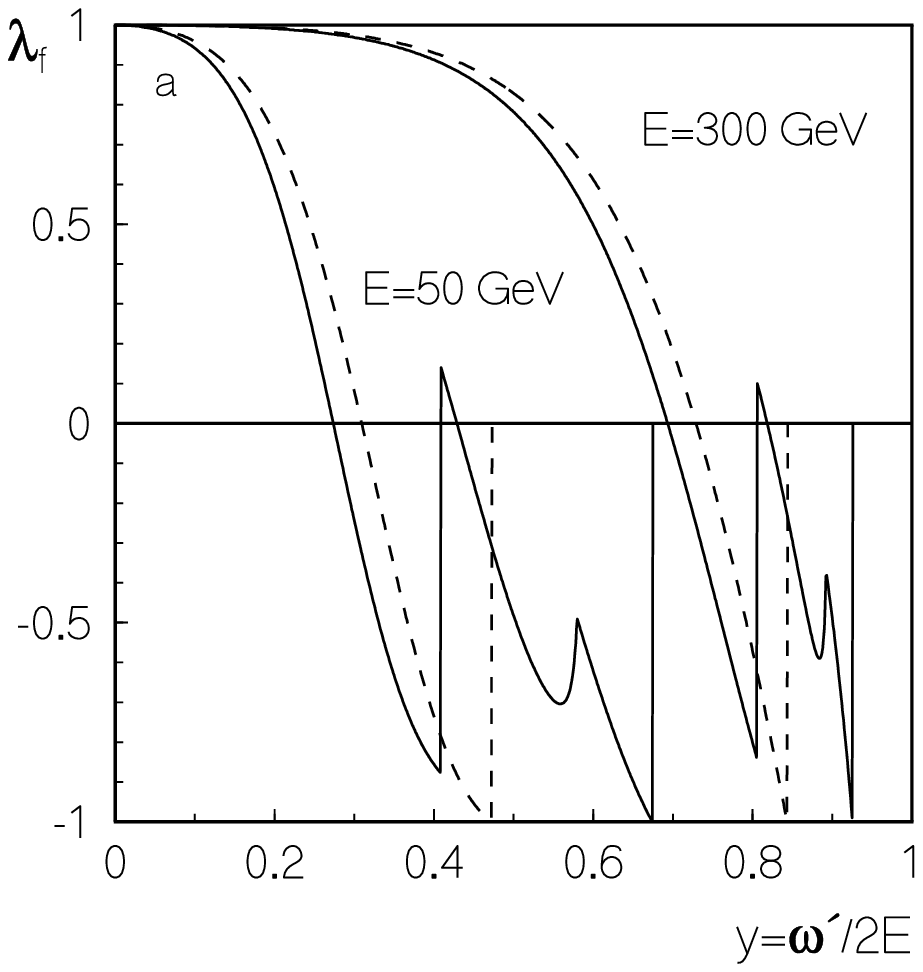}\hspace{-1.5cm}
\leavevmode\epsfxsize=0.4\textwidth\epsfbox{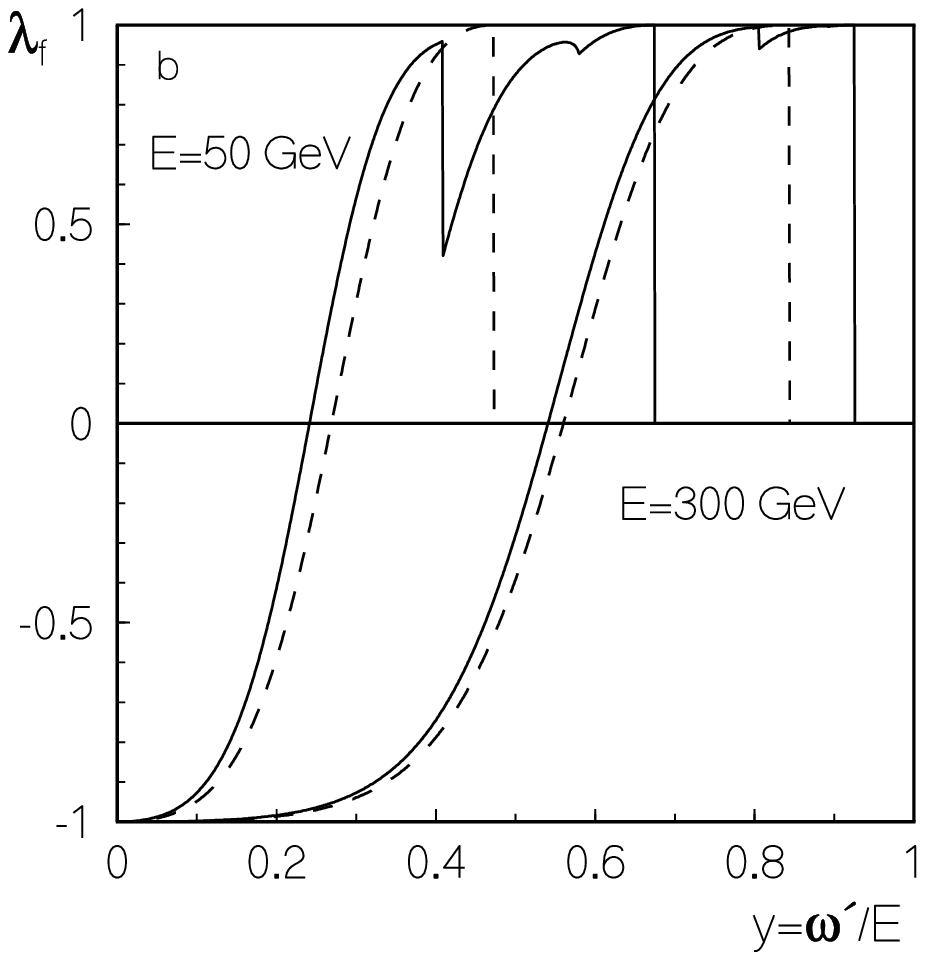}\hspace{-1.5cm}
\leavevmode\epsfxsize=0.4\textwidth\epsfbox{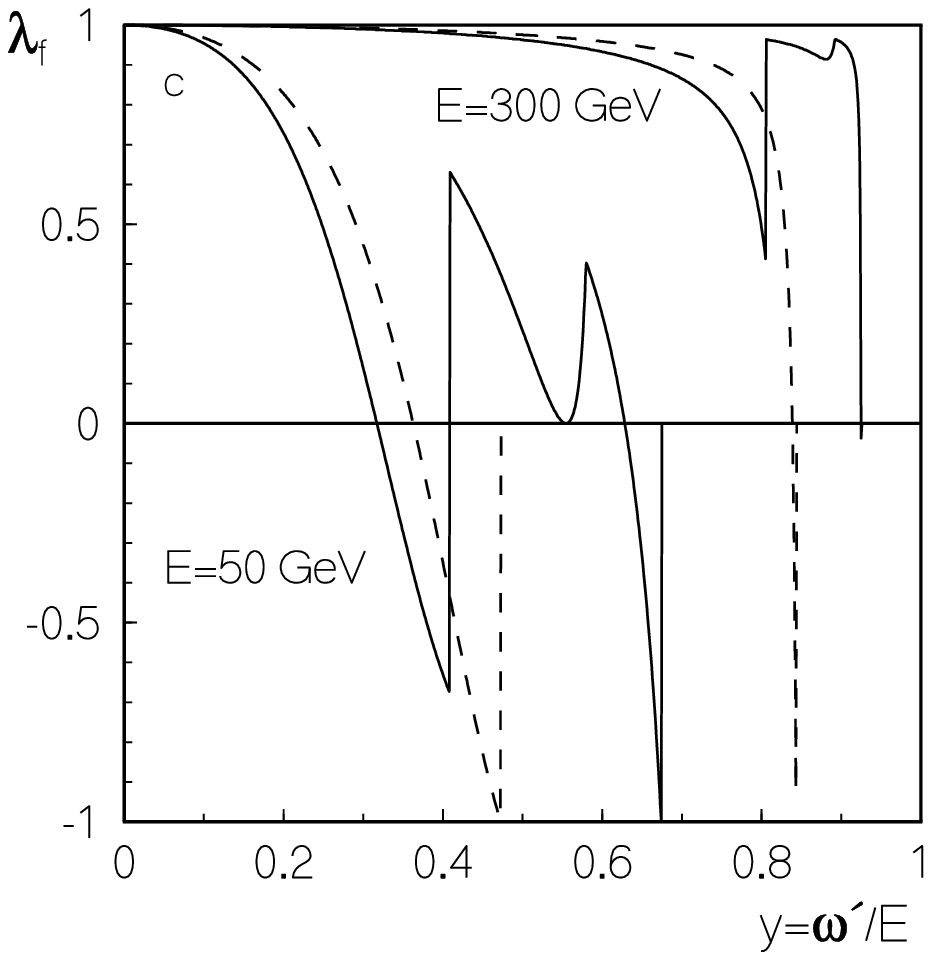} }
\vspace{-1cm}
\caption
{Energy dependence of the degree of circular polarization of high-energy
photon, calculated at $\xi^2=0.3$ for the following polarization states of
the colliding particles:
$(a)\; \lambda_{e} = 0 , \; \lambda = 1 ; \; (b) \; \lambda_{e} = 1 ,
 \; \lambda = - 1; \; (c)\; \lambda_{e} = 1 , \; \lambda = 1$.
The dashed lines correspond to ordinary Compton scattering. }
\end{figure}

\begin{figure}[!hbt]
\hspace{-0.5cm}
\centerline{
\leavevmode\epsfxsize=0.4\textwidth\epsfbox{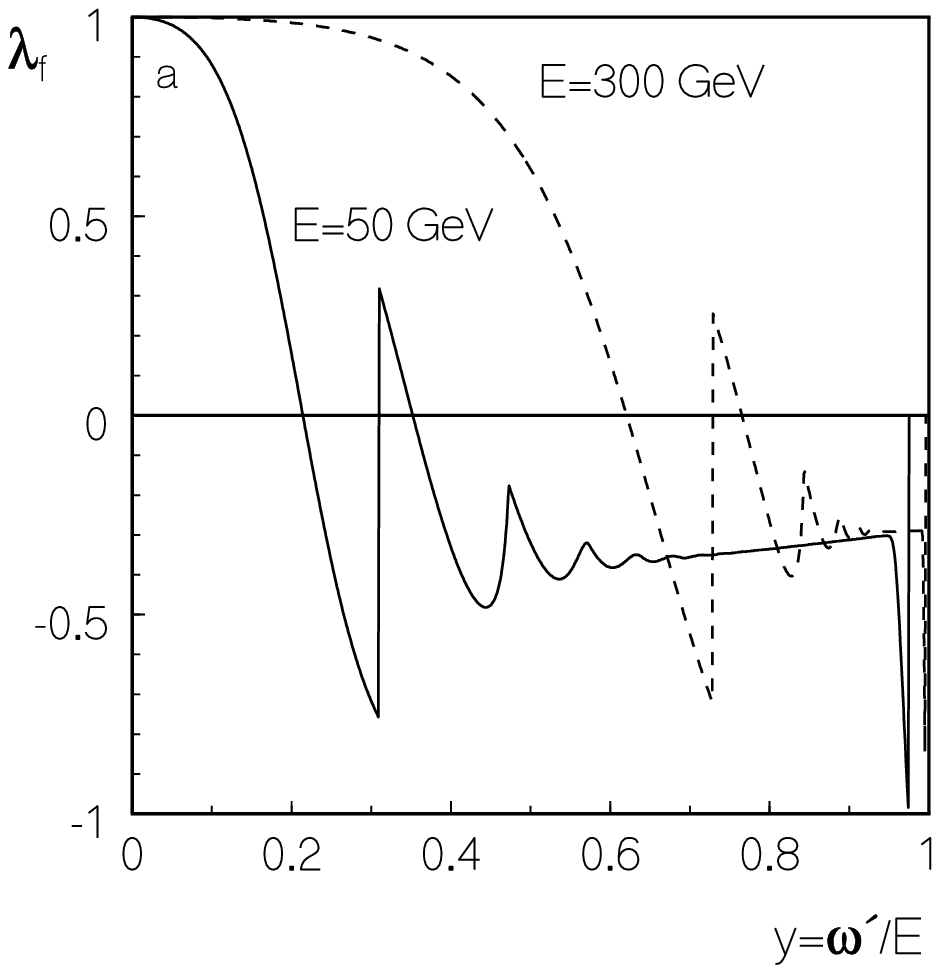}\hspace{-1.5cm}
\leavevmode\epsfxsize=0.4\textwidth\epsfbox{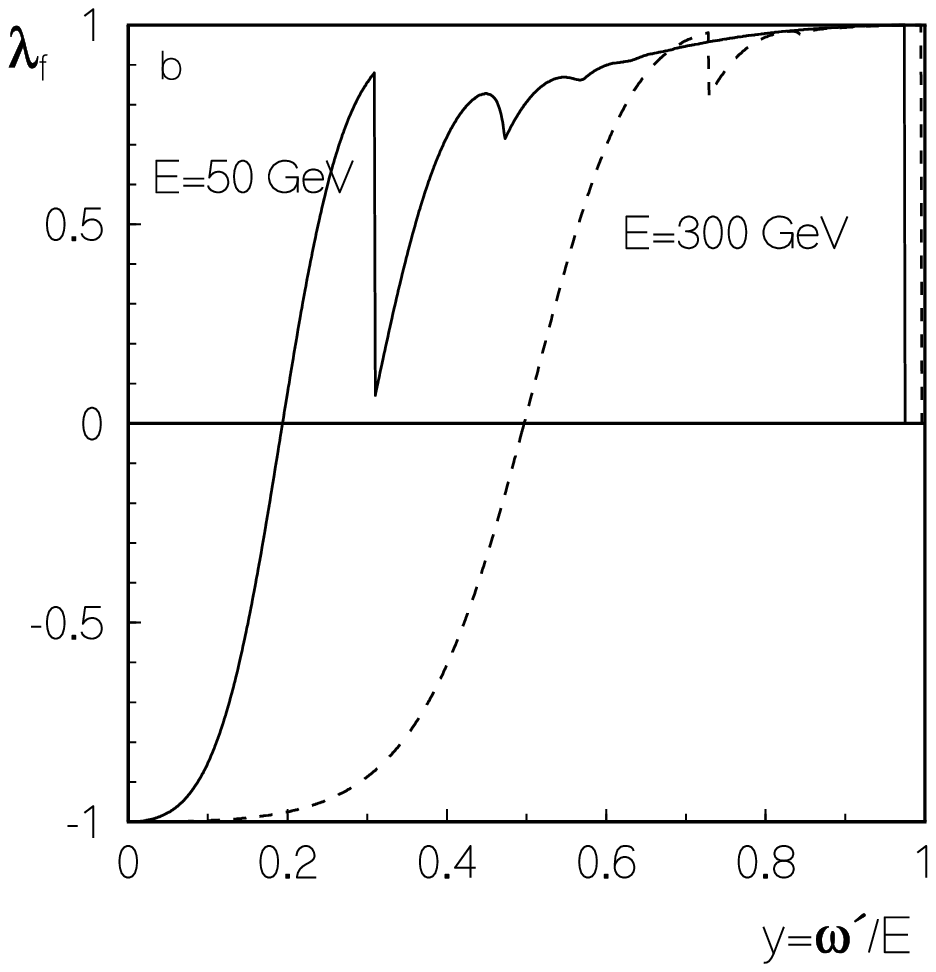}\hspace{-1.5cm}
\leavevmode\epsfxsize=0.4\textwidth\epsfbox{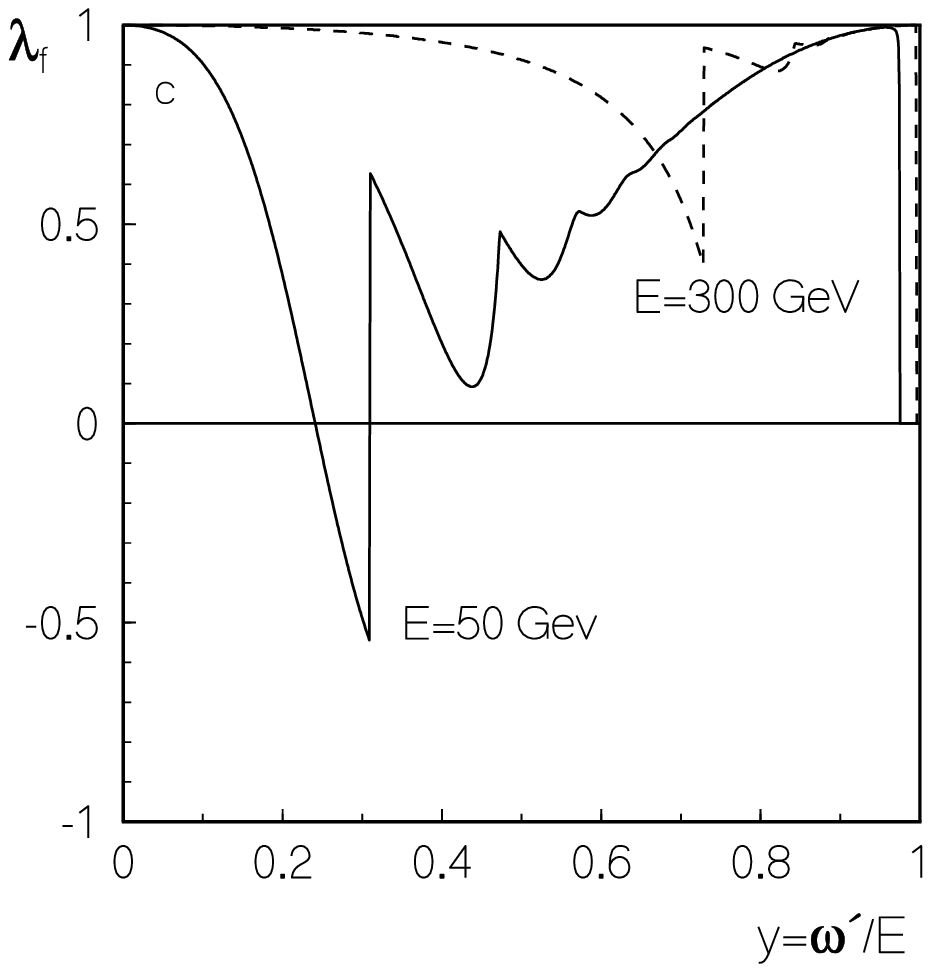} }
\vspace{-1cm}
\caption
{Energy dependence of the degree of circular polarization of high-energy
photon, calculated at $\xi^2=1$ for the following polarization states of
the colliding particles:
$(a)\; \lambda_{e} = 0 , \; \lambda = 1 ; \; (b) \; \lambda_{e} = 1 ,
 \; \lambda = - 1; \; (c)\; \lambda_{e} = 1 , \; \lambda = 1$.
The solid lines correspond to electron energy $E=50$ GeV, and the dashed
lines to $E=300$ Gev.}
\end{figure}

Let us consider the energy
dependence of the degree of circular polarization of a hard $\gamma$ ray,
shown by the graphs in Figs. 4 and 5. For this we first note that the
above-mentioned kinematical features of the behavior of the amplitudes
$M^{~(n)}_{\pm \delta , \delta}$ in (8.7) and (8.8) has a spin origin [86].
In fact, the equation $u = u_{n}$ correspond to photon emission in the
direction of motion of the initial electron beam. In the case of absorption
of $n$ photons ($n > 1 $) from the wave and exact backward scattering of
the hard photon, the total helicity of the $e+n \gamma_{0}$ and $e+\gamma$
system before and after the interaction is not conserved. It is this which
causes all the amplitudes $M^{~(n)}_{\pm \delta , \delta}(u = u_{n})$ for
$n > 1$ and also $M^{~(n)}_{\delta , \delta } (u = u_{1}$)to vanish. The
requirement of helicity conservation also leads to $\lambda_{f}=-\lambda$
for ordinary Compton scattering at the edge of the spectrum [86].

As can be seen from Figs. 4 and 5, the inclusion of nonlinear effects
($\xi^2 \neq 0$) decreases the degree of circular polarization at the first
peak. The contribution of higher harmonics leads to the appearance of
additional peaks, and at the edge of the spectrum (for $n = n_{max}$) we
have $\lambda_{f} = - \lambda$, as in the case of ordinary scattering.
However, it should be noted that the yield of these photons is insignificant,
since the spectra are practically broken off at $y \ll y_{n_{max}}$. The
situation regarding $\lambda \lambda_{e}=-1$ is the most favorable in this
respect, as there is a large range of hard $\gamma$ energies in which the
degree of circular polarization $\mid \lambda_{f} \mid $ is very close to
unity.

\section{\bf $e^{+} e^{-}$-pair production by a hard photon in a \newline
collision with photons of a laser wave}

In Ref. 84 it was shown that a hard photon obtained in the reaction (8.1)
can create $e^{+} e^{-}$ pairs in a collision with photons of the same laser
beam. The threshold for this reaction (8.2) at $s=1$ is very high. The
lowest energy of the Compton photon ($s=1$) in the process (8.2) for a
neodymium laser with $\omega _{0} = 1.17$ eV is $\omega = m^{2} / \omega_{0}
= 223$ GeV. In fact, $e^{+} e^{-}$ pairs will be created in large numbers
and at significantly lower energies owing to collisions of the hard photon
$\gamma$ with several laser photons $\gamma_0$ simultaneously [84].
Observation of the process (8.2) is particularly interesting for verifying
QED in a new parameter region. At the same time, it is an important source
of background for $\gamma e$ and $\gamma \gamma$ collisions, and a possible
method of dealing with it is described in [84].

Like (8.1), the reaction (8.2) is an interaction of electrons and photons
with the field of an electromagnetic wave which is nonlinear in the field
strength. It is easily checked that the inclusion of the influence of the
nonlinear effects in (8.1) on the process (8.2) also leads to a significant
lowering of the $e^{+} e^{-}$-pair production threshold and to an increase
in the number of pairs [90].

The maximum energy of a Compton photon $\gamma $ resulting from the absorption
from the wave of $n$ laser photons of energy $\omega _{0}$ by an electron
of energy $E$ is
$$
\omega _{n} = \; { n x  \over 1 + n x } E \;
,\qquad x  = { 4 \omega _{0} E \over m^{2}}   \; .
\eqnum{9.1}
$$
The threshold value of the $\gamma$ energy for the process (8.2) is given
by
$$
( k + s k_{0} )^{2} = 4 m^{2} \; ,
\eqnum{9.2}
$$
where $k$ and $k_{0}$ are the 4-momenta of the photons $\gamma $ and
$\gamma _{0}$. The corresponding threshold values of the energy of the
electrons in the accelerator beam $E_{ns}$ for $e^{+} e^{-}$-pair production
owing to absorption of $n$ photons from the wave and collisions with $s$
laser photons are determined from (9.1) and (9.2):
$$
E_{ns}  =
{m^{2}  \over 2  \omega _{0} s} (1 + ( 1 + s / n )^{1/2} )\; .
\eqnum{9.3}
$$
For $n=1$ we obtain Eq. (7) of Ref. 84. Using (9.3), we can calculate the
values of $E_{1s}$ and $E_{2s}$ for $\omega _{0} = 1.17$ eV and $1 \leq s
\leq 6$. The results (in GeV) are given in Table 1:

\begin{table}[htb]
\caption{Threshold values of the electron energy in the accelerators beam
$E_{ns}$ (in GeV) for $e^{+}e^{-}$-pair production at various $n$ and $s$ in
the case of neodymium laser}
\begin{center}
\begin{tabular}{|c|c|c|c|c|c|c|} \hline
\hspace*{.4cm} s \hspace*{.4cm} &\hspace*{.4cm} 1\hspace*{.4cm}
&\hspace*{.4cm}  2\hspace*{.4cm}   &\hspace*{.4cm}  3\hspace*{.4cm}
&\hspace*{.4cm}  4\hspace*{.4cm}  & \hspace*{.4cm}  5\hspace*{.4cm}
&\hspace*{.5cm} 6 \hspace*{.5cm}   \\    \hline
 $E_{1s}$ &   269  &  153 & 112 &  90 &  77 & 68   \\    \hline
 $E_{2s}$ &   248  &  135 & 96  &  76 &  64 & 56   \\     \hline
\end{tabular}
\end{center}
\end{table}

These results clearly show that the broadering of the hard-$\gamma$ spectrum
due to nonlinear effects also leads to lowering of the $e^{+} e^{-}$-pair
production threshold.

The matrix elements $M^{(s)}_{\pm \mu \mu }=M^{(s)\lambda \lambda'}_{\pm
\mu \mu}$ and the differential probability for the process (8.2) in the field
of a circularly polarized electromagnetic wave are given by [90]:
$$
d W^{(s)} =
{ e^{2} m^{2} \over 4 \pi \omega } \mid  M^{(s) \lambda \lambda'}_{\pm \mu,
\mu } \mid ^{2} \delta^{4}(s k_{0} + k - q - q'){d^{3} q d^{3} q' \over q_{0}
q_{0}'}
\eqnum {9.4}
$$
$$
M^{(s)}_{\mu \mu } = (-\lambda)^{s} \{-\lambda '
\mu n_{1} n_{3}' J_{s} + {\xi m s \over m^{2}_{*}  u_{s}} (u k n_{0}'-
\lambda' \mu \varepsilon \sqrt{u (u-1)} k n_{3}') J_{s-\lambda \lambda '}
\},~~~~
$$
$$
M^{(s) }_{-\mu \mu }=-\lambda '(-\lambda)^{s}
(n_{1} n_{1}'+ \lambda'\mu)  \{\sqrt{(vv'+ 1)/2}  J_{s} +
{\xi m s u \over m^{2}_{*} u_{s}}
\sqrt{(vv'-1)/2} k n_{1}'J_{s-\lambda \lambda'}\},
$$
where
$$
n_{1} n_{3}' = -{m^{2}_{*} u_{s} \over m^{2} u \sqrt{2(v v'- 1 ) } }
{z \over s \xi}, \; k n_{0}' = {2 m^{2}_{*} u_{s} \over m s \sqrt
{2(v v'+ 1)}} \;  ,
$$
$$
k n'_{3} = - \varepsilon  {2 m^{2}_{*} u_{s} \sqrt{ (u-1)/ u }
\over m s \sqrt{2(vv' - 1) } }, \; n_{1} n'_{1} = \varepsilon \sqrt {{u-1
\over u }} \sqrt {{vv' + 1 \over vv' -1 }} \; ,
$$
$$
k n' _{1} = - {m^{4}_{*} u^{2}_{s} \over s^{2} m^{3} u
\sqrt{ (vv')^{2} - 1 }} {z \over \xi } , \;
\varepsilon  = \hbox{ sign } \sqrt{{u_{s} (u-1) \over u (u_{s}-1)}} \; ,
$$
$$
u = {(k k_{0})^{2} \over 4 k_{0} q \cdot k_{0} q' } , \; u_{s}={s \over s_0}
= {s k k_{0} \over 2 m^{2}_{*}} , \; z = {2 s \xi \over \sqrt{1 + \xi^{2} }}
 \sqrt { { u \over u_{s}} \left (1- {u \over u_{s}} \right ) }\; ,
$$
$$
 q = p + {\xi^{2} m^{2} \over 2 k_{0} p } k_{0} , \;
q'  = p'  + {\xi^{2} m^{2} \over 2 k_{0} p' } k_{0} , \;
q^{2} = (q')^{2} = m^{2}_{*} = m^{2} (1 + \xi^{2})\;  ,
$$
$$
s k_{0} + k = q + q' , \;
 v v'  - 1 = 2 (u_{s} - 1 + \xi^{2} (u_{s} - u) ) \; .
$$
Here $k_{0}, \lambda$ and $k, \lambda '$ are the 4-momenta and helicities
of the laser and hard photons, $\mu$ is the projection of the positron spin
on the axis (1.9), $q$ and $q^{'}$ are the positron and electron
quasimomenta, $s_0$ is the threshold value for the number of absorbed photons,
$J_{s} = J_{s}(z) $ is the Bessel function of argument $z$, and $\xi^{2}$
is the wave intensity parameter (8.3).

The total probability for pair production by a photon in the process (8.2)
per unit volume and unit time is given by [90]:
$$
W={\alpha m^2 \over 4 \; \omega} \sum^{\infty}_{ s>s_0 }\; \int
\limits ^{u_{s}}_{1}\; (F_{0s} + \lambda \lambda' F_{2s} +\mu \lambda
G_{0s} + \mu \lambda' G_{2s})\; {du \over u \sqrt{u(u-1)}}\; ,
\eqnum {9.5}
$$
\begin{eqnarray}
&& F_{0s} = J_{s}^2 + \xi^2 \; (2u-1)\; (-J_{s}^2 +(J_{s-1}^2 +
J_{s+1}^2)/2\; )\; ,\nonumber \\
&& F_{2s} = \xi^2 \; (2u -1) (2u /u_s -1 ) (J_{s-1}^2 - J_{s+1}^2)/2\;
,\nonumber \\
&& G_{0s} = \psi_{+}\psi_{-} \; \xi^2 \; u/u_s (u_s -1) \; (J_{s-1}^2 -
J_{s+1}^2\; )\; , \nonumber \\
&& G_{2s} = \psi_{+}\psi_{-} \; \{ u_s J_s^2 +\xi^2 [u_s J_s^2 +u(u-1)
(J_{s-1}^2 + J_{s+1}^2)\; ] \} \; , \nonumber \\
&&\psi_{\pm} = 1/\sqrt{(vv' \pm 1)/2}\;.  \nonumber
\end{eqnarray}

The total number of $e^{+}e^{-}$-pairs $N_{e^{+} e^{-}}$ created by a hard
photon is obtained by summing over the energy of the Compton photons
[84]:
$$
N_{e^{+}e^{-}} = N_{\gamma }\;
{\tau \over 4} \; \sum^{\infty}_{s_{0}} \; \int \limits^{\omega _{n}}_{ 0}
W^{(s)} (\omega , \omega _{0}, \xi) \; {1\over \sigma _{c}(E)}\;
{d\sigma _{c} \over d\omega }\;  d\omega \; ,
$$
where $N_{\gamma }$ is the total number of hard photons, $\sigma _{c}(E)$
and $d\sigma _{c}/ d\omega $ are the total and differential cross sections
for CBS, $W^{(s)}(\omega , \omega _{0}, \xi)$ is the probability for pair
production by a hard photon per unit time in the process (8.2), and $\tau$
is the duration of the laser flash. The results of numerical calculations
of $\log ( N_{e^{+}e^{-}} / N_{e})$ as a function of the beam electron energy
$E$ for various energies of the laser flash $A$, wave polarizations
$\lambda $, helicity of the initial electron beam $\lambda _{e}$, and
spin projection $\mu $ on the $\vec c_3$ axis (1.9) for positrons are shown
by the graphs in Fig. 6 [90].

\vspace{-1cm}
\begin{figure}[hbt]
\centerline{
\leavevmode\epsfxsize=0.5\textwidth\epsfbox{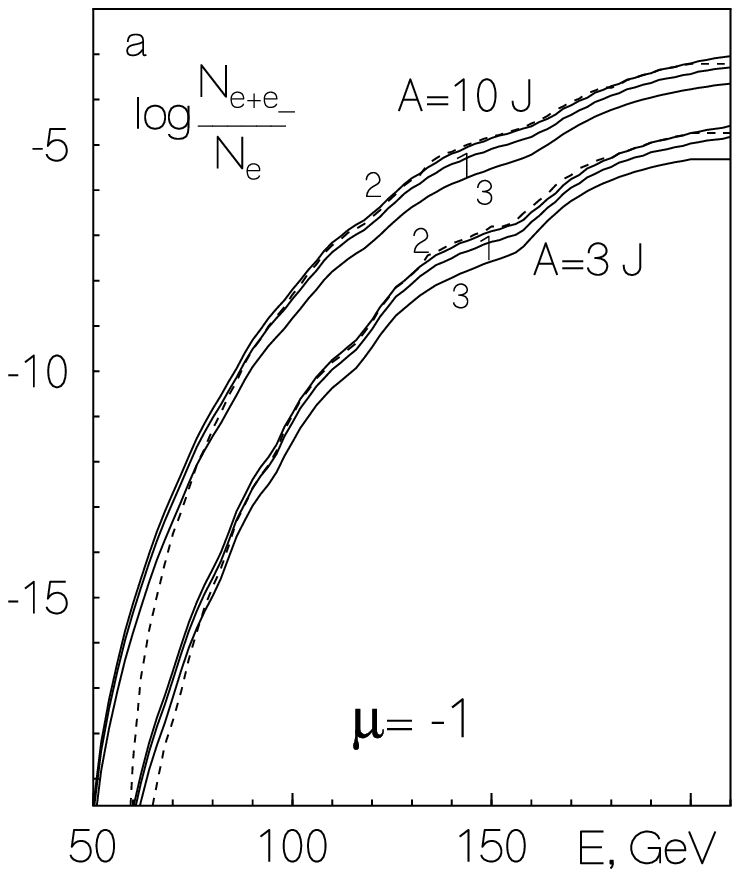}
\leavevmode\epsfxsize=0.5\textwidth\epsfbox{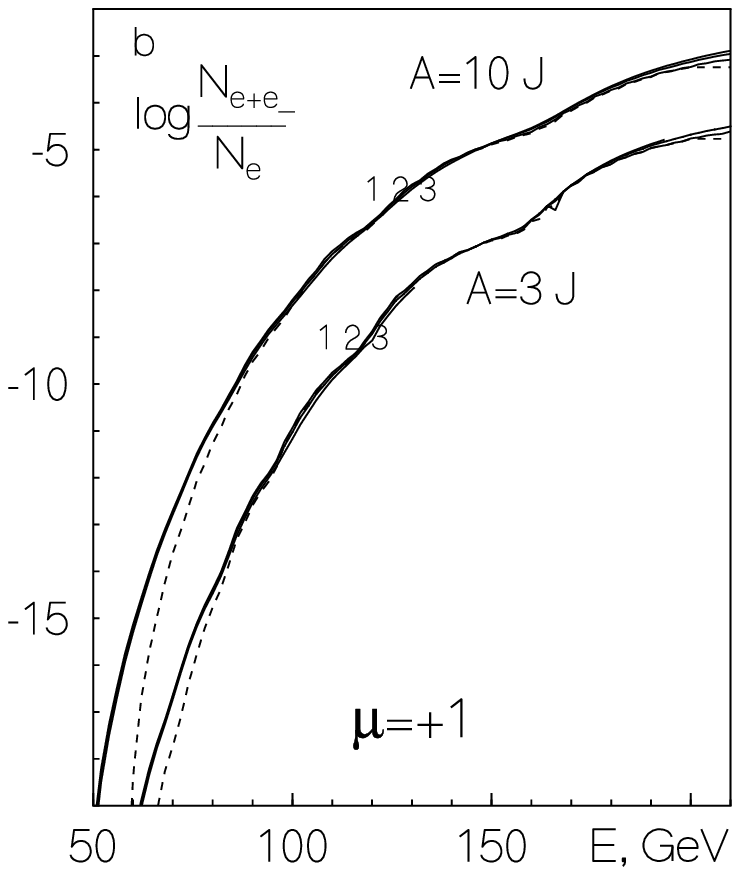} }
\vspace{-1cm}
\caption
{Dependence of the number of $e^+e^-$-pairs created by a hard Compton photon
on the electron beam energy. The lines 1, 2, and 3 correspond to
the following choice of helicities $\lambda$ and $\lambda_e$ of the initial
particles in the reaction (8.1): $(1)\; \lambda \lambda_e = 0, \; (2) \;
\lambda  \lambda_e = - 1, \; (3)\; \lambda \lambda_e = 1$.
The solid lines correspond to $n=2$, and the dashed lines to $n=1$ in (8.1)
Figure (a) corresponds to positron spin projection $\mu=-1$, and (b) to
$\mu=+1$.}
\end{figure}

The lines 1, 2, and 3 correspond to the following choice of helicities:
(1) $\lambda  \lambda _{e} = 0 ; \;(2) \; \lambda \lambda _{e} = -1$; and (3)
$\lambda  \lambda _{e} = 1$. The solid lines correspond to $n = 2$, and the
dashed lines to $n = 1$ in the process (8.1). It follows from Fig. 6 that
nonlinear effects in CBS lead to a significant increase in the number of
$e^{+}e^{-}$ pairs created by a hard photon at current accelerators energies.

\vspace{-1.0cm}
\section*{\bf Conclusion}

The goal of the present review was to explain what the DSB is, what new
contribution it makes to the description of particle spin properties and
also to the development of the covariant method to calculate matrix elements
in the Bogush-Fedorov approach, and how this method is related to others
like that of the CALCUL group.

We have seen that the DSB plays a key role among all the other methods in
that in it the Loretz little group common to particles with 4-momenta
$p_1$ (before the interaction) and $p_3$ (after the interaction) is realized.
The DSB allows the description of the spin states of systems consisting of
two particles (even when they have different masses) by means of the spin
projections on a single common direction. The coincidence of the Loretz little
groups causes the particles before and after the interaction to have a common
set of spin operators which commute with each other, and this allows the
covariant separation of the interactions with and without change of the spin
states of the particles involved in the reaction, so that the dynamics of
the spin interaction can be traced. Thanks to the coincidence of the spin
operators and also the fact that Wigner rotations are singled out, the
mathematical structure of the diagonal amplitudes is maximally simplified.

To calculate matrix elements in the covariant Bogush-Fedorov approach, it
is necessary to know the projection operators of the particle states, the
operator for the transition from the initial to the final state (and its
inverse), and also the raising and lowering spin operators in the case of
spin flip transitions. In the review we have developed this covariant
approach by using the DSB. We have constructed the operators $u^{\delta}
(p_{1}) \;\overline {u}^{~\pm \delta}(p_{3})$ used to calculate the diagonal
amplitudes in the case of transitions without and with spin flip. They are
valid in both the massive and the massless cases. We have obtained three
equivalent representations for them which have a compact form. We have also
studied the transition to the massless case, in which the DSB coincides up
to a sign with the helicity basis.

In the CALCUL method the fermion must be massless. The key feature of that
method is the very convenient choice of photon polarization vectors, in which
the momenta of the fermions from which the photons are emitted are used.
This ensures gauge invariance and simplifies the structure of the amplitudes,
so that they can ultimately be calculated. In the CALCUL method the mass
can be taken into account only in the ultrarelativistic case and only in
the form of awkward mass corrections. As a rule, generalizations of this
method to the massive case require the introduction of auxiliary vectors
unrelated to the kinematics of the problem, which are therefore inconvenient
to work with. Nevertheless, this method contains the attractive idea of
constructing the photon polarization vectors in terms of the 4-momenta of
the particles participating in the reaction. This allows a decrease in the
number of various scalar products in the final expressions for the
amplitudes and thereby simplifies the calculations. Therefore, giving up
on the generality of the treatment makes the solution of the problem more
efficient. This is even more true with regard to the method developed for
calculating diagonal amplitudes, because the construction of the mathematical
formalism for them involves only the 4-momenta of the particles participating
in the reaction. In the DSB this is sufficient, thanks to the use of the
ideas of the covariant Bogush-Fedorov approach.

Let us briefly list the main results of our calculations of several specific
QED processes using the method developed for calculating matrix elements
in the DSB.

We have shown that in the ultrarelativistic (massless) limit, the differential
cross sections for M\"{o}ller and Bhabha bremsstrahlung ($ e^{\pm} e^{-} \to
e^{\pm} e^{-} \gamma $) in the case where not only the initial $e^{\pm}$
and $e^{-}$, but also the photon are helically polarized can be presented
as the product of two factors, one of which is universal and coincides with
that obtained earlier by the CALCUL group when polarization is absent.

The helicity amplitudes of the three-photon annihilation of a free pair
$e^{+} e^{-}\to 3 \gamma$ have been calculated along with the orthopositronium
annihilation amplitudes corresponding to total spin projection $0, \pm 1$.
The differential cross sections taking into account the polarizations of
the various particles were obtained. The annihilation probability was
calculated in the case where one $\gamma$ is linearly polarized and the other
two are unpolarized. The expression obtained for the degree of photon linear
polarization coincides with the results of other authors.

A compact expression was obtained for the differential cross section of the
Bethe-Heitler emission of a linearly polarized photon by an electron, taking
into account the proton recoil and form factors, thanks to the factorization
of the squared electric and magnetic form factors of the proton. In the limit
where the proton is a point particle of infinite mass, this expression becomes
the usual one.

We have studied the reaction $ ep \to ep \gamma$, taking into account the
proton polarizability in the kinematics corresponding to electron scattering
at small angles and photon scattering at fairly large angles, where proton
bremsstrahlung dominates. The results of numerical calculations performed
in the rest frame of the initial proton at electron beam energy $E_{e}=200$
MeV in the chosen kinematics show that the conditions needed to isolate the
subprocess $\gamma p \to \gamma p$ from the reaction $ep \to ep \gamma$
are satisfied, because the relative contribution of the Bethe-Heitler and
interference terms to the reaction cross section is less than 10 \%, and
the cross section for the reaction $ep \to ep \gamma$ is quite sensitive
to the proton polarizability.

A covariant expression has been obtained for the lepton tensor in which the
contribution of states with transverse and longitudinal polarization of the
virtual photon is isolated. It has been shown that inclusion of the lepton
mass tends to increase the degree of linear polarization of the virtual
photon.

We have studied nonlinear effects in Compton back-scattering of photons
by an intense circularly polarized laser wave focused on a beam of
longitudinally polarized ultrarelativistic electrons ($ e + n \gamma_{0}
\to e + \gamma $). We have found that at high intensities the emission of
a hard photon is essentially nonlinear, and the effect of the polarizations
is markedly diminished.

We have shown that the broadering of the spectrum in nonlinear Compton
back-scattering tends to lower the $e^{+}e^{-}$-pair production threshold
and increase the number of pairs in collisions of a hard Compton photon with
several laser photons simultaneously ($\gamma + n \gamma_0 \to e^{+} +
e^{-}$).

Thus, the absence of difficulties associated with inclusion of mass and
calculation of spin flip amplitudes, and also the elegance of the results
obtained, demonstrate the clear superiority of developed method to
calculate matrix elements.

\section*{\bf Acknowledgements}

The authors thank the creators of the covariant approach, professor
A.A. Bogush, and Academician F.I. Fedorov (deceased). Theyr fruitful ideas
and methods have largely determined the direction of these studies and
the nature of the work. The authors are grateful V.I. Kuvshinov for
stimulating the writing of this review, and also to I.F. Ginzburg,
E.A. Kuraev, M.I. Levchuk, A.I. L'vov and V.A. Petrun'kin for asking
questions, for useful discussions, and for theyr interest in the study.


\newpage
\begin{thebibliography}{99}
%
\bibitem{b1} G. Altarelli, in Polarization at LEP, {\bf 1} CERN 88-06, CERN,
Geneva (1988) p.13; ibid., B.W. Lynn, p. 24; W. Hollik, ibid., p.83.
%
\bibitem{b2} HERMES Collaboration, DESY-PRC 93/06, DESY, Hamburg (1993).
%
\bibitem{b3} J.A. Lauber, Report 413, SLAC, Palo Alto (1993), p. 29.
%
\bibitem{b4} H. Haber, in Proceedings of the XXI SLAC Summer Institute
on particle physics, SLAC, Palo Alto (1994).
%
\bibitem{b5} S.B. Nurushev, Preprint 91-103, IHEP, Protvino (1991).
%
\bibitem{b6} C.S. Wu, E. Ambler et al., Phys. Rev. {\bf 105}, (1957) 1413.
%
\bibitem{b7} T.D. Lee, C.N. Yang, Phys. Rev. {\bf 105}, (1957) 1671.
%
\bibitem{b8} I.H. Cristinson et al., Phys. Rev. Lett. {\bf 13}, (1964) 138.
%
\bibitem{b9} A.I. Akhiezer and V.B. Berestetskii, Quantum Electrodynamics
(Wiley, New York, 1965) [Russ: original, 3rd ed., Nauka, Moscow, 1969]
%
\bibitem{b10} V.B. Berestetskii, E.M. Lifshitz, and L.P. Pitaevskii, Quantum
Electrodynamics, 2nd ed. (Pergamon Press, Oxford, 1982) [Russ: original,
Nauka, Moscow, 1989].
%
\bibitem{b11} F. Halsen and A.D. Martin, Quarks and Leptons: an Introductory
Course in Modern Particle Physics (Wiley, New York, 1984) [Russ: transl.,
Mir, Moscow, 1987].
%
\bibitem{b12} J.D. Bjorken and S.D. Drell, Relativistic Quantum Mechanics
(McGraw-Hill, New York, 1965) [Russ: transl., Nauka, Moscow, 1978].
%
\bibitem{b13} F.I. Fedorov, The Loretz group [in Russian] (Nauka, Moscow,
1979).
%
\bibitem{b14} J.L. Powell, Phys. Rev. {\bf 75}, (1949) 32.
%
\bibitem{b15} A.A. Sokolov and I.M. Ternov, Radiation from Relativistic
Electrons (AIP, New York, 1986) [Russ: original, Nauka, Moscow, 1983].
%
\bibitem{b16} Yu.S. Tsai, Phys. Rev. {\bf D48}, (1993) 96.
%
\bibitem{b17} P.A.M. Guichon, G.Q. Lui, and A.W Thomas, Nucl. Phys. {\bf
A591}, (1995) 606.
%
\bibitem{b18} E. Bellomo, Nuovo Cim. {\bf 21}, (1961) 730.
%
\bibitem{b19} A.A. Bogush and F.I. Fedorov, Vestsi Akad. Nauk BSSR,
Ser. Fiz. Tekh. Nauk, No 2, (1962) 26 [in Russian].
%
\bibitem{b20} F.I. Fedorov, Zh. Eksp. Teor. Fiz. {\bf 35} (1958) 493.
%
\bibitem{b21} CALCUL Collaboration, Phys. Lett. {\bf B105}, (1981) 215;
Phys. Lett. {\bf B114}, (1982) 203; Nucl. Phys. {\bf B206}, (1982) 53;
Nucl. Phys. {\bf B206}, (1982) 61; Nucl. Phys. {\bf B239}, (1984) 382;
Nucl. Phys. {\bf B239}, (1984) 395.
%
\bibitem{b22} R. Kleiss and W.J. Stirling, Nucl. Phys. {\bf 262}, (1985) 235.
%
\bibitem{b23} A. Ballestrero and E. Maina, Phys. Lett. {\bf B350}, (1995) 225.
%
\bibitem{b24} F.I. Fedorov, Dokl. Akad. Nauk BSSR {\bf 2}, (1958) 408
[in Russian].
%
\bibitem{b25} F.I. Fedorov, Dokl. Akad. Nauk BSSR, {\bf 5}, (1961) 101
[in Russian].
%
\bibitem{b26} F.I. Fedorov, Dokl. Akad. Nauk SSSR {\bf 143}, (1962) 56
[in Russian].
%
\bibitem{b27} A.A. Bogush and F.I. Fedorov, Dokl. Akad. Nauk BSSR {\bf 5},
(1961) 327 [in Russian].
%
\bibitem{b28} E.E Tkharev and F.I. Fedorov, Yad. Fiz. {\bf 5}, (1962) 1112.
%
\bibitem{b29} A.A. Bogush, Vestsi Akad.Nauk BSSR, Ser.Fiz.Tekh.Nauk, No 2,
(1964) 29.
%
\bibitem{b30} F.I. Fedorov, Vestsi Akad. Nauk BSSR, Ser.Fiz.-Mat.Nauk, No
2 (1974) 58 [in Russian].
%
\bibitem{b31} F.I. Fedorov, Vestsi Akad. Nauk BSSR, Ser. Fiz.-Mat. Nauk,
No 3, (1975) 51 [in Russian].
%
\bibitem{b32} F.I. Fedorov, Izv. Vysh. Ucheb. Zaved. Ser. Fizika, No 2,
(1980) 32 [in Russian].
%
\bibitem{b33} M. Jacob and G. Wick G, Ann. Phys. {\bf 7} (1959) 404.
%
\bibitem{b34} R. Vega and J. Wudka, Phys. Rev. {\bf D53} (1996) 5286.
%
\bibitem{b35}S.M. Sikach,
in Covariant Methods in Theoretical Physics [in Russian],
(Institute of Physics, Belorussian Academy of Sciences, Minsk, 1981), p. 91.
%
\bibitem{b36} S.M. Sikach, Vestsi Akad. Nauk BSSR, Ser. Fiz.-Mat. Nauk, No 2,
(1984) 84 [in Russian].
%
\bibitem{b37} F.I. Fedorov, Teor. Mat. Fiz. {\bf 2} (1970) 343
[Theor. Math. Phys.(USSR)].
%
\bibitem{b38} S.M. Sikach, Candidate's Dissertation, Minsk (1987) [in Russian].
%
\bibitem{b39} M.V. Galynsky and S.M. Sikach,
in Covariant Methods in Theoretical Physics [in Russian],
(Institute of Physics, Belorussian Academy of Sciences, Minsk, 1986), p. 121.
%
\bibitem{b40} M.V. Galynsky et al., Zh. Eksp. Teor. Fiz. {\bf 95} (1989)
1921 [Sov.Phys. JETP. {\bf 68}, (1989) 1111]
%
\bibitem{b41} A.A. Bogush, L.G. Moroz, S.M. Sikach, and F.I. Fedorov,
in Proceedings of the XI Seminar on Problems in High Energy Physics and Field
Theory [in Russian], Protvino, 1988 (Nauka, Moscow, 1989), p. 308.
%
\bibitem{b42} S.M. Sikach, Preprints Nos. 658, 659 Institute of Physics,
Belarusian Academy of Sciences, Minsk (1992); in Covariant Methods in
Theoretical Physics [in Russian], (Institute of Physics, Belarusian
Academy of Sciences, Minsk, 1997), p. 151.
%
\bibitem{b43} Yu.V. Novozhilov, Introduction to Elementary Particle Theory
(Pergamon Press, Oxford, 1975) [Russ. original, Nauka, Moscow, 1972].
%
\bibitem{b44} A.L. Bondarev, Teor. Mat. Fiz. {\bf 101}, (1994) 315 [Theor.
Math. Phys. (USSR)]; e-print hep-ph/9710398; e-print hep-ph/9701332.
%
\bibitem{b45} V. Bargman and E. Wigner, Proc. Ac. Nat. Sci. USA, {\bf 34},
(1948) 211.
%
\bibitem{b46} M.V. Galynsky and S.M. Sikach,
in Covariant Methods in Theoretical Physics [in Russian],
(Institute of Physics, Belorussian Academy of Sciences, Minsk, 1991), p. 52.
%
\bibitem{b47} M.V. Galynsky and S.M. Sikach, Yad. Fiz. {\bf 54}, (1991) 1026.
%
\bibitem{b48} P.D. Gausmaecker, R. Gastmans et al., Phys. Lett. {\bf B105},
(1981) 215.
%
\bibitem{b49} F.A. Berends, R. Gastmanset al., Nucl. Phys. {\bf B206}, (1982)
53, 61.
%
\bibitem{b50} E.A. Kuraev et al., Yad. Fiz. {\bf 32}, (1980) 1059.
%
\bibitem{b51} C.I. Westbrook et al., Phys. Rev. {\bf A40}, (1989) 5489.
%
\bibitem{b52} E.A. Kuraev et al., Yad. Fiz.{\bf 51}, (1990) 1638.
%
\bibitem{b53} G.P. Lepage et al., Phys. Rev. {\bf A28}, (1983) 3090.
%
\bibitem{b54} M.V. Galynsky, O.N. Metelitsa, and S.M. Sikach,
in Covariant Methods in Theoretical Physics [in Russian],
(Institute of Physics, Belorussian Academy of Sciences, Minsk, 1991), p. 43.
%
\bibitem{b55} R.M. Drisco, Phys. Rev. {\bf 102}, (1956) 1542.
%
\bibitem{b56} J.B. Ye, B.Z. Yang et al., Phys. Lett. {\bf A133}, (1988) 309.
%
\bibitem{b57} F.I. Fedorov, The Theory of Gyrotropy [in Russian] (Nauka i
Tekhnika, Minsk, 1976).
%
\bibitem{b58} V.A. Petrun'kin, Fiz. Elem. Chast. At. Yadra, {\bf 12}, (1981)
 692 [Sov. J. Part. Nucl. {\bf 12}, (1981) 278].
%
\bibitem{b59} A.I. L'vov and V.A. Petrun'kin, Lecture Notes in Physics,
{\bf 365}, (1990) 123.
%
\bibitem{b60} A.I. L'vov, V.A. Petrun'kin, S.G. Popov, and
B.B. Wojtsekhowski, Preprint No 91-24, Budker-INP (1991).
%
\bibitem{b61} P.S. Isaev and I.S. Zlatev, Nucl. Phys. {\bf 16}, (1960) 608.
%
\bibitem{b62} B.B. Wojtsekhovski, A.I. L'vov et al.,"Project: Moscow-
Novosibirsk-Gottingen", Preprint Lebedev Physical Institute, Moscow (1992).
%
\bibitem{b63} P. Kroll, M. Schurmann and P.A.M. Guichon, Preprint WU B 95-09
(1995).
%
\bibitem{b64} R.A. Berg and C.N. Lindner, Nucl. Phys. {\bf 26}, (1961) 259.
%
\bibitem{b65} M.V. Galynsky, Preprint 695, Institute of Physics, Belorussian
Academy of Sciences, Minsk, (1994) [in Russian].
%
\bibitem{b66} A.I. L'vov, Yad. Fiz. {\bf 34}, (1981) 1075.
%
\bibitem{b67} A.A. Akhundov, D.Yu. Bardin, and N.M. Shumeiko, Yad. Fiz.
{\bf 44}, (1986) 1517 [Sov. J. Nucl. Phys {\bf 44}, (1986) 988].
%
\bibitem{b68} A.A. Akhundov, D.Yu. Bardin D.Yu. et al., Z. Phys. {\bf C45},
(1990) 645.
%
\bibitem{b69} R.L. Gluckstern , M.H. Hull, and G Breit, Phys. Rev. {\bf 90},
(1953) 1026.
%
\bibitem{b70} H. Olsen and L.C. Maximon, Phys. Rev. {\bf 114}, (1959) 887.
%
\bibitem{b71} J. Asai, H.S. Caplan, and L.C. Maximon,  Can. J. Phys. {\bf 66},
(1988) 1079.
%
\bibitem{b72} P.S. Isaev and I.S. Zlatev, Nuovo Cim. {\bf 13}, (1959) 1.
%
\bibitem{b73} M.V. Galynsky, Yad. Fiz. {\bf 58}, (1995) 701 [Phys. At. Nucl.
{\bf 58}, (1995) 644].
%
\bibitem{b74} C. Audit et.al., CEBAF proposal PR 93-050 (1993).
%
\bibitem{b75} J.F.J. Van den Brand, CEBAF proposal PR 94-011 (1994).
%
\bibitem{b76} S. Scherer, A.Yu. Korchin, and J.H. Koch, Report MKPH-T-96-4,
Mainz (1996).
%
\bibitem{b77} H.W. Fearing and S. Scherer, Report TRI-PP-96-28, MKPH-T-96-18,
Mainz (1996).
%
\bibitem{b78} A.I. Akhiezer and M.P. Rekalo, Electrodynamics of Hadrons [in
Russian] (Naukova Dumka, Kiev, 1977).
%
\bibitem{b79} M.V. Galynsky, D.O. Krimer, and M.I. Levchuk,
in Covariant Methods in Theoretical Physics [in Russian],
(Institute of Physics, Belarusian Academy of Sciences, Minsk, 1997), p. 56.
%
\bibitem{b80} I.F. Ginzburg et al., Yad. Fiz. {\bf 38}, (1983) 372
[Sov. J. Nucl. Phys. {\bf 38} (1983) 222].
%
\bibitem{b81} I.F. Ginzburg et al., Pis'ma Zh. Eksp. Teor. Fiz. {\bf 34},
(1981) 514 [JETP Lett. {\bf 34} (1981) 491].
%
\bibitem{b82} F.R. Arutyunyan and V.A. Tumanyan, Zh. Eksp. Teor. Fiz. {\bf 44},
(1963) 2100 [Sov. Phys. JETP {\bf 44}, (1963) 1412].
%
\bibitem{b83} I.F. Ginzburg et al., Yad. Fiz. {\bf 40}, (1984) 1495 [Sov.
J. Nucl. Phys. {\bf 40}, (1984) 949].
%
\bibitem{b84} I.F. Ginzburg et al., Yad. Fiz. {\bf 37}, (1983) 368
[Sov. J. Nucl. Phys. {\bf 37}, (1983) 222].
%
\bibitem{b85} A.I. Nikishev and V.I. Ritus, Trudy FIAN {\bf 111}, (1979)
[Proc.Lebedev Institute].
%
\bibitem{b86} M.V. Galynsky and S.M. Sikach, Zh. Eksp. Teor. Fiz. {\bf 101},
(1992) 828 [Sov. Phys. JETP {\bf 101}, (1992) 441].
%
\bibitem{b87} C. Bula, K.T. McDonald, et al., SLAC-PUB-7220, 7221 (1996),
SLAC-PUB-7564, (1997); Phys. Rev. Lett. {\bf 79} (1997) 1626.
%
\bibitem{b88} N.B. Narozhnyi and M.S. Fofanov, Zh. Eksp. Teor. Fiz.
{\bf 110}, (1996) 26 [JETP {\bf 83}, (1996) 14].
%
\bibitem{b89} I.F. Ginzburg et al., Yad.Fiz. {\bf 38}, (1983) 1021
[Sov. J. Nucl. Phys. {\bf 38}, (1983) 614].
%
\bibitem{b90} M.V. Galynsky and S.M. Sikach, Advances in Synergetics {\bf 8}
(1997) 60; Minsk, 1997, Edited by V. Kuvshinov \& G. Krylov.
Proceedings of the V Annual seminar "Nonlinear phenomena in complex system".
February 1996, Minsk, Belarus.
\end{thebibliography}
\end{document}